\documentclass[12pt,a4paper,fleqn]{article}
\usepackage{tikz}
\usepackage{caption}
\usepackage{lscape}

\usepackage{fullpage}
\usepackage{amsmath}
\usepackage{amsthm}
\usepackage{amssymb}
\usepackage{amscd}
\usepackage{plain}
\usepackage{graphicx}
\usepackage{t1enc,epsfig}
\usepackage[mathscr]{eucal}
\usepackage{epstopdf}
\usepackage[german,english]{babel}
\usepackage{stmaryrd}

\usepackage{xcolor}

\usepackage{wrapfig}
\usepackage{float}

\newtheorem{theorem}{Theorem}[section]

\newtheorem{corollary}[theorem]{Corollary}
\newtheorem{definition}[theorem]{Definition}
\newtheorem{example}[theorem]{Example}
\newtheorem{lemma}[theorem]{Lemma}
\newtheorem{proposition}[theorem]{Proposition}
\newtheorem{remark}[theorem]{Remark}
\newtheorem{conjecture}[theorem]{Conjecture}
\numberwithin{equation}{section}

\newtheorem*{thmI}{Theorem I}
\newtheorem*{thmII}{Theorem II}
\newtheorem*{thmIII}{Theorem III}
\newenvironment{proofI}{{\bf Proof of Theorem I. }}{\hfill$\rule{1ex}{1ex}$\par\medskip}
\newenvironment{proofII}{{\bf Proof of Theorem II. }}{\hfill$\rule{1ex}{1ex}$\par\medskip}
\newenvironment{proofIII}{{\bf Proof of Theorem III. }}{\hfill$\rule{1ex}{1ex}$\par\medskip}

\newtheorem*{lema}{Lemma 4.5.1}
\newtheorem*{lemb}{Lemma 4.5.2}
\newtheorem*{lemc}{Lemma 4.5.3}

\newtheorem*{lemd}{Lemma 4.6}
\newtheorem*{leme}{Lemma 4.7}
\newtheorem*{lemf}{Lemma 4.8}
\newtheorem*{lemg}{Lemma 4.9}
\newtheorem*{lemh}{Lemma 4.10}

\newtheorem*{thma}{Theorem 3.1A}
\newtheorem*{thmb}{Theorem 3.1B}
\newtheorem*{thmc}{Theorem 3.2A}
\newtheorem*{thmd}{Theorem 3.2B}
\newtheorem*{thme}{Theorem 3.3}
\newtheorem*{thmf}{Theorem 3.4}
\newtheorem*{thmg}{Theorem 4.1A}
\newtheorem*{thmh}{Theorem 4.1B}
\newtheorem*{thmi}{Theorem 6.1A}
\newtheorem*{thmj}{Theorem 6.1B}
\newtheorem*{thmk}{Theorem 6.2A}
\newtheorem*{thml}{Theorem 6.2B}
\newtheorem*{thmm}{Theorem 6.3A}
\newtheorem*{thmn}{Theorem 6.3B}
\newtheorem*{thmo}{Theorem 7A}
\newtheorem*{thmp}{Theorem 7B}
\newtheorem*{thmr}{Theorem 9.1A}
\newtheorem*{thmx}{Theorem 9.1B}
\newtheorem*{thmt}{Theorem 9.2A}
\newtheorem*{thmu}{Theorem 9.2B}
\newtheorem*{thmv}{Theorem 9.3A}
\newtheorem*{thmz}{Conjecture 9.3B}

\newtheorem*{ra}{Remark 3.1}
\newtheorem*{rb}{Remark 4.1}
\newtheorem*{rc}{Remark 4.2}
\newtheorem*{rdd}{Remark 4.3}
\newtheorem*{ree}{Remark 5.1}
\newtheorem*{rff}{Remark 7.1}
\newtheorem*{rgg}{Remark 7.2}
\newtheorem*{remnonumber}{Remark}
\newtheorem*{nnconjecture}{Conjecture}
\newtheorem*{remnonumberr}{Remarks}

\renewenvironment{proof}{{\bf Proof. }}{\hfill$\rule{1ex}{1ex}$\par\vskip 5 truemm}

\begin{document}

\newcommand{\xmark}{{\usefont{U}{eur}{m}{n}\symbol{"63}}}
\newcommand{\xmarkk}{{\usefont{U}{eur}{m}{n}\symbol{"43}}}

\newcommand{\bthm}{\begin{theorem}}
\newcommand{\ethm}{\end{theorem}}
\newcommand{\bd}{\begin{definition}}
\newcommand{\ed}{\end{definition}}
\newcommand{\bs}{\begin{proposition}}
\newcommand{\es}{\end{proposition}}
\newcommand{\bp}{\begin{proof}}
\newcommand{\ep}{\end{proof}}
\newcommand{\be}{\begin{equation}}
\newcommand{\ee}{\end{equation}}
\newcommand{\ul}{\underline}
\newcommand{\br}{\begin{remark}}
\newcommand{\er}{\end{remark}}
\newcommand{\bex}{\begin{example}}
\newcommand{\eex}{\end{example}}
\newcommand{\bc}{\begin{corollary}}
\newcommand{\ec}{\end{corollary}}
\newcommand{\bl}{\begin{lemma}}
\newcommand{\el}{\end{lemma}}
\newcommand{\bfj}{\begin{conjecture}}
\newcommand{\ej}{\end{conjecture}}

%%%%%%%%%%%%%%%%%%%%%%%%%%%%%%%%%%%%%%%%%%%%%%%%%%%%%%%%%
\newcommand{\bthmI}{\begin{thmI}}
\newcommand{\ethmI}{\end{thmI}}
\newcommand{\bthmII}{\begin{thmII}}
\newcommand{\ethmII}{\end{thmII}}
\newcommand{\bthmIII}{\begin{thmIII}}
\newcommand{\ethmIII}{\end{thmIII}}

\newcommand{\bthma}{\begin{thma}}
\newcommand{\ethma}{\end{thma}}
\newcommand{\bthmb}{\begin{thmb}}
\newcommand{\ethmb}{\end{thmb}}
\newcommand{\bthmc}{\begin{thmc}}
\newcommand{\ethmc}{\end{thmc}}
\newcommand{\bthmd}{\begin{thmd}}
\newcommand{\ethmd}{\end{thmd}}
\newcommand{\bthme}{\begin{thme}}
\newcommand{\ethme}{\end{thme}}
\newcommand{\bthmf}{\begin{thmf}}
\newcommand{\bthms}{\begin{thms}}
\newcommand{\ethmf}{\end{thmf}}
\newcommand{\ethms}{\end{thms}}
\newcommand{\bthmg}{\begin{thmg}}
\newcommand{\ethmg}{\end{thmg}}
\newcommand{\bthmh}{\begin{thmh}}
\newcommand{\ethmh}{\end{thmh}}
\newcommand{\bthmi}{\begin{thmi}}
\newcommand{\ethmi}{\end{thmi}}
\newcommand{\bthmj}{\begin{thmj}}
\newcommand{\ethmj}{\end{thmj}}
\newcommand{\bthmk}{\begin{thmk}}
\newcommand{\ethmk}{\end{thmk}}
\newcommand{\bthml}{\begin{thml}}
\newcommand{\ethml}{\end{thml}}
\newcommand{\bthmm}{\begin{thmm}}
\newcommand{\ethmm}{\end{thmm}}
\newcommand{\bthmn}{\begin{thmn}}
\newcommand{\ethmn}{\end{thmn}}
\newcommand{\bthmo}{\begin{thmo}}
\newcommand{\ethmo}{\end{thmo}}
\newcommand{\bthmp}{\begin{thmp}}
\newcommand{\ethmp}{\end{thmp}}
\newcommand{\bthmr}{\begin{thmr}}
\newcommand{\ethmr}{\end{thmr}}
\newcommand{\bthmx}{\begin{thmx}}
\newcommand{\ethmx}{\end{thmx}}
\newcommand{\bthmt}{\begin{thmt}}
\newcommand{\ethmt}{\end{thmt}}
\newcommand{\bthmu}{\begin{thmu}}
\newcommand{\ethmu}{\end{thmu}}
\newcommand{\bthmv}{\begin{thmv}}
\newcommand{\ethmv}{\end{thmv}}
\newcommand{\bthmz}{\begin{thmz}}
\newcommand{\ethmz}{\end{thmz}}

\newcommand{\blema}{\begin{lema}}
\newcommand{\elema}{\end{lema}}
\newcommand{\blemb}{\begin{lemb}}
\newcommand{\elemb}{\end{lemb}}
\newcommand{\blemc}{\begin{lemc}}
\newcommand{\elemc}{\end{lemc}}
\newcommand{\blemd}{\begin{lemd}}
\newcommand{\elemd}{\end{lemd}}
\newcommand{\bleme}{\begin{leme}}
\newcommand{\eleme}{\end{leme}}
\newcommand{\blemf}{\begin{lemf}}
\newcommand{\elemf}{\end{lemf}}
\newcommand{\blemg}{\begin{lemg}}
\newcommand{\elemg}{\end{lemg}}
\newcommand{\blemh}{\begin{lemh}}
\newcommand{\elemh}{\end{lemh}}

\newcommand{\bra}{\begin{ra}}
\newcommand{\era}{\end{ra}}
\newcommand{\brb}{\begin{rb}}
\newcommand{\erb}{\end{rb}}
\newcommand{\brc}{\begin{rc}}
\newcommand{\erc}{\end{rc}}
\newcommand{\brdd}{\begin{rdd}}
\newcommand{\erdd}{\end{rdd}}
\newcommand{\bree}{\begin{ree}}
\newcommand{\eree}{\end{ree}}
\newcommand{\brff}{\begin{rff}}
\newcommand{\erff}{\end{rff}}
\newcommand{\brgg}{\begin{rgg}}
\newcommand{\ergg}{\end{rgg}}
\newcommand{\bremnn}{\begin{remnonumber}}
\newcommand{\eremnn}{\end{remnonumber}}
\newcommand{\bremnnn}{\begin{remnonumberr}}
\newcommand{\eremnnn}{\end{remnonumberr}}

%%%%%%%%%%%%%%%%%%%%%%%%%%%%%%%%%%%%%%%%%%%%%%%%%%%%%%%

\newcommand{\bprfI}{\begin{proofI}}
\newcommand{\eprfI}{\end{proofI}}

\newcommand{\bprfII}{\begin{proofII}}
\newcommand{\eprfII}{\end{proofII}}

\newcommand{\bprfIII}{\begin{proofIII}}
\newcommand{\eprfIII}{\end{proofIII}}

%%%%%%%%%%%%%%%%%%%%%%%%%%%%%%%%%%%%%%%%%%%%%%%%%%%%%%%%%%%%

\newcommand{\eurcc}{{\usefont{U}{eur}{m}{n}\symbol{"63}}}
\newcommand{\eurc}{{\usefont{U}{eur}{m}{n}\symbol{"43}}}

\def\DVC{\text{\eurc}}
\def\BDVC{\text{\eurcc}}

\def\cre{\color{red}}

\def\diy{\displaystyle}

\def\Gam{{\Gamma}} \def\gam {\gamma}
\def\Del{\Delta} \def\del{\delta}
\def\vph{\varphi}
\def\ta {\tau}
\def\eps {\epsilon} \def\veps {\varepsilon}
\def\vrho {\varrho}
\def\Ups{\Upsilon} \def\ups{\upsilon}
\def\Om {\Omega} \def\om{\omega}

\def\pl {\partial}

\def\Ct{\complement}

\def\bbA{{\mathbb A}} \def\bbB{{\mathbb B}} 
\def\bbC{{\mathbb C}} \def\bbD{{\mathbb D}} 
\def\bbE{{\mathbb E}} \def\bbF{{\mathbb F}} \def\bbG{{\mathbb G}}
\def\bbH{{\mathbb H}} \def\bbI{{\mathbb I}} \def\bbJ{{\mathbb J}}
\def\bbL{{\mathbb L}} \def\bbM{{\mathbb M}} \def\bbN{{\mathbb N}}
\def\bbO{{\mathbb O}} \def\bbP{{\mathbb P}} \def\bbQ{{\mathbb Q}}
\def\bbR{{\mathbb R}} \def\bbS{{\mathbb S}} \def\bbT{{\mathbb T}}
\def\bbV{{\mathbb V}} \def\bbW{{\mathbb W}} \def\bbZ{{\mathbb Z}}

\def\cA{{\mathcal A}} \def\cB{{\mathcal B}} \def\cC{{\mathcal C}}
\def\cD{{\mathcal D}} \def\cE{{\mathcal E}} \def\cF{{\mathcal F}}
\def\cG{{\mathcal G}} \def\cH{{\mathcal H}} \def\cJ{{\mathcal J}}
\def\cP{{\mathcal P}} \def\cT{{\mathcal T}} \def\cU{{\mathcal U}}
\def\cV{{\mathcal V}} \def\cW{{\mathcal W}}\def\cN{{\mathcal N}}

\def\sA{\mathscr A} \def\sB{\mathscr B} \def\sC{\mathscr C}
\def\sD{\mathscr D} \def\sE{\mathscr E}
\def\sF{\mathscr F} \def\sG{\mathscr G} 
\def\sH{\mathscr H} \def\sI{\mathscr I}
\def\sL{\mathscr L} \def\sM{\mathscr M}
\def\sN{\mathscr N} \def\sO{\mathscr O}
\def\sP{\mathscr P} \def\sQ{\mathscr Q}  \def\sR{\mathscr R}
\def\sS{\mathscr S} \def\sT{\mathscr T}
\def\sU{\mathscr U} \def\sV{\mathscr V} \def\sW{\mathscr W}
\def\sX{\mathscr X} \def\sY{\mathscr Y} \def\sZ{\mathscr Z}

\def\sPA{{\sP\sA}}

\def\tA{{\tt A}} \def\tB{{\tt B}} \def\tC{{\tt C}}
\def\tD{{\tt D}} \def\tE{{\tt E}} \def\tF{{\tt F}} 
\def\tH{{\tt H}} \def\tK{{\tt K}} \def\tM{{\tt M}} 
\def\tQ{{\tt Q}} \def\tR{{\tt R}} \def\tS{{\tt S}}
\def\tT{{\tt T}} \def\tU{{\tt U}} \def\tV{{\tt V}}

\def\fA{{\mathfrak A}} \def\fB{{\mathfrak B}} \def\fC{{\mathfrak C}}
\def\fD{{\mathfrak D}} \def\fE{{\mathfrak E}} \def\fF{{\mathfrak F}}
\def\fW{{\mathfrak W}} \def\fX{{\mathfrak X}} \def\fY{{\mathfrak Y}}
\def\fZ{{\mathfrak Z}}

\def\frc{{\mathfrak c}}

\def\bfd{{\mathbf d}} \def\bfe{{\mathbf e}}  \def\bfh{{\mathbf h}} 
\def\bfi{{\mathbf i}} \def\bfj{{\mathbf j}} \def\bfk{{\mathbf k}}
\def\bfr{{\mathbf r}}
\def\bfn{{\mathbf n}} \def\bfu{{\mathbf u}} \def\bfv{{\mathbf v}}
\def\bfw{{\mathbf w}}
\def\bfx{{\mathbf x}} \def\bfy{{\mathbf y}} \def\bfz{{\mathbf z}}
\def\Bf1{{\mathbf 1}} \def\co{\complement}

\def\BA{{\mathbf A}} \def\BB{{\mathbf B}} \def\BC{{\mathbf C}} 
\def\BD{{\mathbf D}} \def\BE{{\mathbf E}} \def\BS{{\mathbf S}}
\def\BT{{\mathbf T}} \def\BX{{\mathbf X}} \def\BY{{\mathbf Y}}

\def\BAH{{\bf{AH}}} \def\BBH{{\bf{BH}}} \def\BCH{{\bf{CH}}}
\def\BDH{{\bf{DH}}} \def\BEH{{\bf{EH}}} \def\BSH{{\bf{SH}}}

\def\hatf{}

\def\bmu{{\mbox{\boldmath${\mu}$}}}
\def\bnu{{\mbox{\boldmath${\nu}$}}}
\def\bpi{{\mbox{\boldmath${\pi}$}}}

\def\bPhi{{\mbox{\boldmath${\Phi}$}}}

\def\sB{{\mathscr B}} \def\sL{{\mathscr L}}

\def\rA{{\rm A}}  \def\rB{{\rm B}}  \def\rC{{\rm C}}
\def\rD{{\rm  D}} \def\rE{{\rm  E}} \def\rF{{\rm  F}} 
 \def\rG{{\rm  G}} 
 \def\rlL{{\rm L}} 
\def\rH{{\rm  H}}  \def\rM{{\rm M}} 
\def\rO{{\rm  O}} \def\rR{{\rm  R}} 
 \def\rS{{\rm  S}} \def\rT{{\rm T}} 
\def\rU{{\rm U}}  \def\rV{{\rm  V}} 
\def\rd{{\rm d}} \def\re{{\rm e}} \def\rf{{\rm f}}
\def\rh{{\rm h}} \def\ri{{\rm i}} \def\rj{{\rm j}}
\def\rk{{\rm k}} \def\rl{{\rm l}} 
\def\rmm{{\rm m}} \def\rn{{\rm n}}
\def\rs{{\rm s}} \def\rt{{\rm t}} \def\ru{{\rm u}}
\def\rv{{\rm v}} \def\rx{{\rm x}} \def\ry{{\rm y}}  \def\rz{{\rm z}}

\def\rAC{{\rm{AC}}}
\def\rFCC{{\rm{FCC}}} \def\rBCC{{\rm{BCC}}}
\def\rHC{{\rm{HC}}} \def\rHCP{{\rm{HCP}}}
\def\rPC{{\rm{PC}}}
\def\rPLAC{{\rm{PLAC}}} \def\rEPGM{{\rm{EPGM}}}
\def\rPGS{{\rm{PGS}}} \def\RD{{\rm{R\D}}}
\def\rVC{{\rm{VC}}}

\def\rba{{\rm{(a)}}} \def\rbb{{\rm{(b)}}}
\def\rbc{{\rm{(c)}}} \def\rbd{{\rm{(d)}}} 
\def\rbe{{\rm{(e)}}} \def\rbf{{\rm{(f)}}}
\def\rbg{{\rm{(g)}}} \def\rbh{{\rm{(h)}}}
\def\rbi{{\rm{(i)}}} \def\rbii{{\rm{(ii)}}}
\def\rb3i{{\rm{(iii)}}} \def\rbiv{{\rm{(iv)}}} 
\def\rbv{{\rm{(v)}}} \def\rbvi{{\rm{(vi)}}} 

\def\ov{\overline} \def\ove{{\overline e}}
\def\ovp{{\overline p}} \def\ovv{{\overline v}}
\def\ovx{{\overline x}}
\def\und{\underline} \def\unf{{\underline f}}
\def\unv{{\underline v}}
\def\ubet{{\underline\beta}} \def\ugam{{\underline\gamma}}

\def\es {{\varnothing}}

%t {\bullet}
\def\cc {\circ}

\def\wt{\widetilde} \def\wh{\widehat}
\def\wtx{{\wt x}}

\def\wtD{{\wt D}}
\def\wtm{{\wt m}} \def\wtn{{\wt n}} \def\wtk{{\wt k}}

\def\be{\begin{equation}}
\def\ee{\end{equation}}

\def\beq{\begin{equation}}
\def\eeq{\end{equation}}

\def\beal{\begin{array}{l}} \def\beac{\begin{array}{c}} \def\bear{\begin{array}{r}}
\def\beacl{\begin{array}{cl}} \def\beall{\begin{array}{ll}}
\def\bealllll{\begin{array}{lllll}}
\def\beacr{\begin{array}{cr}}
\def\ena{\end{array}}

\def\bma{\begin{matrix}} \def\ema{\end{matrix}}
\def\bpma{\begin{pmatrix}} \def\epma{\end{pmatrix}}

\def\bcs{\begin{cases}} \def\ecs{\end{cases}}

\def\onwl{\operatornamewithlimits}

\def\diy{\displaystyle}

\def\cF{\mathcal F} \def\cG{\mathcal G} \def\cI{\mathcal I}
\def\cL{\mathcal L} \def\cO{\mathcal O}
\def\cP{\mathcal P} 

\def\alp{{\alpha}} \def\bet{{\beta}}
\def\Gam{{\Gamma}} \def\gam{{\gamma}}
\def\Del{{\Delta}} \def\del{{\delta}}  \def\odel{{\overline\delta}}
\def\kap{{\kappa}}
\def\vphi{{\varphi}}
\def\eps{{\epsilon}} \def\veps{{\varepsilon}}
\def\vrho{{\varrho}} \def\vpi{{\varpi}}
\def\Lam{{\Lambda}} \def\lam{{\lambda}} 
\def\vpi{{\varpi}} \def\vtheta{{\vartheta}}
\def\Om{{\Omega}} \def\om{{\omega}} 

\def\pa {\partial}
\def\comp{\complement}

\def\bfZ{{\mathbf Z}}
\def\b0f{{\mathbf 0}}

\def\rSp {{\rm Supp}}

\def\D{D}
\def\upharp{\hskip-3pt\upharpoonright}

\def\c  {{\mathchar"0\hexnumber@\msafam7B}}
\def\es {{\varnothing}}

%t {\bullet}
\def\cc {\circ}

\def\io{\iota}
\def\rInt{{\rm{Int}}}
\def\rExt{{\rm{Ext}}}
\def\wh{\widehat} \def\wt{\widetilde}

\def\cl{\centerline}

%%%%%%%%%%%%%%%%%%%%%%%%%%%%  201068247927  
%%%%%%%%%%%%%%%%%%%%%%%%%%%%%%%%%%%%%
%% 86602540378443864676372317075294   M251

\def\Sgie{\raise-1.5ex\hbox{${{\displaystyle \sum}^{\sharp} \atop {\scriptstyle \{\g_m\}^{\ex} \in V}}$}}

%options for floating pictures

\makeatletter
 \def\fps@figure{htbp}
\makeatother

%definitions for drawing with tikz
\def\rr{0.86602540378443864676372317075294} %\sqrt{3/4}

\def\hexagongrid{
%\clip[yscale=sqrt(3/4), xslant=0.5] (0, \n) -- (\n,0) -- (\n, -\n) -- (0, -\n) -- (-\n,0) -- (-\n, \n) -- cycle;
\draw [yscale=sqrt(3/4), xslant=0.5] (-\n, -\n) grid (\n, \n);
\draw [yscale=sqrt(3/4), xslant=-0.5] (-\n, -\n) grid (\n, \n);
}

\def\rectangulargrid{
\clip[yscale=sqrt(3/4)] (-\n, -\n) rectangle (\n, \n);
\draw [yscale=sqrt(3/4), xslant=0.5] (-2 * \n, -\n) grid (2 * \n, \n);
\draw [yscale=sqrt(3/4), xslant=-0.5] (-2 * \n, -\n) grid (2 * \n, \n);
}

\def\sublattice{
\clip[yscale=sqrt(3/4), xslant=0.5] (0, \n+\nn) -- (\n+\nn,0) -- (\n+\nn, -\n-\nn) -- (0, -\n-\nn) -- (-\n-\nn,0) -- (-\n-\nn, \n+\nn) -- cycle;
\foreach \x in {-\n,...,\n}
 \foreach \y in {-\n,...,\n}
 {
  \def\xx{\x * \aa + 0.5 * \x * \bb - 0.5 * \y * \bb + 0.5 * \y * \aa}
  \def\yy{\rr * \x * \bb + \rr * \y * \aa + \rr * \y * \bb}

  \shade[shading=ball, ball color=black] (\xx, \yy) circle (.2);
 }
}

\def\sublatticeinrectangle{
\clip[yscale=sqrt(3/4)] (-\nn, -\nn) rectangle (\nn, \nn);
\foreach \x in {-\n,...,\n}
 \foreach \y in {-\n,...,\n}
 {
  \def\xx{\x * \aa + 0.5 * \x * \bb - 0.5 * \y * \bb + 0.5 * \y * \aa}
  \def\yy{\rr * \x * \bb + \rr * \y * \aa + \rr * \y * \bb}

  \shade[shading=ball, ball color=black] (\xx + \dx + 0.5 * \dy, \yy + \rr * \dy) circle (.3);
 }
}

\def\inclinedgrid{
	\draw [yscale=\bb * sqrt(3/4), xslant=\aa + 0.5 * \bb, xscale=\aa * \aa / \bb + \aa  + \bb , yslant=-\aa / \bb, ultra thick] (-\n, -\n) grid (\n, \n);
}

\def\rectsub-lattice{
\clip[yscale=sqrt(3/4)] (-\nn, -\nn) rectangle (\nn, \nn);
\foreach \x in {-\n,...,\n}
 \foreach \y in {-\n,...,\n}
 {
  \filldraw[yscale=sqrt(3/4), xslant=0.5, \cc] (\ss * \x + \dx, \ss * \y + \dy) rectangle (\ss * \x + \ss - 1 + \dx, \ss * \y + \ss - 1 + \dy);

\definecolor{gray1}{gray}{0.1}
\definecolor{gray2}{gray}{0.2}
\definecolor{gray3}{gray}{0.3}
\definecolor{gray4}{gray}{0.4}
\definecolor{gray5}{gray}{0.5}
\definecolor{gray6}{gray}{0.6}
\definecolor{gray7}{gray}{0.7}
\definecolor{gray8}{gray}{0.8}
\definecolor{gray9}{gray}{0.9}
 }
}

%%%%%%%%%%%%%%%%%%%%%%%%%%%%%%%%%%%%%%%%%%%%%%%%%%% 

\def\FigureFourOne
{\begin{figure}[H]
\begin{center}

\begin{tikzpicture}[scale=1]
\def\r32{0.86602540378443864676372317075294}
\def\n{6.5}

%\filldraw [gray] (0, 0) -- (2.5, {sqrt(3/4)}) -- (2, -{sqrt(3)}) -- (-0.5, -{sqrt(27/4)}) -- cycle;

%\def\n{6.5}
%\clip[yscale=sqrt(3/4), xslant=0.5] (0, \n) -- (\n, 0) -- (\n, -\n) -- (0, -\n) -- (-\n, 0) -- (-\n, \n) -- cycle;
%\foreach \x in {-2,...,2}
%\foreach \y in {-2,...,2}
%{\def\xx{2.5 * \x +  0.5 * \y}
%\def\yy{\r32 * \x + \r32 * 3 * \y}
%\shade[shading=ball, ball color=black] (1 + \xx, \yy) circle (.2);
%}

%\shade[shading=ball, ball color=black] (-3.5,{sqrt(147/4)}) circle (.2);
%\shade[shading=ball, ball color=white] (-2.5,{sqrt(147/4)}) circle (.4);

\def\n{8}
\clip[yscale=sqrt(3/4), xslant=0.5] (0, \n) -- (\n, 0) -- (\n, -\n) -- (0, -\n) -- (-\n, 0) -- (-\n, \n) -- cycle;
\draw [yscale=sqrt(3/4), xslant=0.5] (-\n, -\n) grid (\n, \n);
\draw [yscale=sqrt(3/4), xslant=-0.5] (-\n, -\n) grid (\n, \n);

%\draw [scale=sqrt(7), rotate=19.1, yscale=sqrt(3/4), xslant=0.5, color=red, very thick] (0, 0) -- (0, 1) -- (1,1) 
%-- (1,0) -- (0,0);

%\draw [scale=sqrt(7), rotate=19.1, yscale=sqrt(3/4), xslant=0.5, color=purple, line width=1mm] (1, -1) -- (1, -2)
%-- (0,-2) -- (0,-1) -- (1,-1);

%\draw [scale=sqrt(7), yscale=sqrt(3/4), xslant=0.5, color=teal, line width=1mm] (0, 0) -- (-1, 0) -- (-1,-1) 
%-- (0,-1) -- (0,0);
%=======================================

%\draw [scale=sqrt(7), yscale=sqrt(3/4), xslant=0.5, color=red, line width=1mm] (-0.76, 2.64)
%-- (2.66, -0.74) -- (-1.90, -0.74) -- (-1.90, 2.64) -- (-0.76, 2.64);

%%%%%%%%%%%%%%%%%%%%%

\draw [scale=sqrt(7), color=brown, line width=1mm] 
(0,0.66)--(0.57,0.99);

\draw [scale=sqrt(7), color=brown, line width=1mm] 
(0.57,0.99)--(1.14,0.66);

\draw [scale=sqrt(7), color=brown, line width=1mm] 
(0,0.66)--(0,0);

\draw [scale=sqrt(7), color=brown, line width=1mm] 
(1.14,0.66)--(1.14,0);

\draw [scale=sqrt(7), color=brown, line width=1mm] 
(-0.57,-0.33)--(0,0);

\draw [scale=sqrt(7), color=brown, line width=1mm] 
(0,0)--(0.57,-0.33);

\draw [scale=sqrt(7), color=brown, line width=1mm] 
(0.57,-0.33)--(1.14,0);

\draw [scale=sqrt(7), color=brown, line width=1mm] 
(1.14,0)--(1.71,-0.33);

\draw [scale=sqrt(7), color=brown, line width=1mm] 
(0.57,-0.33)--(0.57,-0.99);

%%%%%%%%%%%%%%%%%%%%%

\draw [scale=sqrt(7), color=teal, line width=1mm] 
(-1.90,1.96)--(1.90,1.96);

\draw [scale=sqrt(7), color=teal, line width=1mm] 
(-2.47,0.98)--(2.47,0.98);

\draw [scale=sqrt(7), color=teal, line width=1mm] 
(-3.04,0)--(3.04,0);

\draw [scale=sqrt(7), color=teal, line width=1mm] 
(-2.47,-0.98)--(2.86,-0.98);

\draw [scale=sqrt(7), color=teal, line width=1mm] 
(-1.90,-1.96)--(1.90,-1.96);

%%%%%%%%%%%%%%%%%%%%%

\draw [scale=sqrt(7), color=teal, line width=1mm] 
(-2.66,-0.68)--(-0.76,2.62);

\draw [scale=sqrt(7), color=teal, line width=1mm] 
(-2.08,-1.64)--(0.38,2.62);

\draw [scale=sqrt(7), color=teal, line width=1mm] 
(-1.50,-2.62)--(1.50,2.62);

\draw [scale=sqrt(7), color=teal, line width=1mm] 
(-0.38,-2.62)--(2.08,1.64);

\draw [scale=sqrt(7), color=teal, line width=1mm] 
(0.76,-2.62)--(2.66,0.68);

%%%%%%%%%%%%%%%%%%%%%

\draw [scale=sqrt(7), color=teal, line width=1mm] 
(-2.66,0.68)--(-0.76,-2.62);

\draw [scale=sqrt(7), color=teal, line width=1mm] 
(-2.08,1.64)--(0.38,-2.62);

\draw [scale=sqrt(7), color=teal, line width=1mm] 
(-1.50,2.62)--(1.50,-2.62);

\draw [scale=sqrt(7), color=teal, line width=1mm] 
(-0.38,2.62)--(2.08,-1.64);

\draw [scale=sqrt(7), color=teal, line width=1mm] 
(0.76,2.62)--(2.66,-0.68);

%%%%%%%%%%%%%%%%%%%%%

\draw [scale=sqrt(7), color=red, line width=1mm] 
 (-1.48,2.62)-- (1.48,2.62);

\draw [scale=sqrt(7), color=red, line width=1mm] 
 (-2.09,1.65)-- (2.09,1.65);
 
 \draw [scale=sqrt(7), color=red, line width=1mm] 
 (-2.66,0.66)--(2.66,0.66);
 
 \draw [scale=sqrt(7), color=red, line width=1mm] 
 (-2.82,-0.33)--(2.82,-0.33);

\draw [scale=sqrt(7), color=red, line width=1mm] 
 (-2.26,-1.32)--(2.26,-1.32);

\draw [scale=sqrt(7), color=red, line width=1mm] 
 (-1.68,-2.30)-- (1.68,-2.30);
  
\draw [scale=sqrt(7), color=red, line width=1mm] 
 (-3.02,0)--(-1.51,-2.62);

\draw [scale=sqrt(7), color=red, line width=1mm] 
(-2.46,0.98)--(-0.38,-2.62);

\draw [scale=sqrt(7), color=red, line width=1mm] 
(-1.90,1.96)--(0.76,-2.62);

\draw [scale=sqrt(7), color=red, line width=1mm] 
(-1.14,2.62)--(1.68,-2.30);

\draw [scale=sqrt(7), color=red, line width=1mm] 
(0,2.62)--(2.26,-1.32);

\draw [scale=sqrt(7), color=red, line width=1mm] 
(-2.26,-1.32)--(0,2.62);

\draw [scale=sqrt(7), color=red, line width=1mm] 
(-1.68,-2.30)--(1.14,2.62);

\draw [scale=sqrt(7), color=red, line width=1mm] 
(-0.76,-2.62)--(1.90,1.96);

\draw [scale=sqrt(7), color=red, line width=1mm] 
(0.38,-2.62)--(2.46,0.98);

\draw [scale=sqrt(7), color=red, line width=1mm] 
(1.51,-2.62)--(3.02,0);

%%%%%%%%%%%%%%%%%%%%% 

\filldraw[scale=sqrt(7), color=white] (-1.14,2.62) circle (0.1);
\filldraw[scale=sqrt(7), color=red] (-1.14,2.62) circle (.06);

\filldraw[scale=sqrt(7), color=white] (0,2.62) circle (0.1);
\filldraw[scale=sqrt(7), color=red] (0,2.62) circle (.06);

\filldraw[scale=sqrt(7), color=white] (1.14,2.62) circle (0.1);
\filldraw[scale=sqrt(7), color=red] (1.14,2.62) circle (.06);

\filldraw[scale=sqrt(7), color=white] (-1.71,1.65) circle (0.1);
\filldraw[scale=sqrt(7), color=red] (-1.71,1.65) circle (.06);

\filldraw[scale=sqrt(7), color=white] (-0.57,1.65) circle (0.1);
\filldraw[scale=sqrt(7), color=red] (-0.57,1.65) circle (.06);

\filldraw[scale=sqrt(7), color=white] (0.57,1.65) circle (0.1);
\filldraw[scale=sqrt(7), color=red] (0.57,1.65) circle (.06);

\filldraw[scale=sqrt(7), color=white] (1.71,1.65) circle (0.1);
\filldraw[scale=sqrt(7), color=red] (1.71,1.65) circle (.06);

\filldraw[scale=sqrt(7), color=white] (-2.28,0.66) circle (0.1);
\filldraw[scale=sqrt(7), color=red] (-2.28,0.66) circle (.06);

\filldraw[scale=sqrt(7), color=white] (-1.14,0.66) circle (0.1);
\filldraw[scale=sqrt(7), color=red] (-1.14,0.66) circle (.06);

\filldraw[scale=sqrt(7), color=white] (0,0.66) circle (0.1);
\filldraw[scale=sqrt(7), color=red] (0,0.66) circle (.06);

\filldraw[scale=sqrt(7), color=white] (1.14,0.66) circle (0.1);
\filldraw[scale=sqrt(7), color=red] (1.14,0.66) circle (.06);

\filldraw[scale=sqrt(7), color=white] (2.28,0.66) circle (0.1);
\filldraw[scale=sqrt(7), color=red] (2.28,0.66) circle (.06);

\filldraw[scale=sqrt(7), color=white] (-2.84,-0.33) circle (0.1);
\filldraw[scale=sqrt(7), color=red] (-2.84,-0.33) circle (.06);

\filldraw[scale=sqrt(7), color=white] (-1.71,-0.33) circle (0.1);
\filldraw[scale=sqrt(7), color=red] (-1.71,-0.33) circle (.06);

\filldraw[scale=sqrt(7), color=white] (-0.57,-0.33) circle (0.1);
\filldraw[scale=sqrt(7), color=red] (-0.57,-0.33) circle (.06);

\filldraw[scale=sqrt(7), color=white] (0.57,-0.33) circle (0.1);
\filldraw[scale=sqrt(7), color=red] (0.57,-0.33) circle (.06);

\filldraw[scale=sqrt(7), color=white] (1.71,-0.33) circle (0.1);
\filldraw[scale=sqrt(7), color=red] (1.71,-0.33) circle (.06);

\filldraw[scale=sqrt(7), color=white] (2.84,-0.33) circle (0.1);
\filldraw[scale=sqrt(7), color=red] (2.84,-0.33) circle (.06);

\filldraw[scale=sqrt(7), color=white] (-2.26,-1.32) circle (0.1);
\filldraw[scale=sqrt(7), color=red] (-2.26,-1.32) circle (.06);

\filldraw[scale=sqrt(7), color=white] (-1.14,-1.32) circle (0.1);
\filldraw[scale=sqrt(7), color=red] (-1.14,-1.32) circle (.06);

\filldraw[scale=sqrt(7), color=white] (0,-1.32) circle (0.1);
\filldraw[scale=sqrt(7), color=red] (0,-1.32) circle (.06);

\filldraw[scale=sqrt(7), color=white] (1.14,-1.32) circle (0.1);
\filldraw[scale=sqrt(7), color=red] (1.14,-1.32) circle (.06);

\filldraw[scale=sqrt(7), color=white] (2.26,-1.32) circle (0.1);
\filldraw[scale=sqrt(7), color=red] (2.26,-1.32) circle (.06);

\filldraw[scale=sqrt(7), color=white] (-1.69,-2.30) circle (0.1);
\filldraw[scale=sqrt(7), color=red] (-1.69,-2.30) circle (.06);

\filldraw[scale=sqrt(7), color=white] (-0.57,-2.30) circle (0.1);
\filldraw[scale=sqrt(7), color=red] (-0.57,-2.30) circle (.06);

\filldraw[scale=sqrt(7), color=white] (0.57,-2.30) circle (0.1);
\filldraw[scale=sqrt(7), color=red] (0.57,-2.30) circle (.06);

\filldraw[scale=sqrt(7), color=white] (1.69,-2.30) circle (0.1);
\filldraw[scale=sqrt(7), color=red] (1.69,-2.30) circle (.06);

%%%%%%%%%%%%%%%%%%%%%

\filldraw[scale=sqrt(7), color=white] (-1.14,1.96) circle (0.1);
\filldraw[scale=sqrt(7), color=teal] (-1.14,1.96) circle (.06);

\filldraw[scale=sqrt(7), color=white] (0,1.96) circle (0.1);
\filldraw[scale=sqrt(7), color=teal] (0,1.96) circle (.06);

\filldraw[scale=sqrt(7), color=white] (1.14,1.96) circle (0.1);
\filldraw[scale=sqrt(7), color=teal] (1.14,1.96) circle (.06);

\filldraw[scale=sqrt(7), color=white] (-1.71,0.98) circle (0.1);
\filldraw[scale=sqrt(7), color=teal] (-1.71,0.98) circle (.06);

\filldraw[scale=sqrt(7), color=white] (-0.57,0.98) circle (0.1);
\filldraw[scale=sqrt(7), color=teal] (-0.57,0.98) circle (.06);

\filldraw[scale=sqrt(7), color=white] (0.57,0.98) circle (0.1);
\filldraw[scale=sqrt(7), color=teal] (0.57,0.98) circle (.06);

\filldraw[scale=sqrt(7), color=white] (1.71,0.98) circle (0.1);
\filldraw[scale=sqrt(7), color=teal] (1.71,0.98) circle (.06);

\filldraw[scale=sqrt(7), color=white] (-2.28,0) circle (0.1);
\filldraw[scale=sqrt(7), color=teal] (-2.28,0) circle (.06);

\filldraw[scale=sqrt(7), color=white] (-1.14,0) circle (0.1);
\filldraw[scale=sqrt(7), color=teal] (-1.14,0) circle (.06);

\filldraw[scale=sqrt(7), color=white] (0,0) circle (0.1);
\filldraw[scale=sqrt(7), color=teal] (0,0) circle (.06);

\filldraw[scale=sqrt(7), color=white] (1.14,0) circle (0.1);
\filldraw[scale=sqrt(7), color=teal] (1.14,0) circle (.06);

\filldraw[scale=sqrt(7), color=white] (2.28,0) circle (0.1);
\filldraw[scale=sqrt(7), color=teal] (2.28,0) circle (.06);

\filldraw[scale=sqrt(7), color=white] (-1.71,-0.98) circle (0.1);
\filldraw[scale=sqrt(7), color=teal] (-1.71,-0.98) circle (.06);

\filldraw[scale=sqrt(7), color=white] (-0.57,-0.98) circle (0.1);
\filldraw[scale=sqrt(7), color=teal] (-0.57,-0.98) circle (.06);

\filldraw[scale=sqrt(7), color=white] (0.57,-0.98) circle (0.1);
\filldraw[scale=sqrt(7), color=teal] (0.57,-0.98) circle (.06);

\filldraw[scale=sqrt(7), color=white] (1.71,-0.98) circle (0.1);
\filldraw[scale=sqrt(7), color=teal] (1.71,-0.98) circle (.06);

\filldraw[scale=sqrt(7), color=white] (-1.14,-1.96) circle (0.1);
\filldraw[scale=sqrt(7), color=teal] (-1.14,-1.96) circle (.06);

\filldraw[scale=sqrt(7), color=white] (0,-1.96) circle (0.1);
\filldraw[scale=sqrt(7), color=teal] (0,-1.96) circle (.06);

\filldraw[scale=sqrt(7), color=white] (1.14,-1.96) circle (0.1);
\filldraw[scale=sqrt(7), color=teal] (1.14,-1.96) circle (.06);

%%%%%%%%%%%%%%%%%%%%%

\filldraw[scale=sqrt(7), color=white] (0.57,0.33) circle (0.1);
\filldraw[scale=sqrt(7), color=pink] (0.57,0.33) circle (.06);

%\filldraw[scale=sqrt(7), color=white] (-1.71,0.98) circle (0.1);
%\filldraw[scale=sqrt(7), color=teal] (-1.71,0.98) circle (.06);

\end{tikzpicture}\end{center}
\def\rx{{\rm x}} \def\vphi{{\varphi}} 
\def\sB{{\mathscr B}} \def\sH{{\mathscr H}}
\def\bbZ{{\mathbb Z}}

{\large{\bf Figure 4.1. A dominating excitation for $D^2=5$}}.\smallskip

\small{The figure shows the projection to plane $\rx_1+\rx_2+\rx_3=0$ of a PC $\vphi\in\sH^{(5)}$.
The thin lines in the background form the triangular ${\sqrt{2/3}}$-lattice $T_{0,0}$. The green  lines and 
circles indicate the triangular ${\sqrt{6}}$-mesh $\tau^{(6)}_{0,0,0}\subset T_{0,0}$ and the positions
of particles in $\tau^{(6)}_{0,0,0}$\,. The red lines and circles indicate the projections of the 
triangular ${\sqrt{6}}$-meshes $\tau^{(6)}_{0,\pm 3,1}\subset T_{0,\pm 3}$ and the positions
of particles in $\tau^{(6)}_{0,\pm 3,1}$\,. The brown lines of length $\sqrt{5}$ join sites from neighboring meshes
$\tau^{(6)}_{0,0,0}$ and $\tau^{(6)}_{0,\pm 3,1}$. A hexagon with brown sides encircles the projection of 
an octahedron, and triples of brown segments indicate the projections of sides of tetrahedrons. None 
of the octahedrons or tetrahedrons are equilateral: they are all oblate in the direction 
orthogonal to the plane. The pink circle at the center of the green triangle
indicates the position of a local $u^{-2}$-excitation of type (IIa) which removes three particles
at the vertices of the corresponding green triangle. Such an excitation has the highest frequency 
of occurrence in PC $\vphi\in\sH^{(5)}$ among all periodic PCs.}
\end{figure}
}

%%%%%%%%%%%%%%%%%%%%%%%%%%%%%%%%%%%%%%%%%%%%%%%%%%% 

\def\FigureFourTwo
{\begin{figure}[H]
\begin{center}

\begin{tikzpicture}[scale=1]
\def\r32{0.86602540378443864676372317075294}
\def\n{6.5}

\def\n{8}
\clip[yscale=sqrt(3/4), xslant=0.5] (0, \n) -- (\n, 0) -- (\n, -\n) -- (0, -\n) -- (-\n, 0) -- (-\n, \n) -- cycle;
\draw [yscale=sqrt(3/4), xslant=0.5] (-\n, -\n) grid (\n, \n);
\draw [yscale=sqrt(3/4), xslant=-0.5] (-\n, -\n) grid (\n, \n);

%\draw [scale=sqrt(7), rotate=19.1, yscale=sqrt(3/4), xslant=0.5, color=red, very thick] (0, 0) -- (0, 1) -- (1,1) 
%-- (1,0) -- (0,0);

%\draw [scale=sqrt(7), rotate=19.1, yscale=sqrt(3/4), xslant=0.5, color=yellow, very thick] (0, 0) -- (1, 0) -- (1,-1) 
%-- (0,-1) -- (0,0);

%\draw [scale=sqrt(7), rotate=19.1, yscale=sqrt(3/4), xslant=0.5, color=blue, very thick] (0, 0) -- (-1, 0) -- (-1,1) 
%-- (0,1) -- (0,0);

%\draw [scale=sqrt(7), rotate=19.1, yscale=sqrt(3/4), xslant=0.5, color=brown, very thick] (1, 0) -- (2, 0) -- (2,-1) 
%-- (1,-1) -- (1,0);

%\draw [scale=sqrt(7), rotate=19.1, yscale=sqrt(3/4), xslant=0.5, color=cyan, very thick] (1, -1) -- (2, -1) -- (2,-2) 
%-- (1,-2) -- (1,-1);

%\draw [scale=sqrt(7), rotate=19.1, yscale=sqrt(3/4), xslant=0.5, color=purple, line width=1mm] (1, -1) -- (1, -2)
%-- (0,-2) -- (0,-1) -- (1,-1);

%\draw [scale=sqrt(7), yscale=sqrt(3/4), xslant=0.5, color=teal, line width=1mm] (0, 0) -- (-1, 0) -- (-1,-1) 
%-- (0,-1) -- (0,0);
%=======================================
%-----------------------------------------

\draw [scale=sqrt(7), yscale=sqrt(3/4), xslant=0.5, color=teal, line width=1mm] (0.76, -2.65) 
-- (1.90, -2.65) -- (1.89, 0.76) -- (0.75,1.90) -- (-2.63,1.90) -- (-2.63,0.76) -- (0.76, -2.65); 
\draw [scale=sqrt(7), yscale=sqrt(3/4), xslant=0.5, color=teal, line width=1mm] (-2.63,0.76) 
-- (1.90, 0.76);
\draw [scale=sqrt(7), yscale=sqrt(3/4), xslant=0.5, color=teal, line width=1mm] (-1.52,-0.38)
-- (1.89,-0.38);
\draw [scale=sqrt(7), yscale=sqrt(3/4), xslant=0.5, color=teal, line width=1mm] (-2.64,1.90)
-- (1.90, -2.65);
\draw [scale=sqrt(7), yscale=sqrt(3/4), xslant=0.5, color=teal, line width=1mm] (0.75,1.90)
-- (0.76, -2.65);
\draw [scale=sqrt(7), yscale=sqrt(3/4), xslant=0.5, color=teal, line width=1mm]  (-1.51,1.90)
-- (1.88,-1.5);
\draw [scale=sqrt(7), yscale=sqrt(3/4), xslant=0.5, color=teal, line width=1mm] (-0.37,1.90) 
-- (-0.38,-1.52);

%-----------------------------------------

\draw [scale=sqrt(7), yscale=sqrt(3/4), xslant=0.5, color=red, line width=1mm] (-0.76, 2.64)
-- (2.66, -0.74) -- (-1.90, -0.74) -- (-1.90, 2.64) -- (-0.76, 2.64);
\draw [scale=sqrt(7), yscale=sqrt(3/4), xslant=0.5, color=red, line width=1mm]  (2.66, -0.74) 
-- (2.65,-1.90) -- (-0.76,-1.90)  -- (-1.90, -0.74);
\draw [scale=sqrt(7), yscale=sqrt(3/4), xslant=0.5, color=red, line width=1mm] (-1.90, 2.64)
--  (2.65,-1.90);
\draw [scale=sqrt(7), yscale=sqrt(3/4), xslant=0.5, color=red, line width=1mm] (-0.76, 2.64)
-- (-0.76,-1.90);
\draw [scale=sqrt(7), yscale=sqrt(3/4), xslant=0.5, color=red, line width=1mm] (-1.89,0.38)
--  (1.52,0.38);
\draw [scale=sqrt(7), yscale=sqrt(3/4), xslant=0.5, color=red, line width=1mm] (-1.92,1.52)
-- (1.52,-1.88);
\draw [scale=sqrt(7), yscale=sqrt(3/4), xslant=0.5, color=red, line width=1mm] (0.38,-1.90) 
-- (0.38,1.52);

%-----------------------------------------

\draw [scale=sqrt(7), yscale=sqrt(3/4), xslant=0.5, color=blue, line width=1mm] (-1.51,1.90) -- (-1.90, 2.64) 
-- (-2.64,1.90)--(-1.92,1.52) -- (-1.51,1.90);
\draw [scale=sqrt(7), yscale=sqrt(3/4), xslant=0.5, color=blue, line width=1mm] (-2.63,0.76) -- (-1.90, -0.74)
-- (-1.52,-0.38) -- (-1.89,0.38) -- (-2.63,0.76);
\draw [scale=sqrt(7), yscale=sqrt(3/4), xslant=0.5, color=blue, line width=1mm] (-0.76,-1.90) -- (0.76, -2.65)
-- (0.38,-1.90) -- (-0.38,-1.5) -- (-0.76,-1.88);
\draw [scale=sqrt(7), yscale=sqrt(3/4), xslant=0.5, color=blue, line width=1mm] (1.90, -2.65) -- (2.65,-1.90)
-- (1.88,-1.5) -- (1.52,-1.88) -- (1.90, -2.65);
\draw [scale=sqrt(7), yscale=sqrt(3/4), xslant=0.5, color=blue, line width=1mm] (2.66, -0.74) -- (1.90, 0.76)
-- (1.52,0.38) -- (1.90,-0.38) -- (2.66, -0.74);
\draw [scale=sqrt(7), yscale=sqrt(3/4), xslant=0.5, color=blue, line width=1mm] (0.76,1.90) -- (-0.76, 2.64)
-- (-0.37,1.90) -- (0.38,1.52) -- (0.75,1.90);

%-----------------------------------------

\filldraw[scale=sqrt(7), color=white] (0,1.96) circle (0.1);
\filldraw[scale=sqrt(7), color=red] (0,1.96) circle (.06);

\filldraw[scale=sqrt(7), color=white] (-0.95,1.65) circle (0.1);
\filldraw[scale=sqrt(7), color=green] (-0.95,1.65) circle (.06);

\filldraw[scale=sqrt(7), color=white] (0.95,1.65) circle (0.1);
\filldraw[scale=sqrt(7), color=pink] (0.95,1.65) circle (.06);

\filldraw[scale=sqrt(7), color=white] (0,1.31) circle (0.1);
\filldraw[scale=sqrt(7), color=teal] (0,1.31) circle (.06);

\filldraw[scale=sqrt(7), color=white] (-1.7,0.98) circle (0.1);
\filldraw[scale=sqrt(7), color=teal] (-1.7,0.98) circle (.06);

\filldraw[scale=sqrt(7), color=white] (1.7,0.98) circle (0.1);
\filldraw[scale=sqrt(7), color=teal] (1.7,0.98) circle (.06);

\filldraw[scale=sqrt(7), color=white] (-1.13,0.66) circle (0.1);
\filldraw[scale=sqrt(7), color=red] (-1.13,0.66) circle (.06);

\filldraw[scale=sqrt(7), color=white] (0,0.66) circle (0.1);
\filldraw[scale=sqrt(7), color=red] (0,0.66) circle (.06);

\filldraw[scale=sqrt(7), color=white] (1.13,0.66) circle (0.1);
\filldraw[scale=sqrt(7), color=red] (1.13,0.66) circle (.06);

\filldraw[scale=sqrt(7), color=white] (-1.90,0) circle (0.1);
\filldraw[scale=sqrt(7), color=pink] (-1.90,0) circle (.06);

\filldraw[scale=sqrt(7), color=white] (1.90,0) circle (0.1);
\filldraw[scale=sqrt(7), color=green] (1.90,0) circle (.06);

\filldraw[scale=sqrt(7), color=white] (-1.13,-0.65) circle (0.1);
\filldraw[scale=sqrt(7), color=teal] (-1.13,-0.65) circle (.06);

\filldraw[scale=sqrt(7), color=white] (0,-0.65) circle (0.1);
\filldraw[scale=sqrt(7), color=teal] (0,-0.65) circle (.06);

\filldraw[scale=sqrt(7), color=white] (1.13,-0.65) circle (0.1);
\filldraw[scale=sqrt(7), color=teal] (1.13,-0.65) circle (.06);

\filldraw[scale=sqrt(7), color=white] (-1.7,-0.98) circle (0.1);
\filldraw[scale=sqrt(7), color=red] (-1.7,-0.98) circle (.06);

\filldraw[scale=sqrt(7), color=white] (1.7,-0.98) circle (0.1);
\filldraw[scale=sqrt(7), color=red] (1.7,-0.98) circle (.06);

\filldraw[scale=sqrt(7), color=white] (0,-1.31) circle (0.1);
\filldraw[scale=sqrt(7), color=red] (0,-1.31) circle (.06);

\filldraw[scale=sqrt(7), color=white] (-0.95,-1.65) circle (0.1);
\filldraw[scale=sqrt(7), color=green] (-0.95,-1.65) circle (.06);

\filldraw[scale=sqrt(7), color=white] (0.95,-1.65) circle (0.1);
\filldraw[scale=sqrt(7), color=pink] (0.95,-1.65) circle (.06);

\filldraw[scale=sqrt(7), color=white] (0,-1.96) circle (0.1);
\filldraw[scale=sqrt(7), color=teal] (0,0.-1.96) circle (.06);

\end{tikzpicture}\end{center}
\def\rL{{\rm L}} \def\rU{{\rm U}} 
\def\sB{{\mathscr B}} \def\sL{{\mathscr L}}
\def\bbZ{{\mathbb Z}}

%\caption
{\large{\bf Figure 4.2. The smallest non-removable local excitation for $D^2=5$.}}\smallskip

\small{ The irreducible insertion collection $\xi_0$ is located in $4$ consecutive meshes $T_{i,k}, T_{i,k+1}, T_{i,k+2}, T_{i,k+3}$ %$P_{\rl\rl}$, $P_{\rl\ru}$, $P_{\ru\rl}$ $P_{\ru\ru}$, 
(more precisely, in the intersections $T_{i,k}\cap\bbZ^3$, $T_{i,k+1}\cap\bbZ^3$, 
$T_{i,k+2}\cap\bbZ^3$ and $T_{i,k+3}\cap\bbZ^3$), with the distance between neighboring planes 
$1/{\sqrt 3}$. The repelled configuration $\eta_0$ is located in triangular $\sqrt{6}$-meshes $\tau_\rL\subset T_{i,k},$ and $\tau_\rU\subset T_{i,k+3}$,
(with distance ${\sqrt 3}$ between the planes containing $T_{i,k}$ and $T_{i,k+3}$). The figure shows the 
ortho-projection of $\xi_0$ to the lower mesh $T_{i,k}$ endowed with this basic triangular mesh %$\beta$ 
of size $\sqrt{2/3}$;
the edges/links of the basic mesh are drawn in thin lines. %All sites of lattice $\bbZ^3$ are projected into sites of $\beta$. 
The green and red circles mark insertions of type (IIa) in  $T_{i,k}\cap\bbZ^3$
and  $T_{i,k+3}\cap\bbZ^3$, respectively. Each green or red circle removes the $3$
vertices of the corresponding green or red triangle of side-length ${\sqrt 6}$; the figure contains 
7 green and 7 red such triangles.

The blue edges join a repelled site from mesh  $\tau_{\rm L}$ and a repelled site from 
mesh $\tau_{\rm U}$, the shorter between these edges have length ${\sqrt 5}$ and the longer 
${\sqrt{11}}$. The blue trapezes indicate the ortho-projections of 
tetrahedrons; each tetrahedron, in addition to $4$ blue edges, includes two non-adjacent/skewed edges,
one green and one red. The faint green/pink circles indicate insertions that repel the vertices of the 
tetrahedrons; the faint green circles mark the insertion sites located in $ T_{i,k+1}\cap\bbZ^3$ (there 
are $3$ of them), while the pink circles (another $3$) mark the insertion sites located in 
$T_{i,k+2}\cap\bbZ^3$.}
\end{figure}}

%%%%%%%%%%%%%%%%%%%%%%%%%%%%%%%%%%%%%%%%%%%%%

\def\FigureFiveOne
{\begin{figure}
\begin{center}

\begin{tikzpicture}[scale=.98]
\def\r32{0.86602540378443864676372317075294}
\def\n{6.5}

%\shade[shading=ball, ball color=black] (-3.5,{sqrt(147/4)}) circle (.2);
%\shade[shading=ball, ball color=white] (-2.5,{sqrt(147/4)}) circle (.4);

\def\n{8}
\clip[yscale=sqrt(3/4), xslant=0.5] (0, \n) -- (\n, 0) -- (\n, -\n) -- (0, -\n) -- (-\n, 0) -- (-\n, \n) -- cycle;
\draw [yscale=sqrt(3/4), xslant=0.5] (-\n, -\n) grid (\n, \n);
\draw [yscale=sqrt(3/4), xslant=-0.5] (-\n, -\n) grid (\n, \n);

%\draw [scale=sqrt(7), rotate=19.1, yscale=sqrt(3/4), xslant=0.5, color=red, very thick] (0, 0) -- (0, 1) -- (1,1) 
%-- (1,0) -- (0,0);

%\draw [scale=sqrt(7), rotate=19.1, yscale=sqrt(3/4), xslant=0.5, color=purple, line width=1mm] (1, -1) -- (1, -2)
%-- (0,-2) -- (0,-1) -- (1,-1);

%\draw [scale=sqrt(7), yscale=sqrt(3/4), xslant=0.5, color=teal, line width=1mm] (0, 0) -- (-1, 0) -- (-1,-1) 
%-- (0,-1) -- (0,0);
%=======================================

%\draw [scale=sqrt(7), yscale=sqrt(3/4), xslant=0.5, color=red, line width=1mm] (-0.76, 2.64)
%-- (2.66, -0.74) -- (-1.90, -0.74) -- (-1.90, 2.64) -- (-0.76, 2.64);

%%%%%%%%%%%%%%%%%%%%%

%\draw [scale=sqrt(7), color=brown, line width=1mm] 
%(0,0.66)--(0.57,0.99);

%%%%%%%%%%%%%%%%%%%%%

\draw [scale=sqrt(7), color=teal, line width=1mm] 
(-1.90,1.96)--(1.90,1.96);

\draw [scale=sqrt(7), color=teal, line width=1mm] 
(-2.47,0.98)--(2.47,0.98);

\draw [scale=sqrt(7), color=teal, line width=1mm] 
(-3.04,0)--(3.04,0);

\draw [scale=sqrt(7), color=teal, line width=1mm] 
(-2.47,-0.98)--(2.86,-0.98);

\draw [scale=sqrt(7), color=teal, line width=1mm] 
(-1.90,-1.96)--(1.90,-1.96);

%%%%%%%%%%%%%%%%%%%%%

\draw [scale=sqrt(7), color=teal, line width=1mm] 
(-2.66,-0.68)--(-0.76,2.62);

\draw [scale=sqrt(7), color=teal, line width=1mm] 
(-2.08,-1.64)--(0.38,2.62);

\draw [scale=sqrt(7), color=teal, line width=1mm] 
(-1.50,-2.62)--(1.50,2.62);

\draw [scale=sqrt(7), color=teal, line width=1mm] 
(-0.38,-2.62)--(2.08,1.64);

\draw [scale=sqrt(7), color=teal, line width=1mm] 
(0.76,-2.62)--(2.66,0.68);

%%%%%%%%%%%%%%%%%%%%%

\draw [scale=sqrt(7), color=teal, line width=1mm] 
(-2.66,0.68)--(-0.76,-2.62);

\draw [scale=sqrt(7), color=teal, line width=1mm] 
(-2.08,1.64)--(0.38,-2.62);

\draw [scale=sqrt(7), color=teal, line width=1mm] 
(-1.50,2.62)--(1.50,-2.62);

\draw [scale=sqrt(7), color=teal, line width=1mm] 
(-0.38,2.62)--(2.08,-1.64);

\draw [scale=sqrt(7), color=teal, line width=1mm] 
(0.76,2.62)--(2.66,-0.68);

%%%%%%%%%%%%%%%%%%%%%

%\draw [scale=sqrt(7), color=red, line width=1mm] 
% (-1.48,2.62)-- (1.48,2.62);

%\draw [scale=sqrt(7), color=red, line width=1mm] 
% (-2.09,1.65)-- (2.09,1.65);

%%%%%%%%%%%%%%%%%%%%% 

%\filldraw[scale=sqrt(7), color=white] (0,2.62) circle (0.1);
%\filldraw[scale=sqrt(7), color=red] (0,2.62) circle (.06);

%%%%%%%%%%%%%%%%%%%%%

\filldraw[scale=sqrt(7), color=white] (-1.14,1.96) circle (0.14);
\filldraw[scale=sqrt(7), color=teal] (-1.14,1.96) circle (.06);

\filldraw[scale=sqrt(7), color=white] (0,1.96) circle (0.14);
\filldraw[scale=sqrt(7), color=teal] (0,1.96) circle (.06);

\filldraw[scale=sqrt(7), color=white] (1.14,1.96) circle (0.14);
\filldraw[scale=sqrt(7), color=teal] (1.14,1.96) circle (.06);

\filldraw[scale=sqrt(7), color=white] (-1.71,0.98) circle (0.14);
\filldraw[scale=sqrt(7), color=teal] (-1.71,0.98) circle (.06);

\filldraw[scale=sqrt(7), color=white] (-0.57,0.98) circle (0.14);
\filldraw[scale=sqrt(7), color=teal] (-0.57,0.98) circle (.06);

\filldraw[scale=sqrt(7), color=white] (0.57,0.98) circle (0.14);
\filldraw[scale=sqrt(7), color=teal] (0.57,0.98) circle (.06);

\filldraw[scale=sqrt(7), color=white] (1.71,0.98) circle (0.14);
\filldraw[scale=sqrt(7), color=teal] (1.71,0.98) circle (.06);

\filldraw[scale=sqrt(7), color=white] (-2.28,0) circle (0.14);
\filldraw[scale=sqrt(7), color=teal] (-2.28,0) circle (.06);

\filldraw[scale=sqrt(7), color=white] (-1.14,0) circle (0.14);
\filldraw[scale=sqrt(7), color=teal] (-1.14,0) circle (.06);

\filldraw[scale=sqrt(7), color=white] (0,0) circle (0.14);
\filldraw[scale=sqrt(7), color=teal] (0,0) circle (.06);

\filldraw[scale=sqrt(7), color=white] (1.14,0) circle (0.14);
\filldraw[scale=sqrt(7), color=teal] (1.14,0) circle (.06);

\filldraw[scale=sqrt(7), color=white] (2.28,0) circle (0.14);
\filldraw[scale=sqrt(7), color=teal] (2.28,0) circle (.06);

\filldraw[scale=sqrt(7), color=white] (-1.71,-0.98) circle (0.14);
\filldraw[scale=sqrt(7), color=teal] (-1.71,-0.98) circle (.06);

\filldraw[scale=sqrt(7), color=white] (-0.57,-0.98) circle (0.14);
\filldraw[scale=sqrt(7), color=teal] (-0.57,-0.98) circle (.06);

\filldraw[scale=sqrt(7), color=white] (0.57,-0.98) circle (0.14);
\filldraw[scale=sqrt(7), color=teal] (0.57,-0.98) circle (.06);

\filldraw[scale=sqrt(7), color=white] (1.71,-0.98) circle (0.14);
\filldraw[scale=sqrt(7), color=teal] (1.71,-0.98) circle (.06);

\filldraw[scale=sqrt(7), color=white] (-1.14,-1.96) circle (0.14);
\filldraw[scale=sqrt(7), color=teal] (-1.14,-1.96) circle (.06);

\filldraw[scale=sqrt(7), color=white] (0,-1.96) circle (0.14);
\filldraw[scale=sqrt(7), color=teal] (0,-1.96) circle (.06);

\filldraw[scale=sqrt(7), color=white] (1.14,-1.96) circle (0.14);
\filldraw[scale=sqrt(7), color=teal] (1.14,-1.96) circle (.06);

%%%%%%%%%%%%%%%%%%%%%

%\draw [scale=sqrt(7), color=red, line width=1mm] 
% (-1.48,2.62)-- (1.48,2.62);

\draw [scale=sqrt(7), color=red, line width=.5mm] 
(-1.90,1.96)--(1.90,1.96);

\draw [scale=sqrt(7), color=red, line width=.5mm] 
(-2.47,0.98)--(2.47,0.98);

\draw [scale=sqrt(7), color=red, line width=.5mm] 
(-3.04,0)--(3.04,0);

\draw [scale=sqrt(7), color=red, line width=.5mm] 
(-2.47,-0.98)--(2.86,-0.98);

\draw [scale=sqrt(7), color=red, line width=.5mm] 
(-1.90,-1.96)--(1.90,-1.96);

\draw [scale=sqrt(7), color=red, line width=1mm] 
(-3.03,0)--(-1.51,2.62);

\draw [scale=sqrt(7), color=red, line width=1mm] 
(-2.47,-0.98)--(-0.38,2.62);

\draw [scale=sqrt(7), color=red, line width=1mm] 
(-1.90,-1.96)--(0.76,2.62);

\draw [scale=sqrt(7), color=red, line width=1mm] 
(-1.14,-2.62)--(1.71,2.32);

\draw [scale=sqrt(7), color=red, line width=1mm] 
(0,-2.62)--(2.28,1.31);

\draw [scale=sqrt(7), color=red, line width=1mm] 
(1.14,-2.62)--(2.83,0.33);

\draw [scale=sqrt(7), color=red, line width=1mm] 
(-3.03,0)--(-1.51,-2.62);

\draw [scale=sqrt(7), color=red, line width=1mm] 
(-2.47,0.98)--(-0.38,-2.62);

\draw [scale=sqrt(7), color=red, line width=1mm] 
(-1.90,1.96)--(0.76,-2.62);

\draw [scale=sqrt(7), color=red, line width=1mm] 
(-1.12,2.62)--(1.71,-2.32);

%----------------------------------------------------

\draw [scale=sqrt(7), color=red, line width=1mm] 
(1.12,2.62)--(2.84,-0.33);

%%%%%%%%%%%%%%%%%%%%%

\filldraw[scale=sqrt(7), color=white] (-1.90,1.96) circle (0.1);
\filldraw[scale=sqrt(7), color=red] (-1.90,1.96) circle (.06);

\filldraw[scale=sqrt(7), color=white] (-0.76,1.96) circle (0.1);
\filldraw[scale=sqrt(7), color=red] (-0.76,1.96) circle (.06);

\filldraw[scale=sqrt(7), color=white] (0.38,1.96) circle (0.1);
\filldraw[scale=sqrt(7), color=red] (0.38,1.96) circle (.06);

\filldraw[scale=sqrt(7), color=white] (1.52,1.96) circle (0.1);
\filldraw[scale=sqrt(7), color=red] (1.52,1.96) circle (.06);

\filldraw[scale=sqrt(7), color=white] (-2.47,0.98) circle (0.1);
\filldraw[scale=sqrt(7), color=red] (-2.47,0.98) circle (.06);

\filldraw[scale=sqrt(7), color=white] (-1.33,0.98) circle (0.1);
\filldraw[scale=sqrt(7), color=red] (-1.33,0.98) circle (.06);

\filldraw[scale=sqrt(7), color=white] (-0.19,0.98) circle (0.1);
\filldraw[scale=sqrt(7), color=red] (-0.19,0.98) circle (.06);

\filldraw[scale=sqrt(7), color=white] (0.95,0.98) circle (0.1);
\filldraw[scale=sqrt(7), color=red] (0.95,0.98) circle (.06);

\filldraw[scale=sqrt(7), color=white] (2.09,0.98) circle (0.1);
\filldraw[scale=sqrt(7), color=red] (2.09,0.98) circle (.06);

\filldraw[scale=sqrt(7), color=white] (-3.02,0) circle (0.1);
\filldraw[scale=sqrt(7), color=red] (-3.02,0) circle (.06);

\filldraw[scale=sqrt(7), color=white] (-1.90,0) circle (0.1);
\filldraw[scale=sqrt(7), color=red] (-1.90,0) circle (.06);

\filldraw[scale=sqrt(7), color=white] (-0.76,0) circle (0.1);
\filldraw[scale=sqrt(7), color=red] (-0.76,0) circle (.06);

\filldraw[scale=sqrt(7), color=white] (0.38,0) circle (0.1);
\filldraw[scale=sqrt(7), color=red] (0.38,0) circle (.06);

\filldraw[scale=sqrt(7), color=white] (1.52,0) circle (0.1);
\filldraw[scale=sqrt(7), color=red] (1.52,0) circle (.06);

\filldraw[scale=sqrt(7), color=white] (2.64,0) circle (0.1);
\filldraw[scale=sqrt(7), color=red] (2.64,0) circle (.06);

\filldraw[scale=sqrt(7), color=white] (-2.47,-0.98) circle (0.1);
\filldraw[scale=sqrt(7), color=red] (-2.47,-0.98) circle (.06);

\filldraw[scale=sqrt(7), color=white] (-1.33,-0.98) circle (0.1);
\filldraw[scale=sqrt(7), color=red] (-1.33,-0.98) circle (.06);

\filldraw[scale=sqrt(7), color=white] (-0.19,-0.98) circle (0.1);
\filldraw[scale=sqrt(7), color=red] (-0.19,-0.98) circle (.06);

\filldraw[scale=sqrt(7), color=white] (0.95,-0.98) circle (0.1);
\filldraw[scale=sqrt(7), color=red] (0.95,-0.98) circle (.06);

\filldraw[scale=sqrt(7), color=white] (2.09,-0.98) circle (0.1);
\filldraw[scale=sqrt(7), color=red] (2.09,-0.98) circle (.06);

\filldraw[scale=sqrt(7), color=white] (-1.90,-1.96) circle (0.1);
\filldraw[scale=sqrt(7), color=red] (-1.90,-1.96) circle (.06);

\filldraw[scale=sqrt(7), color=white] (-0.76,-1.96) circle (0.1);
\filldraw[scale=sqrt(7), color=red] (-0.76,-1.96) circle (.06);

\filldraw[scale=sqrt(7), color=white] (0.38,-1.96) circle (0.1);
\filldraw[scale=sqrt(7), color=red] (0.38,-1.96) circle (.06);

\filldraw[scale=sqrt(7), color=white] (1.52,-1.96) circle (0.1);
\filldraw[scale=sqrt(7), color=red] (1.52,-1.96) circle (.06);

%\filldraw[scale=sqrt(7), color=white] (0.57,0.33) circle (0.1);
%\filldraw[scale=sqrt(7), color=pink] (0.57,0.33) circle (.06);
\end{tikzpicture}\end{center}
\def\rx{{\rm x}} \def\vphi{{\varphi}} 
\def\sB{{\mathscr B}} \def\sS{{\mathscr S}}
\def\bbZ{{\mathbb Z}}

\large{\bf Figure 5.1. Neighboring  triangular meshes for $D^2=6$.}\smallskip

\small{The figure shows two neighboring meshes in a PC $\onwl{\bigcup}\limits_{k\in\bbZ}\tau^{(6)}_{i,4k,j_k}$
for $D^2=6$. The thin lines in 
the background form the triangular ${\sqrt{2/3}}$-lattice $T_{i,0}$ in the plane 
$\rx_1+s_2(i)\rx_2+s_3(i)\rx_3= 0$. The green  lines and 
circles indicate the triangular ${\sqrt{6}}$-mesh $\tau^{(6)}_{i,0,0}\subset T_{i,0}$ and 
the positions of occupied sites in this mesh. The red lines and circles indicate the projections of the 
triangular ${\sqrt{6}}$-mesh $\tau^{(6)}_{i, 4,1}\subset T_{i, 4}$ lying in the parallel plane 
$\rx_1+s_2(i)\rx_2+s_3(i)\rx_3= 4$ and of the positions of occupied sites in this mesh.}
%of particles in $\tau^{(6)}_{0,\pm 4,1}$\,.}
\end{figure}}

%%%%%%%%%%%%%%%%%%%%%%%%%%%%%%%%%%%%%%%%%%%%%

\def\FigureFiveTwo
{\begin{figure}
\begin{center}
\begin{tikzpicture}[scale=0.9]
%\clip (-10.5, 1.5) rectangle (7, 10.5);

\begin{scope}
\definecolor{gray3}{gray}{0.3}
\definecolor{gray5}{gray}{0.5}
%\path [fill=gray5, draw=black, very thick] 
%(2, 6) -- (5, 6) -- (5, 9) -- (4, 9) -- (4, 11) -- (1, 11) -- (1, 8) -- (2, 8) -- cycle;

\draw[xstep=.8cm,ystep=0.414,color=black] (-10.5,4) grid (6.4,10.5);

%%%%%%%%%%%%%%%%%%%%%%%
%%%%%%%%%%%%%%%%%%%%%%%

%\draw [scale=sqrt(7), color=brown, line width=1.2mm] 
%(-3.33,3.45)--(-3.025,3.76);

%\draw [scale=sqrt(7), color=brown, line width=1.2mm] 
%(-3.33,3.45)--(-3.025,3.14);

%\draw [scale=sqrt(7), color=brown, line width=1.2mm] 
%(-3.30,2.87)--(-3.025,3.14);

%%%%%%%%%%%%%%%%%%%%%%%

\draw [scale=sqrt(7), color=teal, line width=1mm] 
(-3.93,3.30)--(-2.72,3.92);

\draw [scale=sqrt(7), color=teal, line width=1mm] 
(-3.93,2.67)--(-1.51,3.92);

\draw [scale=sqrt(7), color=teal, line width=1mm] 
(-3.93,2.04)--(-0.30,3.92);

\draw [scale=sqrt(7), color=teal, line width=1mm] 
(-3.63,1.57)--(0.91,3.92);

\draw [scale=sqrt(7), color=teal, line width=1mm] 
(-2.42,1.57)--(2.12,3.92);

\draw [scale=sqrt(7), color=teal, line width=1mm] 
(-1.21,1.57)--(2.42,3.45);

\draw [scale=sqrt(7), color=teal, line width=1mm] 
(0,1.57)--(2.42,2.82);

\draw [scale=sqrt(7), color=teal, line width=1mm] 
(1.21,1.57)--(2.42,2.20);

%%%%%%%%%%%%%%%%%%%%%%%

\draw [scale=sqrt(7), color=teal, line width=1mm] 
(-3.93,1.71)--(-3.63,1.57);

\draw [scale=sqrt(7), color=teal, line width=1mm] 
(-3.93,2.34)--(-2.42,1.57);

\draw [scale=sqrt(7), color=teal, line width=1mm] 
(-3.93,2.97)--(-1.21,1.57);

\draw [scale=sqrt(7), color=teal, line width=1mm] 
(-3.93,3.59)--(0,1.57);

\draw [scale=sqrt(7), color=teal, line width=1mm] 
(-3.325,3.92)--(1.21,1.57);

\draw [scale=sqrt(7), color=teal, line width=1mm] 
(-2.115,3.92)--(2.42,1.57);

\draw [scale=sqrt(7), color=teal, line width=1mm] 
(-0.915,3.92)--(2.42,2.20);

\draw [scale=sqrt(7), color=teal, line width=1mm] 
(0.295,3.92)--(2.42,2.82);

\draw [scale=sqrt(7), color=teal, line width=1mm] 
(1.505,3.92)--(2.42,3.45);

%%%%%%%%%%%%%%%%%%%%%%%
%%%%%%%%%%%%%%%%%%%%%%%

%\draw [scale=sqrt(7), color=brown, line width=2.0mm] 
%(-3.63,2.82)--(-2.42,3.45);

%\draw [scale=sqrt(7), color=brown, line width=0.8mm] 
% (-3.025,3.76) -- (-2.42,2.82);
 
% \draw [scale=sqrt(7), color=brown, line width=2.0mm] 
%(-3.93,3.30)--(-2.72,3.92);

%\draw [scale=sqrt(7), color=brown, line width=0.8mm] 
%(-3.33,3.615) --  (-2.72,2.665);

%\draw [scale=sqrt(7), color=brown, line width=0.8mm] 
%(-3.33,3.615) --  (-2.72,2.665);

%%%%%%%%%%%%%%%%%%%%%%%

\draw [scale=sqrt(7), color=red, line width=0.4mm] 
(-3.93,2.045)--(-3.03,1.57); 

\draw [scale=sqrt(7), color=red, line width=0.4mm]  
 (-3.93,2.665)--(-1.82,1.57);

\draw [scale=sqrt(7), color=red, line width=0.4mm] 
(-3.93,3.295)--(-0.61,1.57);

\draw [scale=sqrt(7), color=red, line width=0.4mm] 
(-3.93,3.915)--(0.605,1.57);

\draw [scale=sqrt(7), color=red, line width=0.4mm] 
(-2.72,3.92)--(1.815,1.57);

\draw [scale=sqrt(7), color=red, line width=0.4mm] 
(-1.51,3.92)--(2.42,1.88);

\draw [scale=sqrt(7), color=red, line width=0.4mm] 
(-0.3,3.92)--(2.425,2.5);

\draw [scale=sqrt(7), color=red, line width=0.4mm] 
(0.91,3.92)--(2.425,3.13);

\draw [scale=sqrt(7), color=red, line width=0.4mm] 
(2.12,3.92)--(2.425,3.73);

%%%%%%%%%%%%%%%%%%%%%%%
%%%%%%%%%%%%%%%%%%%%%%%

\filldraw[scale=sqrt(7), color=white] (-3.025,3.76) circle (0.12);
\filldraw[scale=sqrt(7), color=teal] (-3.025,3.76) circle (.06);

\filldraw[scale=sqrt(7), color=white] (-1.815,3.76) circle (0.12);
\filldraw[scale=sqrt(7), color=teal] (-1.815,3.76) circle (.06);

\filldraw[scale=sqrt(7), color=white] (-0.605,3.76) circle (0.12);
\filldraw[scale=sqrt(7), color=teal] (-0.605,3.76) circle (.06);

\filldraw[scale=sqrt(7), color=white] (0.605,3.76) circle (0.12);
\filldraw[scale=sqrt(7), color=teal] (0.605,3.76) circle (.06);

\filldraw[scale=sqrt(7), color=white] (1.815,3.76) circle (0.12);
\filldraw[scale=sqrt(7), color=teal] (1.815,3.76) circle (.06);

\filldraw[scale=sqrt(7), color=white] (-3.63,3.45) circle (0.12);
\filldraw[scale=sqrt(7), color=teal] (-3.63,3.45) circle (.06);

\filldraw[scale=sqrt(7), color=white] (-2.42,3.45) circle (0.12);
\filldraw[scale=sqrt(7), color=teal] (-2.42,3.45) circle (.06);

\filldraw[scale=sqrt(7), color=white] (-1.21,3.45) circle (0.12);
\filldraw[scale=sqrt(7), color=teal] (-1.21,3.45) circle (.06);

\filldraw[scale=sqrt(7), color=white] (0,3.45) circle (0.12);
\filldraw[scale=sqrt(7), color=teal] (0,3.45) circle (.06);

\filldraw[scale=sqrt(7), color=white] (1.21,3.45) circle (0.12);
\filldraw[scale=sqrt(7), color=teal] (1.21,3.45) circle (.06);

\filldraw[scale=sqrt(7), color=white] (2.42,3.45) circle (0.12);
\filldraw[scale=sqrt(7), color=teal] (2.42,3.45) circle (.06);

\filldraw[scale=sqrt(7), color=white] (-3.025,3.14) circle (0.12);
\filldraw[scale=sqrt(7), color=teal] (-3.025,3.14) circle (.06);

\filldraw[scale=sqrt(7), color=white] (-1.815,3.14) circle (0.12);
\filldraw[scale=sqrt(7), color=teal] (-1.815,3.14) circle (.06);

\filldraw[scale=sqrt(7), color=white] (-0.605,3.14) circle (0.12);
\filldraw[scale=sqrt(7), color=teal] (-0.605,3.14) circle (.06);

\filldraw[scale=sqrt(7), color=white] (0.605,3.14) circle (0.12);
\filldraw[scale=sqrt(7), color=teal] (0.605,3.14) circle (.06);

\filldraw[scale=sqrt(7), color=white] (1.815,3.14) circle (0.12);
\filldraw[scale=sqrt(7), color=teal] (1.815,3.14) circle (.06);

\filldraw[scale=sqrt(7), color=white] (-3.63,2.82) circle (0.12);
\filldraw[scale=sqrt(7), color=teal] (-3.63,2.82) circle (.06);

\filldraw[scale=sqrt(7), color=white] (-2.42,2.82) circle (0.12);
\filldraw[scale=sqrt(7), color=teal] (-2.42,2.82) circle (.06);

\filldraw[scale=sqrt(7), color=white] (-1.21,2.82) circle (0.12);
\filldraw[scale=sqrt(7), color=teal] (-1.21,2.82) circle (.06);

\filldraw[scale=sqrt(7), color=white] (0,2.82) circle (0.12);
\filldraw[scale=sqrt(7), color=teal] (0,2.82) circle (.06);

\filldraw[scale=sqrt(7), color=white] (1.21,2.82) circle (0.12);
\filldraw[scale=sqrt(7), color=teal] (1.21,2.82) circle (.06);

\filldraw[scale=sqrt(7), color=white] (2.42,2.82) circle (0.12);
\filldraw[scale=sqrt(7), color=teal] (2.42,2.82) circle (.06);

\filldraw[scale=sqrt(7), color=white] (-3.025,2.51) circle (0.12);
\filldraw[scale=sqrt(7), color=teal] (-3.025,2.51) circle (.06);

\filldraw[scale=sqrt(7), color=white] (-1.815,2.51) circle (0.12);
\filldraw[scale=sqrt(7), color=teal] (-1.815,2.51) circle (.06);

\filldraw[scale=sqrt(7), color=white] (-0.605,2.51) circle (0.12);
\filldraw[scale=sqrt(7), color=teal] (-0.605,2.51) circle (.06);

\filldraw[scale=sqrt(7), color=white] (0.605,2.51) circle (0.12);
\filldraw[scale=sqrt(7), color=teal] (0.605,2.51) circle (.06);

\filldraw[scale=sqrt(7), color=white] (1.815,2.51) circle (0.12);
\filldraw[scale=sqrt(7), color=teal] (1.815,2.51) circle (.06);

\filldraw[scale=sqrt(7), color=white] (-3.63,2.20) circle (0.12);
\filldraw[scale=sqrt(7), color=teal] (-3.63,2.20) circle (.06);

\filldraw[scale=sqrt(7), color=white] (-2.42,2.20) circle (0.12);
\filldraw[scale=sqrt(7), color=teal] (-2.42,2.20) circle (.06);

\filldraw[scale=sqrt(7), color=white] (-1.21,2.20) circle (0.12);
\filldraw[scale=sqrt(7), color=teal] (-1.21,2.20) circle (.06);

\filldraw[scale=sqrt(7), color=white] (0,2.20) circle (0.12);
\filldraw[scale=sqrt(7), color=teal] (0,2.20) circle (.06);

\filldraw[scale=sqrt(7), color=white] (1.21,2.20) circle (0.12);
\filldraw[scale=sqrt(7), color=teal] (1.21,2.20) circle (.06);

\filldraw[scale=sqrt(7), color=white] (2.42,2.20) circle (0.12);
\filldraw[scale=sqrt(7), color=teal] (2.42,2.20) circle (.06);

\filldraw[scale=sqrt(7), color=white] (-3.025,1.89) circle (0.12);
\filldraw[scale=sqrt(7), color=teal] (-3.025,1.89) circle (.06);

\filldraw[scale=sqrt(7), color=white] (-1.815,1.89) circle (0.12);
\filldraw[scale=sqrt(7), color=teal] (-1.815,1.89) circle (.06);

\filldraw[scale=sqrt(7), color=white] (-0.605,1.89) circle (0.12);
\filldraw[scale=sqrt(7), color=teal] (-0.605,1.89) circle (.06);

\filldraw[scale=sqrt(7), color=white] (0.605,1.89) circle (0.12);
\filldraw[scale=sqrt(7), color=teal] (0.605,1.89) circle (.06);

\filldraw[scale=sqrt(7), color=white] (1.815,1.89) circle (0.12);
\filldraw[scale=sqrt(7), color=teal] (1.815,1.89) circle (.06);

\filldraw[scale=sqrt(7), color=white] (-3.63,1.57) circle (0.12);
\filldraw[scale=sqrt(7), color=teal] (-3.63,1.57) circle (.06);

\filldraw[scale=sqrt(7), color=white] (-2.42,1.57) circle (0.12);
\filldraw[scale=sqrt(7), color=teal] (-2.42,1.57) circle (.06);

\filldraw[scale=sqrt(7), color=white] (-1.21,1.57) circle (0.12);
\filldraw[scale=sqrt(7), color=teal] (-1.21,1.57) circle (.06);

\filldraw[scale=sqrt(7), color=white] (0,1.57) circle (0.12);
\filldraw[scale=sqrt(7), color=teal] (0,1.57) circle (.06);

\filldraw[scale=sqrt(7), color=white] (1.21,1.57) circle (0.12);
\filldraw[scale=sqrt(7), color=teal] (1.21,1.57) circle (.06);

\filldraw[scale=sqrt(7), color=white] (2.42,1.57) circle (0.12);
\filldraw[scale=sqrt(7), color=teal] (2.42,1.57) circle (.06);

%%%%%%%%%%%%%%%%%%%%%%%

\filldraw[scale=sqrt(7), color=white] (-3.93,3.915) circle (0.12);
\filldraw[scale=sqrt(7), color=red] (-3.93,3.915) circle (.06);

\filldraw[scale=sqrt(7), color=white] (-2.72,3.915) circle (0.12);
\filldraw[scale=sqrt(7), color=red] (-2.72,3.915) circle (.06);

\filldraw[scale=sqrt(7), color=white] (-1.515,3.915) circle (0.12);
\filldraw[scale=sqrt(7), color=red] (-1.515,3.915) circle (.06);

\filldraw[scale=sqrt(7), color=white] (-0.305,3.915) circle (0.12);
\filldraw[scale=sqrt(7), color=red] (-0.305,3.915) circle (.06);

\filldraw[scale=sqrt(7), color=white] (0.905,3.915) circle (0.12);
\filldraw[scale=sqrt(7), color=red] (0.905,3.915) circle (.06);

\filldraw[scale=sqrt(7), color=white] (2.115,3.915) circle (0.12);
\filldraw[scale=sqrt(7), color=red] (2.115,3.915) circle (.06);

\filldraw[scale=sqrt(7), color=white] (-3.33,3.615) circle (0.12);
\filldraw[scale=sqrt(7), color=red] (-3.33,3.615) circle (.06);

\filldraw[scale=sqrt(7), color=white] (-2.12,3.615) circle (0.12);
\filldraw[scale=sqrt(7), color=red] (-2.12,3.615) circle (.06);

\filldraw[scale=sqrt(7), color=white] (-0.91,3.615) circle (0.12);
\filldraw[scale=sqrt(7), color=red] (-0.91,3.615) circle (.06);

\filldraw[scale=sqrt(7), color=white] (0.30,3.615) circle (0.12);
\filldraw[scale=sqrt(7), color=red] (0.30,3.615) circle (.06);

\filldraw[scale=sqrt(7), color=white] (1.51,3.615) circle (0.12);
\filldraw[scale=sqrt(7), color=red] (1.51,3.615) circle (.06);

\filldraw[scale=sqrt(7), color=white] (-3.93,3.295) circle (0.12);
\filldraw[scale=sqrt(7), color=red] (-3.93,3.295) circle (.06);

\filldraw[scale=sqrt(7), color=white] (-2.72,3.295) circle (0.12);
\filldraw[scale=sqrt(7), color=red] (-2.72,3.295) circle (.06);

\filldraw[scale=sqrt(7), color=white] (-1.515,3.295) circle (0.12);
\filldraw[scale=sqrt(7), color=red] (-1.515,3.295) circle (.06);

\filldraw[scale=sqrt(7), color=white] (-0.305,3.295) circle (0.12);
\filldraw[scale=sqrt(7), color=red] (-0.305,3.295) circle (.06);

\filldraw[scale=sqrt(7), color=white] (0.905,3.295) circle (0.12);
\filldraw[scale=sqrt(7), color=red] (0.905,3.295) circle (.06);

\filldraw[scale=sqrt(7), color=white] (2.115,3.295) circle (0.12);
\filldraw[scale=sqrt(7), color=red] (2.115,3.295) circle (.06);

\filldraw[scale=sqrt(7), color=white] (-3.33,2.975) circle (0.12);
\filldraw[scale=sqrt(7), color=red] (-3.33,2.975) circle (.06);

\filldraw[scale=sqrt(7), color=white] (-2.12,2.975) circle (0.12);
\filldraw[scale=sqrt(7), color=red] (-2.12,2.975) circle (.06);

\filldraw[scale=sqrt(7), color=white] (-0.91,2.975) circle (0.12);
\filldraw[scale=sqrt(7), color=red] (-0.91,2.975) circle (.06);

\filldraw[scale=sqrt(7), color=white] (0.30,2.975) circle (0.12);
\filldraw[scale=sqrt(7), color=red] (0.30,2.975) circle (.06);

\filldraw[scale=sqrt(7), color=white] (1.51,2.975) circle (0.12);
\filldraw[scale=sqrt(7), color=red] (1.51,2.975) circle (.06);

\filldraw[scale=sqrt(7), color=white] (-3.93,2.665) circle (0.12);
\filldraw[scale=sqrt(7), color=red] (-3.93,2.665) circle (.06);

\filldraw[scale=sqrt(7), color=white] (-2.72,2.665) circle (0.12);
\filldraw[scale=sqrt(7), color=red] (-2.72,2.665) circle (.06);

\filldraw[scale=sqrt(7), color=white] (-1.515,2.665) circle (0.12);
\filldraw[scale=sqrt(7), color=red] (-1.515,2.665) circle (.06);

\filldraw[scale=sqrt(7), color=white] (-0.305,2.665) circle (0.12);
\filldraw[scale=sqrt(7), color=red] (-0.305,2.665) circle (.06);

\filldraw[scale=sqrt(7), color=white] (0.905,2.665) circle (0.12);
\filldraw[scale=sqrt(7), color=red] (0.905,2.665) circle (.06);

\filldraw[scale=sqrt(7), color=white] (2.115,2.665) circle (0.12);
\filldraw[scale=sqrt(7), color=red] (2.115,2.665) circle (.06);

\filldraw[scale=sqrt(7), color=white] (-3.33,2.355) circle (0.12);
\filldraw[scale=sqrt(7), color=red] (-3.33,2.355) circle (.06);

\filldraw[scale=sqrt(7), color=white] (-2.12,2.355) circle (0.12);
\filldraw[scale=sqrt(7), color=red] (-2.12,2.355) circle (.06);

\filldraw[scale=sqrt(7), color=white] (-0.91,2.355) circle (0.12);
\filldraw[scale=sqrt(7), color=red] (-0.91,2.355) circle (.06);

\filldraw[scale=sqrt(7), color=white] (0.30,2.355) circle (0.12);
\filldraw[scale=sqrt(7), color=red] (0.30,2.355) circle (.06);

\filldraw[scale=sqrt(7), color=white] (1.51,2.355) circle (0.12);
\filldraw[scale=sqrt(7), color=red] (1.51,2.355) circle (.06);

\filldraw[scale=sqrt(7), color=white] (-3.93,2.045) circle (0.12);
\filldraw[scale=sqrt(7), color=red] (-3.93,2.045) circle (.06);

\filldraw[scale=sqrt(7), color=white] (-2.72,2.045) circle (0.12);
\filldraw[scale=sqrt(7), color=red] (-2.72,2.045) circle (.06);

\filldraw[scale=sqrt(7), color=white] (-1.515,2.045) circle (0.12);
\filldraw[scale=sqrt(7), color=red] (-1.515,2.045) circle (.06);

\filldraw[scale=sqrt(7), color=white] (-0.305,2.045) circle (0.12);
\filldraw[scale=sqrt(7), color=red] (-0.305,2.045) circle (.06);

\filldraw[scale=sqrt(7), color=white] (0.905,2.045) circle (0.12);
\filldraw[scale=sqrt(7), color=red] (0.905,2.045) circle (.06);

\filldraw[scale=sqrt(7), color=white] (2.115,2.045) circle (0.12);
\filldraw[scale=sqrt(7), color=red] (2.115,2.045) circle (.06);

\filldraw[scale=sqrt(7), color=white] (-3.33,1.725) circle (0.12);
\filldraw[scale=sqrt(7), color=red] (-3.33,1.725) circle (.06);

\filldraw[scale=sqrt(7), color=white] (-2.12,1.725) circle (0.12);
\filldraw[scale=sqrt(7), color=red] (-2.12,1.725) circle (.06);

\filldraw[scale=sqrt(7), color=white] (-0.91,1.725) circle (0.12);
\filldraw[scale=sqrt(7), color=red] (-0.91,1.725) circle (.06);

\filldraw[scale=sqrt(7), color=white] (0.30,1.725) circle (0.12);
\filldraw[scale=sqrt(7), color=red] (0.30,1.725) circle (.06);

\filldraw[scale=sqrt(7), color=white] (1.51,1.725) circle (0.12);
\filldraw[scale=sqrt(7), color=red] (1.51,1.725) circle (.06);

%%%%%%%%%%%%%%%%%%%%%%%

%\filldraw[scale=sqrt(7), color=white] (-0.30,2.82) circle (0.12);
%\filldraw[scale=sqrt(7), color=pink] (-0.30,2.82) circle (.08);

%\draw [scale=sqrt(7), color=red, line width=0.4mm] 
%(-3.63,1.57)--(0.905,3.92);
%\draw [scale=sqrt(7), color=red, line width=0.4mm] 
%(-2.42,1.57)--(2.13,3.92);

\draw [scale=sqrt(7), color=red, line width=0.4mm] 
(-3.93,3.30)--(-2.72,3.92);

\draw [scale=sqrt(7), color=red, line width=0.4mm] 
(-3.93,2.67)--(-1.51,3.92);

\draw [scale=sqrt(7), color=red, line width=0.4mm] 
(-3.93,2.04)--(-0.30,3.92);

\draw [scale=sqrt(7), color=red, line width=0.4mm] 
(-3.63,1.57)--(0.91,3.92);

\draw [scale=sqrt(7), color=red, line width=0.4mm] 
(-2.42,1.57)--(2.12,3.92);

\draw [scale=sqrt(7), color=red, line width=0.4mm] 
(-1.21,1.57)--(2.42,3.45);

\draw [scale=sqrt(7), color=red, line width=0.4mm] 
(0,1.57)--(2.42,2.82);

\draw [scale=sqrt(7), color=red, line width=0.4mm] 
(1.21,1.57)--(2.42,2.20);
%\draw (-10,0) grid (34,28);
\end{scope}
%\path [fill=white, draw=white] (6,21) -- (6.5,21.5) -- (9.5,21.5) -- (10,21) -- (9.5,20.5) -- (6.5,20.5) -- (6,21) -- cycle;
%\draw (-2.5, 21) node   {\Large {${\bf{Ext}}({\mbox{\boldmath$\Gamma$}})$}};
\end{tikzpicture} \end{center}
\def\bbZ{{\mathbb Z}} \def\rx{{\rm x}}
%\vskip 1cm
\large{\bf Figure 5.2. Neighboring rhombic meshes for $D^2=6$.}\smallskip

\small{ The figure shows two neighboring rhombic meshes in a PC 
$\onwl{\bigcup}\limits_{k\in\bbZ}\alp^{(8,16)}_{i,3k,j_k}$ for $D^2=6$. 
The thin lines in the background form the rectangular $(1\times {\sqrt 2}/2)$-lattice
$Q_{i,0}$ in the plane $s_1(i)\rx_1+s_2(i)\rx_2+s_3(i)\rx_3=0$ orthogonal to the non-main diagonal 
$(s_1(i),s_2(i),s_3(i))$. The green lines and circles indicate
the positions of particles in the rhombic $({\sqrt 8},{\sqrt{16}})$-mesh $\alpha^{(8,16)}_{i,0,0}\subset Q_{i,0}$ 
in this plane, with rhombus 
diagonals of lengths ${\sqrt 8}$ and ${\sqrt{16}}$. The red lines and circles indicate the projections of the 
rhombic $({\sqrt 8},{\sqrt{16}})$-mesh $\alpha^{(8,16)}_{i,3,1}\subset Q_{i, 3}$ lying in the parallel plane
$s_1(i)\rx_1+s_2(i)\rx_2+s_3(i)\rx_3=3$ and of the positions of occupied sites in this mesh.

}
\end{figure}}

%%%%%%%%%%%%%%%%%%%%%%%%%%%%%%%%%%%%%%%%%%%%
%%%%%%%%%%%%%%%%%%%%%%%%%%%

\title{\bf Kepler's conjecture and phase transitions\\ in the high-density hard-core model on  $\mathbb{Z}^3$}

\author{\bf A. Mazel$^1$, I. Stuhl$^2$, Y. Suhov$^{2,3}$}

\date{}
\footnotetext{2010 {\em Mathematics Subject Classification:\; primary 60G60, 82B20, 82B26}}
\footnotetext{{\em Key words and phrases:} hard-core model, sphere-packing, cubic lattice, FCC lattice, HCP structure, Pirogov-Sinai theory, phase transition, extreme periodic Gibbs measure, high-density/large fugacity, periodic ground state, perfect configuration, m-potential, local repelling forces, Peierls bound, dominance, computer-assisted 
enumeration, sliding, crystallographic point group $O_h$, integer quaternions, quadratic integer rings, sums of three squares

\noindent
$^1$ AMC Health, New York, NY, USA;\;\;
$^2$ Math Dept, Penn State University, PA, USA;,\;\;
$^3$ DPMMS, University of Cambridge and St John's College, Cambridge, UK.}

\maketitle

\begin{abstract}
We perform a rigorous study of the identical sphere packing problem in $\bbZ^3$ and of phase transitions in the corresponding hard-core model. The sphere diameter $D>0$ and the 
fugacity $u\gg 1$ are the varying parameters of the model. We solve the sphere packing problem for values $D^2= 2, 3, 4, 5, 6, 8, 9, 10, 11, 12, 2\ell^2$, $\ell\in\bbN$. For values $D^2=2, 3, 5, 8, 9, 10, 12, 2\ell^2$, $\ell\in\bbN$ and $u>u^0(D)$ we establish the diagram of periodic pure phases, completely or partially. For the case $D^2=2\ell^2$, $\ell\in\bbN$ we use results from Hales' proof of Kepler's conjecture.

%We perform a rigorous study of dense packings and the high-density  (or large-fugacity) 
%Gibbs distributions for hard spheres on a unit cubic lattice $\bbZ^3$, for varying values 
%%of the (Euclidean) diameter $D>0$. Only attainable values of $D$ are relevant, for which 
%$D^2=m^2+n^2+k^2$, $m,n,k \in\bbZ$.  In the first part of the work, we focus on values
%$D^2=2\ell^2$, $\ell\in\bbN$: here we identify natural collections of periodic maximally-dense sphere 
%packings (periodic ground states) and analyze their symmetries. Such dense-packings
%are constructed from FCC sub-lattices; for $\ell$ divisible by $3$ there is a rich family of dense-packings
%of a distinctive `layered' structure. A particular example is an HCP dense-packing.

%In the second part, we consider attainable values with $D^2\leq 11$, for which we 
%i%dentify all periodic ground states and characterize the large-fugacity extreme periodic 
%Gibbs distributions (periodic pure phases). The characterization is based on the Pirogov--Sinai 
%theory and its modifications;  in addition to the identification of the periodic ground states, it requires 
%the proof of a Peierls bound.  Some of the presented arguments are computer-assisted. 
\end{abstract}

\section{Introduction}\label{Sec1}

In this work we study the hard-core (HC) model of statistical mechanics on the integer lattice $\bbZ^3$. Lattice hard-core models attracted a considerable interest \cite{AGMS, AMMGPS, Ba1, Ba2, Bu, DKBP, Dob1, FAL, Ga, GF, HaPe, HePr, JMTR, JTMSR, JL1, JL2, NR1, NR2, TF, VMDDR}. The model is characterized by the value of fugacity $u$ and the (Euclidean) hard-core 
diameter $D$. The latter specifies the space of admissible configurations of the model. 
A configuration in $\bbZ^3$ (or simply, a configuration)
is a map  $\phi\,:\;x\in\bbZ^3\mapsto\{0,1\}$ which is identified with the set $\big\{x\in\bbZ^3:\;
\phi (x)=1\big\}$. It is convenient to think that a site $x\in\bbZ^3$ with $\phi (x)=1$ is occupied by 
a particle and a site $x\in\bbZ^3$ with $\phi (x)=0$ is vacant in configuration $\phi$. Given $D>0$, a configuration $\phi$ is called {\it $D$-admissible} (or {\it admissible}, for short)  if $\rho (x,x')\geq D$
for any pair of distinct occupied sites $x,x'\in\phi$ where $\rho$ stands for the Euclidean distance. 
An admissible configuration is referred to as AC or $D$-AC; it is interpreted as a configuration of non-overlapping open hard balls (HC particles) of diameter $D$. The set of ACs in $\bbZ^3$ is denoted by $\sA_D(\bbZ^3)$. 

It suffices to consider {\it attainable} $D\geq 1$ for which $D^2$ is integer, with $D^2=m^2+n^2+k^2$ 
where $m,n,k\in\bbZ$, as the model with a non-attainable $D$ is equivalent to the one with the nearest
larger
attainable value. A distance is attainable if it can be realized on $\bbZ^3$; throughout the paper we assume attainability without stressing it every time again.  

The other parameter characterizing the HC model is the {\it fugacity} (or activity) $u>0$.
The energy of a finite configuration $\phi\in\{0,1\}^{\bbZ^3}$ for fugacity $u$ is defined as 
\beq\label{eq:ener} H (\phi )=\begin{cases}
\big(-\ln u\big)\cdot \sharp (\phi), & \phi\in\sA_D(\bbZ^3),\\
+\infty, & \phi\not\in\sA_D(\bbZ^3).
\end{cases}\eeq 
Here and below, $\sharp (\phi )$ stands for the number of particles/occupied sites in $\phi$; a configuration $\phi$ with $\sharp (\phi )<\infty$ is referred to as finite.

A standard question in statistical physics is to describe a complete 
{\it phase diagram} (the structure of {\it pure phases}) for all
$u$ and $D$, which seems to be a problem beyond the reach for the existing methods.
In this paper we study {\it periodic} pure phases and focus on the case $u\gg 1$ when the system shows a tendency to have high 
particle density. This question is intrinsically related to the dense-packing problem of spheres of 
diameter $D$ on lattice $\bbZ^3$. Note that in the literature exist several terms: dense-packing, 
optimal dense-packing, close-packing, densest packing. In Section \ref{Sec2} we provide a formal 
definition of an equivalent  notion of a {\it perfect configuration}; this term stresses the fact that such an AC has no local defects. For $u>1$, a periodic dense-packing gives a {\it periodic ground state} (PGS) of the model \eqref{eq:ener}. 

The concept of a PGS carries a particular importance from the statistical physics point of view, as it is used in the Pirogov--Sinai theory \cite{PiS, Si, Za} for generating periodic pure phases (i.e., {\it extreme periodic Gibbs measures}, EPGMs for short) by means of 
{\it polymer expansions}. The identification of EPGMs is one of the principal goals of this paper. This identification describes the high-density periodic phase diagram of the model. The phenomenon of non-uniqueness of EPGMs corresponds to a phase transition \cite{Ge} which is a central question in the equilibrium statistical mechanics. 

Compared to the continuous cases of $\bbR^2$ and $\bbR^3$, the respective
dense-packing problems on lattices $\bbZ^2$ and $\bbZ^3$ are more complex, 
as the structure of dense-packings on 
$\bbZ^2$ and $\bbZ^3$ depends on arithmetic properties of $D$, 
%apparently, offers a richer variety of periodic dense-packing ACs.  
%depends on the value $D$ in a non-trivial way. In general, the 
whereas on $\bbR^2$ and $\bbR^3$ 
one can simply let $D=1$. The disk-packing problem on $\bbZ^2$ and other 2D lattices has been 
recently solved in \cite{MSS1, MSS2, MSS3}. The problem in $\bbZ^3$ appears to be harder than in $\bbZ^2$; it is related to famous Kepler's conjecture in $\bbR^3$ solved in 
\cite{Ha1, Ha2, LHF}. %There is no analog of the Kepler conjecture for $\bbZ^3$, and, as the current 
%paper shows, it is a non-trivial task to conjecture the structure of dense-packings for all values of $D$. 
At present, there is no analog of Kepler's conjecture for $\bbZ^3$, and, as the 
current paper 
shows, it is a non-trivial task to suggest the structure of dense-packings for an arbitrary value of $D$. Our work could be considered as an initial step in studying the problem of sphere-packings in $\bbZ^3$ and their random 
perturbations. %use the term the Kepler conjecture to mark this area of research. 

We begin with available analogies between the HC model in $\bbZ^3$ and those in $\bbZ^2$ and $\bbR^3$; one such analogy is the appearance
of  a sub-lattice as a PGS. 
Up to Euclidean motions, the only dense-packing of identical spheres in $\bbR^3$ which is a lattice is 
the FCC (face-centered cubic) lattice. Accordingly, the FCC lattice gives rise to a continuum family 
of layered dense-packing configurations in $\bbR^3$; see \cite{Ha1, 
Ha2, LHF}. %This family includes a countable subset of  periodic dense-packing 
%configurations, which is complete, up to Euclidean motions. 
One of these layered 
dense-packing configurations is known as HCP (hexagonal closed-packing).

A natural question is: for what values of $D$ there exist $D$-FCC sub-lattices in $\bbZ^3$, since in this case the entire family of $\bbZ^3$-PGSs is inherited from $\bbR^3$. This question can be answered by means 
of algebraic number theory. In particular, a dense-packing AC in $\bbZ^3$ which is a $D$-FCC 
sub-lattice exists iff $D^2=2\ell^2$ where $\ell\in\bbN$ \cite{Iona2009,Iona2007, Iona2016}.
%Among other things, it establishes a connection between sphere-packing problems on $\bbZ^3$ 
%and on an FCC-lattice. 
Moreover, for $D^2=2\ell^2$ there are finitely many $D$-FCC 
sub-lattices in $\bbZ^3$, typically more than one. In both $\bbR^3$ and $\bbZ^3$ it is natural to partition dense-packing ACs in general and dense-packing sub-lattices in particular into equivalence classes generated by isometries of $\bbR^3$ and $\bbZ^3$, respectively. In contrast to FCC in $\bbR^3$, not every $D$-FCC sub-lattice in $\bbZ^3$ generates a continuum of layered dense-packing ACs. It occurs iff $D^2=2\ell^2$ where $\ell\in\bbN$ and $\ell=0 \hskip-2pt\mod 3$. Furthermore, for a given $\ell$ there may be one or several equivalence classes of $D$-FCC sub-lattices, and their number
depends again on arithmetic properties of $\ell$. 
%As we said, for $D={\sqrt 2}l$ there are equivalence classes of 
%$D$-FCC sub-lattices of $\bbZ^3$ which cannot be taken into each other by $\bbZ^3$-symmetries. 
In the present work these classes and their symmetries have been studied in detail. % in terms of the $\bbZ^3$-symmetry group $O_\rh$.

%But, as in the case of $\bbR^3$, the absence of periodic dense-packing 
%configurations in $\bbZ^3$ different from the above ones has not been proven. 
A direct analogy with the case of $\bbR^3$ stops here; the structure of $\bbZ^3$-PGSs for a general $D^2\neq 2\ell^2$ remains unclear. Some of our examples show that $\bbZ^3$-PGSs are 
not necessarily layered. Nevertheless, one would like to believe that
for any attainable $D$ there are PGSs that are sub-lattices in $\bbZ^3$ (as it is in the case of $\bbZ^2$ \cite{MSS1}) 
but this remains an open (and probably, rather difficult) question. In the case of an affirmative answer, the
next question would be how to specify the PGS sub-lattices in terms of $D$.
 %even for $D={\sqrt 2}l$ there are difficulties in analyzing the phase 
%diagram of the model: 

As was mentioned before, in statistical physics the EPGMs are typically generated by some PGSs, 
but not necessarily by all of them. 
Due to the symmetries of the HC model, if a PGS generates an EPGM, then each PGS from the 
same equivalence class also generates an EPGM. 
Therefore, an important question is: which of those classes generate 
EPGMs? This question is naturally connected with the theory of {\it dominant} PGSs 
\cite{BS, Sl1, Za}. Our expectation is that in general only 
one PGS-equivalence class is dominant (i.e., generates EPGMs), but such a conjecture is far 
from being proven. 

A popular tool for identifying EPGMs is the Pirogov-Sinai (PS) theory. 
If there are only finitely many PGSs -- which happens in a number of cases considered in this paper --  
%then it suffices to use the original version of the PS theory. %In this situation it
then it suffices to verify the Peierls bound in a standard form proposed in \cite{PiS, Si, Za}; 
see \eqref{hamexcitation} below. However, we also discuss more subtle cases
where there are countably many PGSs (and a continuum of dense packings in total). Here we use 
the approach proposed in \cite{BS}. %to discuss the dominant PGS class for some $D\leq{\sqrt{12}}$. 

The problem of identifying the dense-packing ACs and the resulting high-density phase diagram for the 
HC model on $\bbZ^3$ seems to be rather involved. This work makes steps in this direction: for 
$D^2=2,3,5,8,9,10,12$, and for $D^2=2^{2n+1}$, $n\in\bbN$, we establish the complete diagram of 
periodic pure phases when $u$ is large enough ($u>u^0(D^2)$). The considered cases already 
demonstrate a richer variety of answers, compared with $\bbZ^2$ and $\bbR^3$.

For $D^2=3$ and $D^2=12$, the PGSs in $\bbZ^3$ turn out to be $D$-BCC (body-centered cubic) 
sub-lattices and their $\bbZ^3$-shifts; such structures are inherited by the EPGMs. However, this 
pattern seems exceptional, and we think that $D$-BCC sub-lattices do not arise as PGSs for larger 
values of  $D$. For $D^2=8, 9, 10$ the PGSs in $\bbZ^3$ turn out to be deformed $D$-FCC 
sub-lattices and their $\bbZ^3$-shifts; again, such structures persist in the corresponding EPGMs. 
In contrast to $\bbR^3$, these sub-lattices  do not generate additional layered dense-packings.  

However, for $D^2=5$ the deformed $D$-FCC sub-lattices do generate additional layered 
dense-packings. Consequently, there is an infinite degeneracy of PGSs. The 
ordered EPGMs are generated by deformed $D$-HCP configurations which form the only dominant PGS class. Surprisingly, it is deformed $D$-HCPs, rather than deformed 
$D$-FCCs, that generate EPGMs. 
This is due to the fact that $D$-HCPs have a larger density of a specific low-energy excitation. Such a phenomenon seems to be of a generic nature and is observed in other cases discussed below.

For $D^2=6$ the situation resembles the case of $D^2=5$ but is more involved. Similarly to 
$D^2=5$, there is a single class of dense-packing sub-lattices: it consists of deformed $D$-FCC 
sub-lattices and 
their $\bbZ^3$-shifts. However, in contrast to $D^2=5$, it generates two rather than one continuum 
families of layered dense-packings. 

For $D^2=4, 11$ we have a phenomenon 
of {\it sliding}: here there are countably many PGSs that are not separated enough from each other. 
The sliding phenomenon has been extensively studied 
for the HC model in 2D (\cite{MSS1, MSS2, Krachun, HaPe}); in particular, it was established that on $\bbZ^2$
there are only  39 values of $D$ with sliding. A relatively short analytic argument proves sliding for the case $D^2=4$. The case of $D^2=11$ is considerably more complicated. It requires a chain of analytical constructions which reduce the problem to a non-trivial computer enumeration. A complete list of sliding values of $D$ on 
$\bbZ^3$ remains open. %We conjecture that on $\bbZ^3$ for the values of $D$ with sliding, the 
%EPGM is unique, at least for large values of fugacity $u$, and use the term disordered when referring to this EPGM.

For $D^2=2\ell^2$, $\ell\in\bbN$, the structure of PGSs is inherited from $\bbR^3$. t is the proof of Kepler's conjecture by Hales \cite{Ha1, 
Ha2, LHF} which allows us to obtain the results listed below. In particular, for
$D^2=2^{2n+1}$, $n\in \bbN\cup\{0\}$, there is a single PGS class: it consists solely of 
$D$-FCC sub-lattices and their $\bbZ^3$-shifts, and there are no other dense-packings. Consequently, 
for $D^2=2^{2n+1}$ each PGS generates an EPGM. For all remaining values of $D^2=2\ell^2$ there 
exist at least two but finitely many PGSs classes, each consisting of $D$-FCC sub-lattices and their 
$\bbZ^3$-shifts. 

If $\ell\neq 0\mod 3$ then at least one 
of these classes generates EPGMs. 
We conjecture that there is always a unique dominant class, but we do not have a generic argument
covering all possible values of $\ell$. 

If $\ell= 0\mod 3$ then each class of $D$-FCC sub-lattices gives rise to a continuum family of layered dense-packings. For each class of $D$-FCC sub-lattices the corresponding family contains a class of $D$-HCP configurations. The type of excitations removing infinite degeneracy (similar to the one for $D^2=5$) does 
exist in each of these layered families, and therefore we expect that the 
corresponding $D$-HCP configurations are the only PGSs that can generate EPGMs when $u$ is large enough. We conjecture that among them only one class of $D$-HCP configurations is dominant.

All in all, we observe two types of infinite degeneracy of PGSs 
in $\bbZ^3$. In the first case, all dense-packings are split into one or more layered continuum
families. In this case we expect the system to be ordered 
and have a non-unique EPGM for $u$ large enough. In the current paper this fact is proved for $D^2=5$, 
but we believe that our proof reveals the core of the phenomenon and in principle can be extended to all values of $D$ with 
this type of PGS degeneracy. In the second case, the set of dense-packings is considerably wider 
and contains both layered and non-layered configurations. For example, such a 
situation occurs for the value $D^2=4, 11$ with sliding. Note that on 
$\bbZ^2$ the purely layered infinite degeneracy already constitutes sliding. 

A summary of our results/conjectures on PGSs and EPGMs is presented 
in Table 1.

\bigskip

In our opinion, a further progress in the understanding the HC model in $\bbZ^3$ may include answering 
the following open questions.

\begin{description}
\item{1)} Is it true that for every attainable $D$ there exists at least one dense-packing 
AC in $\bbZ^3$ which is periodic?
\item{2)} Is it true that for every attainable $D$ there exists at least one dense-packing 
AC which is a sub-lattice of $\bbZ^3$?
\item{3)} In the case of an affirmative answer to the previous question, is it possible to develop a 
number-theoretical description of those sub-lattices, similarly to the two-dimensional case?
\item{4)} What are the exact values of $D$ with sliding on $\bbZ^3$? Are there finitely many of them?
%\item{5)} Is the Gibbs measure unique (for all $u>0$) for sliding values of $D^2$?
\item{5)} Does there exist an m-potential representation of the energy in \eqref{eq:ener}
for all attainable values of $D$? 
\item{6)} Does some sort of a Peierls bound hold true for all values of $D$? 
\item{7)} Is it true that the dominant class of ground states is always unique? 
\end{description}

\begin{table}[H]\label{table1}
\centering
\begin{displaymath}
%\begin{array}{|rrrrl|}
%\boxed{
\begin{array}{|c|c|c|c|c|c|c|}
\hline
D^2 &\bma\sharp\;\;{\rm{of}}\\{\rm{PGS}}\\ {\rm{classes}}\ema &\bma{\rm{total}}\;\,\sharp\;{\rm{of}}\\ 
{\rm{PGSs}}\ema
&\bma{\rm{PGS}}\\ {\rm{type}}\ema&{\rm{density}}&\bma{\rm{total}}\;\,\sharp\;{\rm{of}}\\
{\rm{EPGMs}}\ema&\bma{\rm{EPGM}}\\ {\rm{type}}\ema\\ 
\hline
1&1&1&{\mathbb Z}^3&1&1&{\mathbb Z}^3\\ 
\hline
2&1&2&{\rm{FCC}}&1/2&2&{\rm{FCC}}\\
\hline
3&1&4&{\rm{BCC}}&1/4&4&{\rm{BCC}}\\
\hline
4&\aleph_0&\aleph_0&{\rm{sliding}}&1/8& &\\
\hline
5&\aleph_0&\aleph_0&{\rm{layered}}&1/9&72&{\rm{dHCP}}\\
\hline
6&\aleph_0&\aleph_0&{\rm{layered}}&1/12& &\\
\hline
8&1&16&{\rm{FCC}}&1/16&16&{\rm{FCC}}\\
\hline
9&1&120&{\rm{dFCC}}&1/20&120&{\rm{dFCC}}\\
\hline
10&1&208&{\rm{dFCC}}&1/26&208&{\rm{dFCC}}\\
\hline
11&\aleph_0&\aleph_0&{\rm{sliding}}\,&1/32& &\\
\hline
12&1&32&{\rm{BCC}}&1/32&32&{\rm{BCC}}\\
%\hline
%13&1&140&{\rm{dFCC}}&1/35&140*&{\rm{dFCC}}\,*\\
\hline\hline
2^{2n+1}&1&2^{3n+1}&{\rm{FCC}}&1/2^{3n+1}&2^{3n+1}&{\rm{FCC}}\\
\hline
2l^2,3\hskip-5pt\not|l&{\rm{finite}}&{\rm{finite}}&{\rm{FCC}}&1/(2l^3)& \rm{finite} &{\rm{FCC}}\\
\hline
2l^2,3|l&\aleph_0&\aleph_0&{\rm{layered}}&1/(2l^3)&\rm{finite}\,*&{\rm{HCP}}\,*\\
\hline
\end{array}
\end{displaymath}
\caption{{\bf A summary of results and conjectures on $\bbZ^3$}\newline
Here $\sharp$ stands for cardinality, $*$ marks a conjectured 
prediction, prefix d means deformed and an empty cell means that the question is open. The table shows 
the number of PGS equivalence classes, followed by the total number of PGSs and their type, 
together with the density of occupied sites in PGSs (the maximal packing density). Next, 
it indicates the total 
number of periodic pure phases/EPGMs and their types.} %The acronyms and technical terms featured in the table have been introduced above; their detailed meaning is provided in the corresponding sections of the paper. 
\end{table}

This paper includes $11$ sections. In Section \ref{Sec2} we introduce necessary technical concepts. In Sections \ref{Sec:D=234} - \ref{SecD=11} we consider the cases
of values of $D^2\leq 12$. % for which we establish our results on periodic dense-packing 
%configurations and the phase diagram.  
Section \ref{SecD=2llEPGM} contains the analysis for $D^2=2\ell^2$, $\ell\in\bbN$. Section \ref{AppendixA} is an appendix containing the description of cubic and FCC $\ell$-sub-lattices in $\bbZ^3$, depending on arithmetic properties of $\ell$. %of the cubic 
%and FCC $l$-sub-lattices in $\bbZ^3$ and related PGSs, for $D={\sqrt 2}l$, $l\in\bbN$, while in Section 
% we collect results on the EPGMs for these values of $D$. 
 Section \ref{AppendixB} is an appendix containing a note on the PS theory for an infinite/hard-core potential.

\section{The HC model: m-potentials, local repelling forces, perfect configurations,
sub-lattices and meshes}\label{Sec2}

%\subsection

%{\large{\bf 2.1. Gibbs distributions}}\vskip 1cm 

We start with the definition of the HC model of statistical mechanics. The model is defined on the unit cubic lattice $\bbZ^3$. The configuration space is 
\beq\sA_D\big(\bbZ^3\big):=\big\{\phi\in\{0, 1\}^{\bbZ^3} : \rho(x',x'')\geq D,\; \hbox{whenever} \;\phi(x')\phi(x'')=1,  \; x', x''\in \bbZ^3 \big\}.\eeq
The energy $H(\phi )$ of a finite configuration $\phi\in\sA_D(\bbZ^3)$ is given by \eqref{eq:ener}, and the model is characterized by two varying parameters, $u$ and $D$. If $\phi\in\sA_D(\bbZ^3)$ is not finite then \eqref{eq:ener} should be understood as a formal sum. 

%Throughout the paper we consider (unit) lattices $\bbZ^s$, $s=1, 2, 3$, canonically embedded in 
%$\bbR^s$ with 
%the standard Euclidean distance $\rho$. As was said earlier, the hard-core model is
%determined by the configuration space $\sA_D(\bbZ^3)$ 
%and energy $H(\phi )$, $\phi\in\sA_D(\bbZ^3)$, as in \eqref{eq:ener}. 
%Let us stress again that the model is characterized by two varying parameters, $u$ and $D$. 

We analyze the phase diagram of the model focusing on periodic pure phases in a large fugacity regime. %A standard goal is to analyze the {\it phase diagram} of the model, i.e., 
The goal is to identify the set $\sE(D, u)$ of EPGMs \cite{Ge} for given values of $u$ and $D$. The main assumption throughout the paper is that the value $u$ is large: $u\geq u^0$
where $u^0=u^0(D^2)\in (1,\infty )$.  Our approach follows the PS theory \cite{PiS, Si} and its 
developments \cite{BS, HS, Za}. The PS theory is based on the concept of a {\it ground state} (GS).  For the model \eqref{eq:ener}, a GS is an AC  $\vphi\in\sA_D(\bbZ^3)$ such that the energy $H(\vphi)$ cannot be diminished by 
admissible perturbations localized in some lattice ball $\rB_s(x)=\{y\in\bbZ^3:\;\rho (x,y)<s\}$. 
%  in $\bbZ^3$ of radius $s>0$,  centered at $x$, and $o$ marks the origin. 
Observe, that for $u>1$, a GS $\vphi$ is an AC in which one cannot increase the number of particles in any ball $\rB_s(x)$.

Typically, the PS theory works with PGSs and aims at constructing EPGMs 
with the help of absolute convergent {\it polymer expansions} around a PGS \cite{Za}. In this case we say that a PGS generates 
an EPGM and an EPGM is generated by a PGS. We want to stress that the EPGMs generated by different 
PGSs are mutually singular \cite{Ge}.

%We denote the set of the GSs by 
%$\sG =\sG (u,D)$ and its subset formed by the PGSs by $\sG^{\rm{per}} =\sG^{\rm{per}} (u,D)$.

A convenient way to identify GSs
is to represent the energy in terms of a suitable {\it m-potential} \cite{HS}: for any finite $\phi\in\{0,1\}^{\bbZ^3}$   
\beq\label{hamiltonian} H (\phi )=\sum_{x\in\bbZ^3}U \big(\phi\upharp_{\rB_s(x)}\big).\eeq
%where $\sA^\rf$ is the set of finite ACs, with $\sharp (\phi )<\infty$.
Here and below $\upharpoonright$ stands for restriction. The constant $s\geq D$  is independent of $\phi$ and $x$ but may vary with $D$. By definition, in the HC model $U \big(\phi\upharp_{\rB_s(x)}\big)=+\infty$ if $\phi\upharp_{\rB_s(x)}$ is not admissible. It takes finite values if $\phi\in\sA_D\big(\bbZ^3\big)$. We would like to stress that the summation in \eqref{hamiltonian} is over all $x\in\bbZ^3$, including the vacant sites in AC $\phi$. Following \cite{HS}, we say that $U(\cdot)$ is an m-potential (for the HC model with given $u,D$) 
if there exists a configuration $\vphi\in\{0,1\}^{\bbZ^3}$, such that 
\beq\label{m-pot} U \big(\vphi \upharp_{\rB_s(x)}\big) = U^0\;\;\;\forall\;x\in \bbZ^3,\; \hbox{ where }\;
 U^0:=\min_{x\in\bbZ^3,\, \phi\in \{0,1\}^{\bbZ^3}} U \big(\phi\upharp_{\rB_s(x)}\big)<+\infty.
% \min\left[ U (\psi ):\,\psi\in\sA_D(\rB_s(o))\right].
\eeq
% (Here m- means minimal.) 
In particular, $\vphi\in\sA_D\big(\bbZ^3\big)$. In what follows we always choose $U(\cdot)$ to be translation invariant. Hence, $U^0=\min\left[ U (\psi ):\,\psi\in\sA_D\big(\rB_s(o)\big)\right]$, where $o$ denotes the origin, and 
\beq\beal \sA_D\big(\rB_s(x)\big):=\big\{\phi\in\{0, 1\}^{\rB_s(x)} : \rho(x',x'')\geq D,\\ \qquad\qquad\qquad\qquad\qquad\qquad\qquad\hbox{whenever} \;\phi(x')\phi(x'')=1,  \; x', x''\in \bbZ^3 \big\}.\ena\eeq
Equivalently, $\psi\in\sA_D\big(\rB_s(x)\big)$ is expressed as $\psi\subseteq\rB_s (x)$ as an AC is  identified with the corresponding set of occupied sites. 
% $\sA_D(\rB_s(x))$ refers to the set of the ACs $\psi$ with 
 %$\psi\subseteq\rB_s (x)$. 

We call an AC $\varphi$ satisfying \eqref{m-pot} a {\it perfect configuration} (PC). It is clear that if 
$\vphi$ is a PC then all its $\bbZ^3$-shifts $\vphi + x$, $x\in\bbZ^3$, are PCs. Next, if $\vphi$ is a 
periodic PC then the collection of its shifts is finite and consists of $v(\sP(\vphi))$ distinct PCs, 
where $\sP(\vphi)$
denotes a (lattice) fundamental parallelepiped for $\vphi$ and $v(\sP(\vphi))$ stands for the Euclidean 
volume of % number of $\bbZ^3$-sites in 
$\sP(\vphi)$. Observe, that for the HC model, the notion of a PC coincides with the notion of a 
dense-packing. Consequently, the notion of a periodic PC coincides with the notion of a PGS. 

The representation 
\eqref{hamiltonian} of $H$ in terms of an m-potential $U(\cdot)$ (if it exists) is non-unique, but the set of 
PCs does not depend on the specific choice of the m-potential. Furthermore,
if an m-potential exists then a PC is always 
a GS. Conversely, if an m-potential exists then a PGS is always a PC. Owing to this fact, we will freely pass 
from periodic PCs to PGSs and back. However, we note that a non-periodic GS is not necessarily a PC. 

An advantage of representing $H$ 
in terms of an m-potential $U(\cdot)$ is that it leads to a convenient description of 
possible perturbations of a given PC $\vphi$. Suppose an AC $\phi$ differs from $\vphi$ on 
a finite set  in $\bbZ^3$ and consider the union $V$ of balls $\rB_s(x)$ such that \\
$U \big(\phi\upharp_{\rB_s(x)}\big)>U^0$. %Next, take $V$ and 
%unite it with all balls $\rB_s(x)$ that intersect $V$. The resulting set is denoted by $W$. 
Then, according to \cite{BS}, the restriction $X:=\phi \upharp_V$ is called an 
{\it elementary excitation}, when $V$ has a single connected component. In this case, the set $V$ 
is called the {\it support} of $X$ and is denoted by $Supp(X)$; its cardinality is denoted by $v\big(Supp(X)\big)$.
(See definitions on pp. 104-105 and comment 5 on p. 118 in \cite{BS}.) This matches the definition of a 
{\it contour} in \cite{PiS, Si, Za}. Owing to a discrete character of our HC model, the quantity 
$$U^1:= \min\Big[ U (\psi ):\,\psi\in\sA_D(\rB_s(o)),\; U (\psi )> U^0\Big]$$ 
yields $U^1 >U^0$.

For an elementary excitation $X$, we 
denote by $v(X)$ the cardinatily of the set of $x\in Supp(X)$ such that $U\big(\phi\upharp_{\rB_s(x)}\big) > U^0$; %(the number of $\bbZ^3$-sites in) the
%union of the corresponding balls $\rB_s(x)$ such that
%$U\big(\phi\upharpoonright_{\rB_s(x)}\big) > U^0$;
 see p. 106 in \cite{BS}. Also, denote by $v(\rB_s(o))$ 
the cardinality of the lattice ball $\rB_s(o)$. Consider a PGS $\varphi$ and an 
AC $\phi$ which differs from $\varphi$ in a single elementary excitation $X$. Then (cf. the last equation 
on p. 106 in \cite{BS}):
\beq\label{hamexcitation}\beal
H(X) := H(\phi) - H(\varphi) = \sum\limits_{\rB_s(x)\subseteq X} \Big( U\big(\phi  \upharp_{ \rB_s(x)}\big) 
-U\big(\varphi  \upharp_{ \rB_s(x)}\big)\Big)\\ \qquad\qquad\qquad
\geq\dfrac{\big(U^1 - U^0\big) v(X)}{v(\rB_s(o))}\geq\dfrac{\big(U^1 - U^0\big) v\big(Supp(X)\big)}{(v(\rB_s(o)))^2}.\ena\eeq
The bound \eqref{hamexcitation} is known as the {\it Peierls bound} and the value $\dfrac{U^1 - U^0}{(v(\rB_s(o)))^2} > 0$ 
as the {\it Peierls constant} (in our case it is proportional to $\ln\,u$). The Peierls bound is a key ingredient in the PS theory \cite{PiS, Si, Za}. In this paper, the form of Peierls bound \eqref{hamexcitation} suffices 
for all considered cases, except for $D^2=5$. For $D^2=5$, the PGS family is countably infinite 
which requires a more involved version of the Peierls bound; cf. (2.9) in \cite{BS}.

Let $\vphi'$ and $\vphi''$ be different PCs. Take a
parallelepiped $\sP$ and consider $\vphi' \upharp_{\bbZ^3\setminus \sP}\cup\, \vphi'' \upharp_{\sP}$. 
The resulting configuration is not necessarily admissible and some uniquely defined set of particles 
$\alpha=\alpha(\vphi', \vphi'') \in \vphi' \hskip-1pt\upharp_{\bbZ^3\setminus \sP}$ needs to be removed from this 
configuration to restore the admissibility. Suppose there exist PCs $\vphi'$ and $\vphi''$ such that one can construct 
a monotonically increasing sequence of parallelepipeds $\sP_n$, with $\sharp (\alpha_n) <C$, where $\alpha_n$ 
is the corresponding set of removed particles, $\sharp (\alpha_n)$ is its cardinality and $C$ does not depend on $n$. 
Then we say that the phenomenon of {\it sliding} is exhibited for the corresponding value $D$. %{\color{red}The presence of sliding indicates that the Peierls bound fails even in the form of (2.9) from \cite{BS}. To be more 
%precise, the retouch property required for (2.9) is violated. 
A typical scenario of sliding is when some straight 
line of occupied sites in a PC can be shifted (along itself) without breaking the admissibility of the configuration 
(and consequently resulting in another PC).

With m-potential at hand without lost of generality one can restrict all considerations to admissible configurations only. Having this in mind throughout this paper we use the following formalism to construct m-potentials $U(\cdot)$. Given an 
attainable value $D^2$, a {\it local repelling force family} (LRFF) is defined as a real function 
$(x,y)\in\bbZ^3\times\bbZ^3\mapsto {\wh f}(x,y)\geq 0$ such that $\forall$ $x,y\in\bbZ^3$: (i)  ${\wh f}(x,y)\not\equiv0$, 
(ii) ${\wh f}(x,y)={\wh f}(y,x)$, %(iii) ${\wh f}(0):=1$, 
(iii) ${\wh f}(x,y)=0$ if $\rho (x,y)\geq D$.  In all 
cases under consideration in this paper, ${\wh f}(x,y)=f({\rho(x, y)}^2)$, i.e., the local force depends only on 
the Euclidean distance between $\bbZ^3$-sites. %; in addition ${\wh f}(x,y)\geq 0$, i.e., the local force is {\it repelling}. 
Consequently, an LRFF is identified with the collection of real 
values $\unf^{(D^2)}=\{f(\rho(x,y)^2):\;0\leq\rho (x,y)<D\}$.  

Working with balls $\rB_s(x)$ we omit subscript $s$ when its value is irrelevant or clear from the context.
Given $x\in\bbZ^3$ and an AC $\psi\in\sA_D\big(\rB (x)\big)$, define
\beq F (\psi):=\sum\limits_{y\in\psi}f({\rho(x,y)}^2) \equiv \sum\limits_{y\in \rB (x)}\psi(y)f({\rho(x,y)}^2),\eeq
the total force acting on site $x$ in $\psi$.
Next, set
\beq F^*(x):=\max\;\Big[F (\psi):\;\psi\in\sA_D\big(\rB (x)\big)\Big].\eeq
If there exists an AC $\varphi \in\sA_D\big(\bbZ^3\big)$ such that 
$F \left(\vphi\upharp_{\rB (x)}\right)=F^*(x)$ for all $x\in\bbZ^3$ then
 $-F(\phi\hskip-1pt\upharp_{\rB(x)})$ is an m-potential. In this case we say that an LRFF $\unf^{(D^2)}$ 
generates an m-potential $F(\phi\upharp_{\rB(x)})$ or, in brief, that $\unf=\unf^{(D^2)}$ is an 
m-LRFF (for a given $D^2$). 

Observe that for any finite AC $\phi$
\beq\beacl\label{F} \;\sum\limits_{x\in\bbZ^3}F\left(\phi\hskip-2pt\upharpoonright_{\rB(x)}\right)
&=\sum\limits_{x\in\bbZ^3}\sum\limits_{y\in\phi}f({\rho(x,y)}^2)=\sum\limits_{y\in\phi}\sum\limits_{x\in\bbZ^3}f({\rho(x,y)}^2)
=C \sharp (\phi ),\ena\eeq
where
\beq\label{eq:C} C:=\sum\limits_{x\in\bbZ^3}f({\rho(x,y)}^2)
=\sum\limits_{x\in\rB (y)}f({\rho(x,y)}^2).\eeq
Let
\begin{equation}
U\left(\phi\hskip-2pt\upharpoonright_{\rB(x)}\right) := - \diy \ln(u)\frac{1}{C} F\left(\phi\upharpoonright_{\rB(x)}\right)
\end{equation}
then $U\left(\phi\hskip-2pt\upharpoonright_{\rB(x)}\right)$ gives an m-potential for $H(\phi)$ from 
\eqref{eq:ener}. 

It is convenient to select $\unf$ in such a way that $F^\ast=1$. The motivation comes from the 
following way of constructing elementary excitations of a PC. Take a PC $\vphi$, 
insert particles at a finite collection of sites 
$\xi =\{x_i\}\in\sA_D\big(\bbZ^3\big)$, and remove those particles from $\vphi$ that have been repelled. 
(A repelled particle $y\in\vphi$ is the one at distance $< D$ from an inserted site $x\in\xi$.) 
Let $\eta$ denote the collection of sites where the repelled particles are located: 
$\eta =\eta (\vphi,\xi)\subseteq\vphi$. Then the resulting AC is $\phi =(\vphi\setminus\eta)\cup\xi$.

Let $\sharp  (\xi )$ and $\sharp (\eta )$ be the numbers of inserted and repelled 
particles in a PC $\vphi$, respectively. The difference $\sharp  (\xi ) - \sharp (\eta )$, i.e. the energy of the 
excitation, is counted as follows. Let $x\in\xi$ be an inserted site and suppose it repels particles located 
in $Y(x)=Y(x,\vphi )\subseteq\vphi$,
so that $\eta =\onwl{\cup}\limits_{x\in\xi}Y(x)$. Take an LRFF $\unf=\{f(\rho (x,y)^2)\}$,  then 
$$\sum\limits_{y\in\eta}\left[1-\sum\limits_{x\in\xi}f(\rho (x,y)^2) \right]=\sharp (\eta )-\sharp  (\xi )$$
because for a PC $\vphi$ one has $\sum\limits_{y\in\eta}f(\rho (x,y)^2)=F^\ast=1$. 
The quantity 
\beq\label{E} E(y):=1-\sum\limits_{x\in\xi}f(\rho(x,y)^2)\geq 0 \eeq
is called an {\it excess} for the repelled particle $y\in\vphi$ (under insertion $\xi$). In the bound  \eqref{E} we use the symmetry of function $\wh{f}$ and, consequently, of $f$. Correspondingly,  
\beq\sharp (\eta )-\sharp  (\xi )=\sum\limits_{y\in\eta}E(y).\eeq
The above argument provides a motivation for calling the quantity $f(\rho (x, y)^2)$ a local  repelling force. 

The majority of PCs considered below have specific {\it layered} structures. One of them is the densest FCC sub-lattice of $\bbZ^3$: which we refer to as  $\bbA_3$ (with a slight mishandle of 
notation): 
\beq\label{eq:A3i}\beac\bbA_3:=\big\{m\,(1,1,0) + n\,(1,0,1) + k\,(0,1,1) :\, m, n, k\in\bbZ\big\}.  %\\
\ena\eeq
Here and below we write a point $x\in\bbR^3$ as a vector with a triple of Cartesian coordinates
$(\rx_1,\rx_2,\rx_3)$ where $\rx_i\in\bbR$. The addition of vectors and multiplication by scalars are done in the usual manner, and the shift of a set $\chi\subset\bbR^3$ by a vector $x\in\bbR^3$
is represented as $\chi +x$. We use the term a {\it mesh} for a subset of $\bbR^3$, 
congruent to a two- or three-dimensional lattice but not necessarily containing the origin. A mesh 
congruent to a two-dimensional lattice where generating vectors form an equilateral
triangle is referred to as triangular. If the equilateral triangle has side-length $b$, we call it
a triangular $b$-mesh or -- if it is a lattice -- a triangular $b$-lattice. More generally, a mesh 
congruent to a
lattice with two generating vectors of equal length is called rhombic.  A mesh with 
orthogonal generating vectors is referred to as rectangular; if in addition the generating vectors 
are of equal length, we say it is a square mesh. The term a square $b$-mesh is used when 
generating vectors have length $b$.

Layered PCs considered in this paper are unions of equidistant parallel two-dimensional meshes. The first kind of such PCs uses planes orthogonal to a 
{\it main diagonal}
in $\bbR^3$. %Given $s_2, s_3 \in \{-1,1\}$, we obtain the 4 planes  $\rx_1+s_2 \rx_2 + s_3 \rx_3 = 0$ 
%orthogonal to the 4 main diagonals. Each main diagonal is parallel to a vector $(1,s_2,s_3)$.
We use index $i=0, 1, 2, 3$ to enumerate the 4 main diagonals $(1, s_2(i), s_3(i))$: 
\beq\label{eq:1s2s3}\beal (1, s_2(0), s_3(0))=(1, 1, 1), \qquad(1, s_2(1), s_3(1))=(1, -1, 1),\\
(1, s_2(2), s_3(2))=(1, -1, -1), \;\quad (1, s_2(3), s_3(3))=(1, 1 , -1).\ena\eeq

%The index 
%\beq\label{eq:is1s2} i =i(s_1,s_2)= (3-s_2-2 s_3) /2\eeq
%takes values $0,1,2,3$ and is used to identify 
%these planes and diagonals. Conversely, given $i \in \{0,1,2,3\}$, we have 
%\beq\label{eq:s1s2i}s_2(i) =1- 2\;[i/2]\;\hbox{ and }\;s_3(i) =1- 2\;(i \bmod 2),\eeq
%where $[\;\cdot\; ]$ stands for the integer part.

Given 
$k \in \bbZ$, denote by $T_{i, k}$ the projection of $\bbZ^3$ into the affine plane 
$$\rx_1+s_2(i)\rx_2+s_3(i)\rx_3=k,\quad k\in\bbZ,$$ 
orthogonal to $(1, s_2(i), s_3(i))$. Then 
\beq\label{eq:Tik1}\beal T_{i, k}=\left\{m\left(\dfrac{1}{3},-\dfrac{2s_2}{3},\dfrac{s_3}{3}\right)
+n\left(-\dfrac{1}{3},-\dfrac{s_2}{3},\dfrac{2s_3}{3}\right),\;m,n\in\bbZ\right\}+\dfrac{k}{3}(1,s_2,s_3),\,\ena\eeq
is a triangular $\sqrt{2/3}$-mesh; we refer to it as basic. Given $q\in\bbN$ 
and $i$, $k$ as above, we work with a finite family of triangular ${\sqrt q}$-sub-meshes 
$\tau^{(q)}_{i, k, j}\subset T_{i, k}$, labelled by $j=0,1,\ldots ,r$. Sub-meshes $\tau^{(q)}_{i, k, j}$ forming this
collection are obtained as some $\bbZ^3$-shifts of each other. In addition,  we fix an 
$h=h(q)\in \bbN$, and, given  a double-infinite sequence 
$\{j_k,\,k\in\bbZ\}$ with digits $j_k=0,1,... ,r$, consider the layered AC
\beq\label{layered1}%\om_{i, \{j_k\}}=
\bigcup_{k \in \bbZ} \tau^{(q)}_{i, hk, j_k}\eeq
and its $\bbZ^3$-shifts. %This construction can be used to obtain PCs for all values of 
%$D^2$ considered in this paper, although in a number of cases we provide a 
%simpler PC design. 
In case when we use the above construction we specify 
the corresponding values $q$, $h(q)$, the family of meshes $\tau^{(q)}_{i, hk, j}$, $0\leq j\leq r$, 
and the type of sequence $\{j_k\}$ allowing us to obtain a PC for a given $D^2$ via \eqref{layered1}. An 
important class is formed by periodic layered PCs where the corresponding sequence $\{j_k\}=\ov{j_0...j_p}$ is obtained by repeating a finite string $j_0...j_p$.

The second kind of layered PCs uses planes orthogonal to {\it non-main diagonals}. As 
above, we use index $i=0,...,5$ to enumerate the 6 non-main diagonals $(s_1(i), s_2(i), s_3(i))$: 
\beq\label{eq:s1s2i2}\beal (s_1(0), s_2(0), s_3(0))=(1, 1, 0), \quad (s_1(1), s_2(1), s_3(1))=(-1, 1, 0);\\
(s_1(2), s_2(2), s_3(2))=(1, 0, 1), \quad (s_1(3), s_2(3), s_3(3))=(-1, 0 , 1),\\
(s_1(4), s_2(4), s_3(4))=(0, 1, 1), \quad (s_1(5), s_2(5), s_3(5))=(0, -1 , 1).\ena\eeq
%Then the planes $s_1(i)\rx_1+s_2(i)\rx_2+s_3(i)\rx_3=0$ are orthogonal to the corresponding 
%non-main diagonal $i$. In the current paper, this kind of layered PCs is used for $D^2=6$.

Similarly to the previous construction, given $k\in\bbZ$, we denote by $Q_{i, k}$ the projection of $\bbZ^3$ to the affine 
plane 
$$s_1(i)\rx_1+s_2(i)\rx_2+s_3(i)\rx_3=k,\quad k\in\bbZ,$$ orthogonal to non-main diagonal $(s_1(i), s_2(i), s_3(i))$. Then 
\beq\label{eq:Qik1}\beal Q_{i, k}=\left\{m\cdot a
+n\cdot b,\;m,n\in\bbZ\right\}+\dfrac{k}{2}(s_1(i),s_2(i),s_3(i)),\,\ena\eeq
where \begin{equation}\label{a} a:=\left(1-|s_1(i)|,1-|s_2(i)|,1-|s_3(i)|\right)\end{equation} and 
\begin{equation}\label{b}\beal b:=
\diy\frac{1}{2}\Big(s_2(i)-s_2(i)|s_3(i)|-s_3(i)+s_3(i)|s_2(i)|, \\ \qquad\qquad\qquad s_3(i)-s_3(i)|s_1(i)|-s_1(i)+s_1(i)|s_3(i)|, \\ \qquad\qquad\qquad\qquad s_1(i)-s_1(i)|s_2(i)|-s_2(i)+s_2(i)|s_1(i)|\Big).\ena\end{equation}

It is not hard to see that $Q_{i, k}$ is a rectangular $(1\times\sqrt{2}/2)$-mesh %congruent to $\bbZ\times\left(\bbZ/\sqrt{2}\right)$. 
which we again call basic. Given $q_1, q_2\in\bbN$ and $i,k$ as above, we work with a finite family of rhombic $\sqrt{q_1},\sqrt{q_2}$-sub-meshes $\alp^{(q_1,q_2)}_{i, k, j}\subset Q_{i,k}$, 
$j=0,1,\ldots, t$. Here ${\sqrt{q_1}},{\sqrt{q_2}}$ are the lengths of the diagonals in a rhombus with side-length
$\sqrt{q_1 + q_2}/2$ emerging in the analysis. As above, sub-meshes $\alp^{(q_1,q_2)}_{i, k, j}$ are obtained 
as some $\bbZ^3$-shifts of each other. Next, we fix an $h=h(q_1, q_2)\in \bbN$ and, given a double-infinite 
sequence $\{j_k,\,k\in\bbZ\}$ with digits $j=0,1,\ldots, t$, consider the layered AC
\beq\label{layered2}%\varpi_{i, \{j_k\}}=
\bigcup_{k \in \bbZ} \alpha^{(q_1, q_2)}_{i, hk, j_k}\eeq
and its $\bbZ^3$-shifts. The value $h$ and the precise form of meshes $\alp^{(q_1,q_2)}_{i, hk, j}$
are chosen so that the ensuing layered configuration \eqref{layered2} gives a PC. In case when we use the above construction we specify 
the corresponding values $q_1, q_2$, $h(q_1, q_2)$, the family of meshes $\alpha^{(q_1, q_2)}_{i, hk, j}$, $0\leq j\leq t$, 
and the type of sequence $\{j_k\}$ allowing us to obtain a PC for a given $D^2$ via \eqref{layered2}.

Finally, an analogous construction can be performed when we choose planes $\rx_i=0$, $i=1,2,3$,
orthogonal to the {\it coordinate axes}. %This version is used for $D^2=9$. 
The projection $Z_{i,k}$ of 
$\bbZ^3$ to the affine plane $\rx_i=k$ is congruent to the two-dimensional square lattice $\bbZ^2$, and, 
given $q\in\bbN$,  we work with a finite family of square $\sqrt{q}$-sub-meshes $\theta^{(q)}_{i, k, j}\subset Z_{i,k}$, 
$j=0,1,\ldots, u$. As before, sub-meshes $\theta^{(q)}_{i, k, j}$ are obtained as some 
$\bbZ^3$-shifts of each other. Again, we fix an $h=h(q)\in \bbN$ and, given a double-infinite sequence 
$\{j_k,\,k\in\bbZ\}$ with digits $j_k=0,1,\ldots ,u$, consider the layered AC 
\beq\label{layered3}%\zeta_{i, \{j_k\}}=
\bigcup_{k \in \bbZ} \theta^{(q)}_{i, hk, j_k},\eeq
and its $\bbZ^3$-shifts. The value $h$ and the precise form of meshes $\theta^{(q)}_{i, hk, j_k}$
are chosen so that the ensuing layered configuration \eqref{layered3} gives a PC. In case when we use 
the above construction we specify 
the corresponding values $q$, $h(q)$, the family of meshes $\theta^{(q)}_{i, hk, j}$, $0\leq j\leq u$, 
and the type of sequence $\{j_k\}$ allowing us to obtain a PC for a given $D^2$ via \eqref{layered3}.

In the forthcoming sections we establish 
a number of theorems and put forward some conjectures on PCs and EPGMs by following a 
standardized scheme. The main step in the analysis of PCs
is the identification of an LRFF $\unf^{(D^2)}$ generating an m-potential $U(\cdot)$ for a 
given $D^2$. First, we select the values $f(\rho(x,y)^2)$. Next, we verify that $F^\ast = 1$;
it includes -- for $D^2=5, 6, 8, 9, 10, 12$ -- the use of a computer routine {\tt{VerifyForces.java}}. 
% which needs a finite (and at times massive) enumeration. 
Then we present a single PC which justifies the m-potential property of the proposed family 
$\unf^{(D^2)}$. Finally, we use results of {\tt VerifyForces.java} combined with appropriate analytic arguments 
to enlist all PCs for $\unf^{(D^2)}$ and, consequently, for the value $D^2$. This solves the sphere-packing problem on
$\bbZ^3$ for the values of $D^2=2, 3, 4, 5, 6, 8 ,9, 10, 12$. %; in simple cases it is a straightforward inspection. 
The case of $D^2=11$ uses an alternative computer-assisted approach. A different scheme is used for 
the values $D^2=2\ell^2$, $\ell\in\bbN$, %considered in Sections \ref{SecD=2llPGS}, \ref{SecD=2llEPGM}, 
where the m-potential emerges from results of \cite{Ha1, Ha2, LHF}. In both cases the ensuing analysis of the EPGMs 
is based on the PS theory and its extensions. 
%Peierls bound \eqref{hamexcitation} which is verified in each case individually.

Finally, we would like to note a difference between HC models on $\bbZ^2$ and $\bbZ^3$ which
is that on $\bbZ^3$ -- in contrast to 
$\bbZ^2$ -- there are non-periodic extreme Gibbs measures. The analysis of such Gibbs measures can be 
carried similarly to \cite{Dob2}, but this is outside of the scope of this paper.

%Typically, in a situation with finitely many PGSs one cannot pass, in 
%an AC $\phi$, from a domain occupied by one PGS to that occupied by another PGS, without the presence 
%of a separating boundary formed by $\rB_s(x)$'s with $U\big(\phi  \upharpoonright_{\rB_s(x)}\big) > U^0$. 
% The condition
%that the set of PGSs is finite is a key ingredient of the PS theory, although it can be relaxed under
%suitable requirements.

%####################################################
%                                                                                                                D^2=2,3,4
%###################################################

\section{Cases $D^2=2,3,4$}\label{Sec:D=234}

\subsection{$D^2=2$.}\label{subsec:D=2} 
This is historically the first case, considered in \cite{Dob1}. %In terms
%of this work, the case $D^2=2$ is defined by the m-potential. 
For $D^2=2$, an LRFF family $\unf=\unf^{(2)}$ has  
\begin{equation}\label{fD=2}f(0)=1, f(1)=\dfrac{1}{6}.
\end{equation} 
Here the balls $\rB (x)=\rB_{\sqrt{2}}(x)$ contain $7$ sites. Each $\psi\in\sA_{\sqrt{2}}\big(\rB (x)\big)$ 
%We work with $D$-ACs 
%$\psi\subset\rB (x)$ for $D={\sqrt 2}$: each such 
%$\psi$ 
contains at most $6$ particles. Consider the potential function $U(\psi)=U^{(2)}(\psi)$:
\beq\label{eq:U01}U (\psi)=\dfrac{1}{2}\sum_{y\in B (x)}\psi (y)f\big(\rho (x,y)^2\big).
 %\quad\hbox{where $\psi\in\sA_{\sqrt{2}}\big(\rB (x)\big)$.}
\eeq%{\mathbf 1}(\psi (y)=1)  
The normalizing constant $C$ from \eqref{eq:C} equals $2$. For any finite $\phi\in\sA_{\sqrt{2}}\big(\bbZ^3\big)$,
we have $\sharp  (\phi )=\sum\limits_{x\in\bbZ^3}U (\phi\upharp_{\rB (x)})$, 
in agreement with \eqref{F}. 

A direct calculation shows that the lattice 
\beq\beal\label{eq:PCD=2}
\vphi^{(2)}:=\{m\,(1,1,0) + n\,(1,0,1) + k\,(0,1,1) :\, m, n, k\in\bbZ \} = \bbA_3
\ena\eeq
is a PC. %: it is an example of an FCC-sub-lattice in $\bbZ^3$. 
The fundamental parallelepiped for $\vphi^{(2)}$
has volume $2$. Consider the collection $\sS^{(2)}$ of $\bbZ^3$-shifts of $\vphi^{(2)}$. 

\bthma\label{Theorem 3.1A.}
Set $\sS^{(2)}$ exhausts all\ {\rPC}s for $D^2=2$. The cardinality of\ $\sS^{(2)}$ is $2$. 
Every \ \rPC \ is periodic. The {\rPC}s\ are $\bbZ^3$-symmetric to each other. The particle density of any\ {\rPC}\ equals $1/2$.\ethma

\bp
The assertion follows directly from \eqref{fD=2} and \eqref{eq:PCD=2}.
\ep

\bthmb\label{Theorem 3.1B.}\hskip-4pt{\rm \cite{Dob1}}
Let $u$ be large enough: $u\geq u^0(2)$. Then there are 
$2$ {\rEPGM}s, i.e., $\sharp (\sE ({\sqrt 2},u))=2$, and each \rEPGM\ is generated by
a \rPGS\ from \ $\sS^{(2)}$.
\ethmb

\bp
Follows from Theorem 3.1A, the Peierls bound \eqref{hamexcitation} and the PS theory. 
\ep

The original proof of Theorem 3.1B was given in \cite{Dob1} before the notion of an m-potential and the PS 
theory had been invented.

%#########################################
%                                                                                             D^2=3
%#########################################

\subsection{$D^2=3$.}\label{subsec3.2} 

For $D^2=3$, an LRFF family $\unf=\unf^{(3)}$ has 
\begin{equation}\label{fD=3}f(0)=1, f(1)=f(2)=\dfrac{1}{6}.
\end{equation} 
Here the balls $\rB (x)=\rB_{\sqrt 3}(x)$ contain $19$ sites.
%$\rB (x)=\{y\in\bbZ^3:\;\rho (x,y)\leq{\sqrt 2}\}$. 
Each $\psi\in\sA_{\sqrt{3}}\big(\rB (x)\big)$ contains at most $6$ particles. Consider 
the potential function $U(\psi )=U^{(3)}(\psi )$:
\beq\label{eq:U3}U (\psi )=\dfrac{1}{4}\sum_{y\in B (x)}\psi (y)f\big(\rho (x,y)^2\big).
 % \quad\hbox{where $\psi\in\sA_{\sqrt{3}}\big(\rB (x)\big)$.}
\eeq%{\mathbf 1}(\psi (y)=1)  
The normalizing constant $C$ from \eqref{eq:C} equals $4$. 
For any finite $\phi\in\sA_{\sqrt{3}}\big(\bbZ^3\big)$,
we have $\sharp  (\phi )=\sum\limits_{x\in\bbZ^3}U (\phi\upharp_{\rB (x)})$, in agreement with  \eqref{F}.

A direct calculation shows that the lattice 
\beq\label{eq:BCC3}\beal
\vphi^{(3)}:=\{m\,(2,0,0) + n\,(0,2,0) + k\,(1,1,1) :\, m, n, k\in\bbZ \} \\
\ena\eeq
is a PC: it is a $\sqrt{3}$-BCC sub-lattice in $\bbZ^3$. The fundamental parallelepiped for $\vphi^{(3)}$
has volume $4$. Consider the collection $\sS^{(3)}$ of $\bbZ^3$-shifts of $\vphi^{(3)}$. 

\bthmc\label{Theorem 3.2A.} 
Set $\sS^{(3)}$ exhausts all \ {\rPC}s for $D^2=3$. The cardinality of $\sS^{(3)}$ is $4$. 
Every \ \rPC \ is periodic. The {\rPC}s\ are $\bbZ^3$-symmetric to each other. The particle density of any\ \rPC \ equals \ $1/4$.
\ethmc

\bp
The assertion follows directly from \eqref{fD=3} and \eqref{eq:BCC3}.
\ep

\bthmd\label{Theorem 3.2B.} 
Let $u$ be large enough: $u\geq u^0(3)$. Then there are 
$4$ {\rEPGM}s, i.e., $\sharp (\sE ({\sqrt 3},u))=4$, and each\ \rEPGM\ is generated by
a \rPGS\ from $\sS^{(3)}$.
\ethmd

\bp
Follows from Theorem 3.2A, the Peierls bound \eqref{hamexcitation} and the PS theory. 
\ep

%#########################################
%                                                                                             D^2=4
%#########################################

\subsection{$D^2=4$.}\label{subsec3.3} 

For $D^2=4$, an LRFF family $\unf=\unf^{(4)}$ has 
\beq\label{fD=4}f(0)=1, f(1)=\dfrac{1}{2},\;\;f(2)=\dfrac{1}{4},\;\;f(3)=\dfrac{1}{8}.\eeq
Here the balls $\rB (x)=\rB_{2}(x)$ contain $27$ sites.
%$\rB (x)=\{y\in\bbZ^3:\;\rho (x,y)\leq{\sqrt 2}\}$. 
Each $\psi\in\sA_{2}\big(\rB (x)\big)$ contains at most $8$ particles. Consider 
the potential function $U(\psi )=U^{(4)}(\psi )$:
\beq\label{eq:U4}U (\psi )=\dfrac{1}{8}\sum_{y\in B (x)}\psi (y)f\big(\rho (x,y)^2\big).
% \quad\hbox{where $\psi\in\sA_2\big(\rB (x)\big)$.}
\eeq%{\mathbf 1}(\psi (y)=1)  
The normalizing constant $C$ from \eqref{eq:C} equals $8$. 
For any finite $\phi\in\sA_2\big(\bbZ^3\big)$,
we have $\sharp  (\phi )=\sum\limits_{x\in\bbZ^3}U (\phi\upharp_{\rB (x)})$, 
in agreement with  \eqref{F}.

Straightforward examples of PCs are a cubic sub-lattice $\vphi^{(4)}_{\varnothing}=2\bbZ^3$ 
and its shifts $\vphi^{(4)}_{\varnothing}+x$, $x\in\bbZ^3$, giving $8$ PCs in total, of
particle density $1/8$. However, there is a possibility of constructing a continuum of PCs. For 
example, take PC $\vphi^{(4)}_{\varnothing}$ and let $S$ be a finite subset in 
$\{ x=(\rx_1,\rx_2,\rx_3)\in \vphi^{(4)}_{\varnothing} : 
\rx_3=0\}$. Denote by $\vphi^{(4)}_S$ an AC obtained from $\vphi^{(4)}_{\varnothing}$ by shifting 
all occupied sites $y=(\ry_1,\ry_2,\ry_3)$ with $(\ry_1,\ry_2, 0)\in S$ to $y=(\ry_1,\ry_2,\ry_3+1)$. It 
is not hard to see that $\vphi^{(4)}_S$ is also a PC. 

\bthme\label{Theorem:3.3.} 
The\, \rHC\ model for $D^2=4$ exhibits sliding.
\ethme

\bp
Let $\vphi'=\vphi^{(4)}_{\varnothing}$, take a square $S$ of fixed side-length $l$, and let 
$\vphi''=\vphi^{(4)}_S$. Denote by 
$\sP_n$ the parallelepiped with the base $S$ and the height $n\in\bbN$. Then the sequence $\sP_n$ is 
monotonically increasing,  while $\sharp (\alpha_n)\leq 2 l^2$. \ep

Let us now describe the set of all PCs for $D^2=4$. Fix one of the coordinate directions, say along the 
$\rx_3$-axis. The construction uses two-dimensional PCs $\vtheta_k$ in the horizontal affine planes $\rx_3=k$;
let us first recall that such PCs are described in the following way, in general, not uniquely. See \cite{MSS1}.
We take a square $2$-mesh in $Z_{3,k}:=\bbZ^2+(0,0,k)$ and put the particles in every site of the mesh. 
Next, choose a direction, $\rx_1$ or $\rx_2$. Then select a collection of one-dimensional $2$-meshes
parallel to the chosen direction,  with particles on them, and shift them in this direction  by a unit length.
In total, there is a continuum of two-dimensional PCs $\vtheta_k$.

Let us now describe the set of all PCs for $D^2=4$. Fix one of the co-ordinate directions, say along the 
$\rx_3$-axis. The construction uses two-dimensional PCs $\vtheta_k$ in the horizontal affine planes 
$\rx_3=k$. Such PCs are described in the following way, in general, not uniquely (cf. \cite{MSS1}). We take a square $2$-mesh in $Z_{3,k}:=\bbZ^2+(0,0,k)$ and put the particles in every site of the mesh. 
Next, choose a direction, $\rx_1$ or $\rx_2$. Then select a collection of one-dimensional $2$-meshes
parallel to the chosen direction,  with particles on them, and shift the selected $2$-meshes in this direction by a unit length. This gives a continuum of two-dimensional PCs $\vtheta_k$.

The obtained PCs on $\bbZ^3$ form two disjoint categories: 
$\rx_3$-even-complete and 
$\rx_3$-odd-complete (e-complete and o-complete for short). An e-complete PC has, at each level 
$\rx_3=2k$, $k\in\bbZ$, a two-dimensional PC $\vtheta_k$ of the above form.
The resulting PC on $\bbZ^3$ is denoted by ${\ov\vphi}:=\onwl{\bigcup}\limits_{k\in\bbZ} \vtheta_k$.  
An o-complete PC is constructed in a similar manner, with odd layers $\rx_3=2k+1$ in place of even 
ones. In addition, every e-complete or o-complete PC $\ov\vphi$ can generate a family  of 
descending PCs obtained by shifting $\rx_3$-directed copies of $\bbZ$, with particles at sites of 
$\ov\vphi$ on it, by a unit length.
The same construction can be done for other co-ordinate directions. 
 
In short, we (i) fix one of the co-ordinate directions; (ii) select an e-complete or 
o-complete PC corresponding to this co-ordinate axis; (iii) finally, construct a PC 
descending from the selected e-complete or o-complete one. Let $\sS^{(4)}$ 
denote the set of all PCs obtained by 
the above construction.

\bthmf\label{Theorem 3.4.} 
Set \ $\sS^{(4)}$ exhausts all\, {\rPC}s for $D^2=4$. The cardinality of \ $\sS^{(4)}$ is continuum. 
There are countably many periodic \ {\rPC}s. The particle density of any\, \rPC\  equals $1/8$.
\ethmf

\bp Note that open  $2\times 2\times 2$-cubes $\rC (x)$ centered at occupied sites 
$x\in \phi\in \sA_2\big(\bbZ^3\big)$ are pair-wise disjoint. Correspondingly, a PC emerges iff the 
union of the closures of these cubes covers the entire $\bbR^3$.  

Take an arbitrary PC $\vphi$ and choose one of the co-ordinate directions, say along the $\rx_3$-axis.
Denote by $\Pi(y)$ the ortho-projection of cube $\rC (y)$ to the horizontal plane\ $P_{3,0}:\;\rx_3=0$  %$P$
in $\bbR^3$.
Let us partition the occupied sites $y=(\ry_1,\ry_2,\ry_3)\in\vphi$ into even and odd categories: even if $\ry_3$
is even and odd if $\ry_3$ is odd. We will refer to them as even and odd particles in $\vphi$ and also
call the corresponding cubes even or odd, respectively. 
We claim that two occupied sites, $y_1,y_2\in\vphi$ such that one of them is even and the other 
is odd cannot have the intersection $\Pi (y_1)\cap \Pi (y_2)$ with a positive area. In fact, 
suppose that $\Pi (y_1)\cap \Pi (y_2)$ covers an open unit square $\rS$. Then either the (infinite) vertical prism $\sR$ projected to $\rS$ has 
no intersection with a cube $\rC (y)$ or their intersection is a vertical parallelepiped of height $2$ 
(a vertical 2-brick, for brevity). Thus, we have to cover the piece of $\sR$ between   $\rC (y_1)$
and $\rC (y_2)$ with vertical 2-bricks while the distance between $y_1$ and $y_2$ is odd. This is 
impossible without having an empty space which implies  that $\vphi$ is not a PC. This justifies the 
claim.

As a result, we get that even and odd cubes are projected into disjoint open $2\times 2$-squares 
in plane \ $P_{3,0}$.  In other words, $P_{3,0}$ is partitioned into a pair of even and odd subsets, 
$V_\rE$ and $V_\rO$, and a collection of continuous broken lines representing the boundary 
between $V_\rE$ and $V_\rO$. 

Consider a $\bbZ^3$-site $y\in P_{3,0}$ and the two lines containing $y$ and parallel 
to the $\rx_1$- and $\rx_2$- coordinate 
axes. If both these lines intersect the boundary between $V_\rE$ and $V_\rO$ then for any two cubes 
$\rC (\wh{y}_1)$, $\rC (\wh{y}_2)$ with $\Pi (\wh{y}_i) \ni y$ one has  $\Pi (\wh{y}_1)=\Pi (\wh{y}_2)$. 
This is because the projection $\Pi (\wh{y}_i)$ is at even distances (along the $\rx_1$- and 
$\rx_2$- coordinate axes) from the boundary. 

If site $y\in P_{3,0}$ does not have the above property then we have the following cases. 
(i) None of the above two lines intersects the boundary. (ii)  Only one of these lines intersects the 
boundary. 

In case (i), for any site $z\in P_{3,0}$ of the opposite parity to $y$, both corresponding lines intersect the 
boundary. Consequently, for any such $z$ the projection $\Pi ({\wh z})$ containing site $z$ is uniquely
determined.

In case (ii), assume for definiteness that the line through $y$ which does not
intersect the boundary is parallel to the 
$\rx_2$-axis. Then for any $z\in P_{3,0}$ of the opposite parity,  the line through $z$ parallel to the 
$\rx_1$-axis does intersect the boundary. If for any site $z\in P_{3,0}$ of the 
opposite parity the line through 
$z$ parallel to the $\rx_2$-axis also intersects the boundary then the projection $\Pi ({\wh z})$ 
containing site $z$ is uniquely determined for such $z$. 

In the opposite case we have at least one $z'=(\rz'_1,\rz'_2,0)\in P_{3,0}$ of the opposite parity 
to $y$ such that the line through $z'$ and parallel to the $\rx_2$-axis does not intersect the boundary. 
Moreover,  there exists an entire strip of non-uniqueness sites
\beq\label{eq:S3last}\{z''=(\rz'_1 , \rz''_2 , 0)\in P_{3,0}\},\;\;\rz''_2 \in \bbZ.\eeq
Then
take all occupied sites $\wh{z}\in \varphi$ such that $\Pi (\wh{z})\ni z$ where $z\in P_{3,0}$ 
has the projection-uniqueness property, shift $\wh{z}\mapsto \wh{z}+(0, 0, 1)$, and recalculate the boundary 
(as such sites $z$ will change their parity). Observe that after this operation the non-uniqueness strip 
\eqref{eq:S3last} remains intact. Among all such strips consider the one which is closest to $y$. Then the vertical 
line separating this strip from the half of $P_{3,0}$ containing $y$ is necessarily a part of the new boundary.
Consequently, the PC $\vphi$ consists of layers orthogonal to the $\rx_1$-axis. \ep

%The failure of the Peierls bound indicated in Theorem 3.3 and the multitude of PCs in Theorem 3.4 lead us to 
%the following conjecture. 

%\begin{nnconjecture}
%For $D^2=4$ there is a unique Gibbs measure. 
%\end{nnconjecture}

%#########################################
%                                                                                                      D^2=5                          
%#########################################

\section{Case $D^2=5$}\label{Sec:D=5}

For $D^2=5$, an LRFF $\unf=\unf^{(5)}$ has  
\begin{equation}\label{fD=5}f(0)=1,\;\;f(1)=\dfrac{2}{3},\;\;f(2)=\dfrac{1}{3},\;\;f(3)=f(4)=0.
\end{equation} 
Here the balls $\rB (x)=\rB_{\sqrt{3}}(x)$ contain $19$ sites.  Each $\psi\in\sA_{\sqrt{5}}\big(\rB (x)\big)$ contains at most $3$ particles. Consider the potential function $U(\psi)=U^{(5)}(\psi)$:
\beq\label{eq:U1}U (\psi)=\dfrac{1}{9}\sum_{y\in B (x)}\psi (y)f\big(\rho (x,y)^2\big).
% \quad\hbox{where $\psi\in\sA_{\sqrt{5}}\big(\rB (x)\big)$.}
\eeq%{\mathbf 1}(\psi (y)=1)  
The normalizing constant $C$ from \eqref{eq:C} equals $9$. For any finite $\phi\in\sA_{\sqrt{5}}\big(\bbZ^3\big)$,
we have $\sharp  (\phi )=\sum\limits_{x\in\bbZ^3}U (\phi\upharp_{\rB (x)})$, 
in agreement with \eqref{F}. 

The PCs for $D^2=5$ are layered ACs emerging from the first construction proposed 
in Section \ref{Sec2}; cf. \eqref{layered1}. Here $q=6$, $h=3$ and $r=2$, and every 
$\tau^{(6)}_{i, hk, j}$ is a triangular ${\sqrt 6}$-sub-mesh in $\bbZ^3\cap T_{i,3k}$ where $T_{i,3k}$
is the basic ${\sqrt{2/3}}$-mesh defined in \eqref{eq:Tik1}. Further, for $i=0,1,2,3$ and $s_1=s_1(i),s_2=s_2(i)$ 
given by \eqref{eq:1s2s3} we define %\eqref{eq:is1s2}, \eqref{eq:s1s2i},
\beq\label{eq:tau45} \beac\tau^{(6)}_{i,3k,0}:=\big\{m(1,-2s_2,s_3)+n(-1,-s_2,2s_3),\;m,n\in\bbZ\big\}+k(1,s_2,s_3)\\
\tau^{(6)}_{i,3k,1}:=\tau^{(6)}_{i,3k,0}+(0,s_2,-s_3),\quad 
\tau^{(6)}_{i,3k,2}:=\tau^{(6)}_{i,3k,0}+(0,-s_2,s_3).\ena \eeq
It is instructive to note that the occupied sites in meshes $\tau^{(6)}_{i,3k,1}$ and $\tau^{(6)}_{i,3k,2}$
cover the centers of triangles in mesh $\tau^{(6)}_{i,3k,0}$ in an alternating manner: the centers of two 
neighboring triangles in $\tau^{(6)}_{i,3k,0}$ (sharing a common side) are occupied by particles from 
two different meshes. In fact, the same is true for any two meshes: their occupied sites cover the centers
of triangles in the third mesh, alternately.
The allowed sequences $\{j_k,\;k\in\bbZ\}$ have digits $j_k=0,1,2$ with $j_0=0$ and $j_k\neq j_{k+1}$.
The set $\sS^{(5)}$ consists of all layered PCs 
$\onwl{\bigcup}\limits_{k\in\bbZ}\tau^{(6)}_{i,3k,j_k}$ (see \eqref{layered1}) with the allowed sequences 
$\{j_k\}$ and $\bbZ^3$-shifts of these configurations. We also introduce the subset $\sS^{(5)}_{\rm{per}}$ 
consisting of PCs $\vphi^{(5)}_{i,\{j_k\}}$ with
periodic allowed sequences $\{j_k\}$ and the $\bbZ^3$-shifts of these PCs. (To prevent a confusion, let us emphasize that the superscript $(5)$ for $\vphi$ refers to $D^2$ while the superscript $(6)$ for $\tau$ refers to the mesh size.) 

\bthmg\label{thm:4.1.} 
Set \ $\sS^{(5)}$ exhausts all\, {\rPC}s\ for $D^2=5$. The cardinality of\ $\sS^{(5)}$ is continuum. 
Set \ $\sS^{(5)}_{\rm{per}}$ is countable and exhausts all periodic\, {\rPC}s\ for $D^2=5$.
The particle density of any\, \rPC\  equals $1/9$.
\ethmg

%\bl\label{lem:4.1}%{\bf Lemma 4.1.} {\sl 
%Each $\vphi^{j,k_0}_\alpha$ is an \rPC.
%\el

%\bl\label{lem:4.2}%{\bf Lemma 4.2.} {\sl 
%The {\rAC}s $\{\vphi^{j,k_0}_\alpha\} $ exhaust all {\rPC}s:
%there are no other {\rPC}s.
%\el

\bp 
The set of ACs $\psi$ on $\rB(x)$ which give  $U(\psi)=1/9$  is partitioned into $3$ subsets where each 
subset is characterized by a fixed collection of values $f(\,\cdot\,)$ participating in the sum $U(\psi)$
\beq\beac\label{list:dist5}\{f(0)\}, \{f(1), f(2)\}, \{f(2), f(2), f(2)\}.\ena\eeq

Assume that a PC $\vphi$ does not contain a pair of particles at distance $\sqrt{6}$. According to \eqref{list:dist5}, 
the only possibility is $\{f(1), f(2)\}$ and therefore there are two particles at distance $\sqrt{5}$ from each other. 
For definiteness, take particles at sites $x_1=(0, 0,0)$ and $x_2=(1, 2, 0)$. Consider the vacant site 
$x_3=(1, 1,1)$. The only way to implement the set $\{f(1), f(2)\}$ is to place a particle at site $(1, 1, 2)$ which is at 
distance $\sqrt{6}$ from $x_1$ and hence contradicts the assumption. The possible ways to implement the subset 
$\{f(2), f(2), f(2)\}$ are to place a particle either at the site $(2, 0, 1)$ or at the site $(2, 1, 2)$. However, both $(2, 0, 1)$ 
and $(2, 1, 2)$ are at distance $\sqrt{6}$ from $x_2$, which again contradicts the assumption.  

Now, take a PC $\vphi$ and a pair of particles at distance $\sqrt{6}$, for definiteness, say $x_1=(0, 0, 0)$ and $x_2=(2, 1, 1)$.
Consider the vacant site $x_3=(1, 0, 1)$ which implies $\{f(2), f(2), f(2)\}$. The only possibility to implement this subset is to 
place a particle at site $(1, -1, 2)$ that forms an equilateral triangle with $x_1$ and $x_2$. Repeating this 
argument we obtain a triangular mesh. The densest combination of such triangular meshes leads to our family $\sS^{(5)}$. 
\ep

To state our results on EPGMs for $D^2=5$, we introduce a subset 
$\sH^{(5)}\subset \sS^{(5)}_{\rm{per}}$ formed by the deformed HCP configurations. These are periodic
PCs $\vphi^{(5)}_{i,\overline{01}}$, $\vphi^{(5)}_{i,\overline{02}}$ and their $\bbZ^3$-shifts.%, for  periodic sequences $\{j_k\}$,
%with period $\ov{j_0j_1}$, where $j_0=0$, $j_1=1,2$. %The cardinality of 
%$\sH^{(5)}$ equals $4\cdot 3\cdot 3\cdot 2=72$.

\bthmh\label{thm:4.2.} 
Let $u$ be large enough: $u\geq u^0(5)$. Then there are  
$72$ {\rEPGM}s, i.e., $\sharp (\sE ({\sqrt 5},u))=72$, and each \rEPGM\ is generated by
a \rPGS\ from $\sH^{(5)}$.
\ethmh

{\bf Proof.} %\bp
The proof is an application of results of \cite{BS}. Applicability of these results needs verification 
of several conditions, including the bound (2.9) from \cite{BS}.  

We begin this verification starting with  Sect 2.1 of \cite{BS}.
In the definition of the $l$-boundary we use $l={\sqrt 2}$ in contrast to $l=2$ as chosen in \cite{BS}, because 
\cite{BS} uses the max-distance metric. Next, in the terminology of Sect 2.2 in \cite{BS}, a local GS in a domain $\Lam$
is a configuration $\vphi\in\sA_{\sqrt{5}}(\Lam)$ %an AC $\vphi$ in $\Lam$ 
such that $U(\vphi \upharp_{\rB (x)})=\diy\frac{1}{9}$ for every $x$ with 
$\rB (x)\subset\Lam$.

Next, we need to check the property that two local ground states coinciding on the $l$-boundary 
coincide on the whole of $\Lam$. This follows from the argument in the proof of Theorem 4.1A.

For the HC model the statistical weight of any local excitation is of the form
$u^{- n}$ where $n$ is a positive integer. Furthermore, the inverse temperature $\beta =\ln\,u$,
and possible values of the excitation energy are positive integers. We choose $E_D=2$, 
where $E_D$ is a notation taken from \cite{BS}; the subscript $D$ is unrelated to our admissibility 
distance $D$.

Now, we need to verify the retouch property from \cite{BS}. To this end, we have to list all elementary 
excitations of energy $\leq E_D$, i.e., of energy  $1$ and $2$.
Removing a single particle yields an elementary excitation $\gamma_1$ of energy $1$. Removing two particles 
%at distance ${\sqrt 5}$ or ${\sqrt 6}$ 
close enough from each other (see the definition of $Supp(X)$ in Section 2) gives an elementary excitation $\gamma_2$ of energy $2$.

\bl\label{lem:4.4}%{\bf Lemma 4.4.} 
Elementary excitations $\gamma_1$ have the same density in all \ {\rPC}s. Elementary excitations $\gamma_2$ also have the same density in all \ {\rPC}s. 
\el

\bp A direct calculation shows that the density of $\gam_1$ is \ $1/9$.  The density of 
$\gam_2$ is the same for all PCs $\vphi^{(5)}_{i,\{j_k\}}$ by construction. Namely, if two removed particles are located in the same mesh $\tau^{(6)}_{i,3k,j_k}$ then the density is the same for all meshes and therefore for all PCs. Similarly, if the removed particles belong to two different meshes at a fixed distance from each other then the density does not depend on the choice of the pair of meshes. 
\ep

The next %(and final) 
excitation of energy $2$, $\gamma_2^{\ast}$, is constructed as follows. Consider three subsequent meshes $\tau':=\tau^{(6)}_{i,3(k-1),j_{k-1}}, \tau:=\tau^{(6)}_{i,3k,j_k}, \tau'':=\tau^{(6)}_{i,3(k+1),j_{k+1}}$ and
assume that in the middle mesh $\tau$ there is a triangle $\triangle_0$ and in meshes $\tau', \tau''$ there are 
triangles $\triangle_{\pm 1}$, and the centers of $\triangle_{\pm 1}$ are projected to the center of
$\triangle_0$. Then, if we place a particle at the center of $\triangle_0$ and remove the particles 
from the vertices of $\triangle_0$, we obtain excitation $\gamma_2^{\ast}$. See Figure 4.1.

\FigureFourOne

The most elaborate part 
of our argument is

%{\bf PC=perfect configuration /PPC=periodic PC}
%\Figure4.2

\bl\label{lem:4.3}%{\bf Lemma 4.3.} {\sl 
The excitations of types $\gamma_1, \gamma_2$ and $\gamma_2^{\ast}$ exhausts all elementary excitations of energies $1$ and $2$.
\el

\bp %A trivial way of producing a local excitation (elementary or not) of an AC $\phi\in\sA$ is 
%to remove particles from $\phi$. Another way is to insert particles at a finite collection of sites 
%$\xi =\{x_i\}\in\sA_D(\bbZ^3)$ and remove those particles from $\phi$ that have been repelled. 
%(A repelled particle $y\in\phi$ is the one at distance $<{\sqrt 5}$ from an inserted site $x\in\xi$.) 
%Let $\eta$ denote the collection of sites where the repelled particles are located: 
%$\eta (=\eta (\phi,\xi))\subseteq\phi$. 
%For simplicity, assume we deal with a global PC in $\bbZ^3$. 
%\newpage
%The balance between $\sharp  (\xi )$ and $\sharp (\eta )$, the numbers of inserted and repelled 
%particles in an AC $\phi$, can be counted as follows.
%Let $x\in\xi$ be an inserted site and suppose it repels particles located at $Y(x)=Y(x,\phi )\subseteq\phi$,
%so that $\eta =\onwl{\cup}\limits_{x\in\xi}Y(x)$. Assume
%that we have a collection of numbers $f(x,y)\in [0,1]$ with the property
%$\sum\limits_{y\in Y(x)}f(x,y)\leq 1$, $\forall$ $x\in\phi^\comp$. Then the sum 
%$$\sum\limits_{y\in\eta}\left[1-\sum\limits_{x\in\xi}f(x,y) \right]\geq\sharp (\eta )-\sharp  (\xi ).$$
%If $\phi$ is a PC, this sum is always $\geq 0$. The quantity 
%$$E(y)=1-\sum\limits_{x\in\xi}f(x,y)$$
%is called an {\it excess} for the repelled particle $y\in\phi$ (under insertion $\xi$).
%That is, they coincide with our LRFF family $\unf^{(5)}$. 
%It is not hard to check that for a PC $\vphi$, the sum $\sum\limits_{y\in Y(x,\vphi )}f(x,y)=1$, $\forall$
%$x\in\vphi^\comp$. (If $\phi$ is an AC and $x\in\phi$, we have the equality
%$\sum\limits_{y\in Y(x,\phi )}f(x,y)=f(x,x)=1$ by construction.)
To describe possible exitations we use the corresponding inserted collection $\xi$ and repelled collection $\eta$. 
An insertion of a single particle into a PC can be done in several ways. 

(I) A particle is inserted between two 
neighboring triangular meshes,  $\tau$ and $\tau'$. Such an inserted particle repels four particles
at the vertices of a unique tetrahedron $\sT$ containing the site of insertion. Two vertices of tetrahedron $\sT$
belong to mesh $\tau$ and other two to $\tau'$. Accordingly, one edge of $\sT$ belongs to 
 $\tau$ and the other to $\tau'$; these two edges have length ${\sqrt 6}$ each and the angle between them
 equals $\pi/3$. Furthermore, the length of the segment that is orthogonal to each of them equals
 ${\sqrt 3}$ and the end-points of this segment divide the edges at the ratio $2:1$.
 Among the remaining four edges of $\sT$ (which join $\tau$ and $\tau'$), three
 have length ${\sqrt 5}$ and one has length $\sqrt{11}$. See a tetrahedron with 4 blue edges on Figure 4.2.
 
(II)  Another situation emerges when the site of insertion belongs to the plane containing $\tau$; it happens 
iff this site is a center of a triangle in $\tau$. Then consider three subsequent meshes $\tau',\tau,\tau''$ as above. Correspondingly, three situations can occur, depending upon the mutual position of  $\tau',\tau$ and $\tau''$.
(IIa) The inserted particle repels only three particles at the vertices of the triangle in $\tau$. See the pink circle on Figure 4.1.  
(IIb) The inserted particle repels four particles at the vertices of a tetrahedron whose base is a triangle 
from $\tau$ and the fourth vertex belongs to $\tau'$ or $\tau''$; three base edges of the tetrahedron  
have length ${\sqrt 6}$ and three other length ${\sqrt 5}$.  (IIc) The inserted particle repels five particles 
at the vertices of a triangular bi-pyramid with one vertex in $\tau'$, one in $\tau''$ and three at the vertices
of a triangle in $\tau$. 

The inserted collection $\xi$ is called {\it reducible} if in the corresponding repelled collection $\eta$ there is a site $y$ repelled by a single site $x\in\xi$. For the reduced collection $\xi'=\xi\setminus\{x\}$, the 
corresponding set $\eta'\subseteq
(\eta\setminus\{y\})$, and hence $E(\xi'):=\sharp (\eta' )-\sharp (\xi')\leq \sharp (\eta)-\sharp (\xi)=:E (\xi)$. 

The next observation is that any reducible collection $\xi$ of energy $2$ can be 
reduced, by  subsequently 
throwing away sites $x_m, ..., x_1$ from $\xi$, either to an irreducible inserted collection
$\xi_0$ or to a reducible collection $\xi_1$ of type (IIa), both of energy $2$. The second 
option is non-feasible, as one cannot add a new insertion site to $\xi_1$ without increasing the 
energy. We now show that the first option is also non-feasible, by verifying that any 
irreducible $\xi$ has energy at least $8$. The inserted collection $\xi_0$ with the minimal  energy 
contains 24 inserted particles and leads to 36 repelled  particles. See Figure 4.2.

Furthermore, $\xi_0$ is located in $4$ consecutive basic $\sqrt{3}/2$-meshes 
$T_{i,k}, T_{i,k+1}, T_{i,k+2}, T_{i,k+3}$ %$P_{\rl\rl}$, $P_{\rl\ru}$, $P_{\ru\rl}$ $P_{\ru\ru}$, 
(more precisely, in the intersections $T_{i,k}\cap\bbZ^3$, $T_{i,k+1}\cap\bbZ^3$, 
$T_{i,k+2}\cap\bbZ^3$ and $T_{i,k+3}\cap\bbZ^3$), with the distance $1/\sqrt{3}$ between any two neighboring 
planes containing these meshes. The repelled collection $\eta_0$ is located in triangular meshes 
$\tau_{\rm L}\subset T_{i,k},$ and $\tau_{\rm U}\subset T_{i,k+3}$,
(with distance ${\sqrt 3}$ between the planes containing $T_{i,k}$ and $T_{i,k+3}$). Figure 4.2
shows the ortho-projection of $\xi_0$ to the plane containing the lower mesh $T_{i,k}$ endowed with this basic 
triangular mesh of size $\sqrt{2/3}$;
the edges of the basic mesh are drawn in thin lines.  
The green and red circles mark insertions of type (IIa) in  $T_{i,k}\cap\bbZ^3$
and  $T_{i,k+3}\cap\bbZ^3$, respectively. Each such circle is located at the center of a triangle
of the same color with the side-length ${\sqrt 6}$ and repels the $3$ vertices of this triangle. In the figure
there are 7 green and 7 red such triangles. \def\rL{{\rm L}} \def\rU{{\rm U}}

\FigureFourTwo

The blue edges join a repelled site from mesh  $\tau_{\rm L}$ and a repelled site from 
mesh $\tau_{\rm U}$, the shorter between these edges have length ${\sqrt 5}$ and the longer 
${\sqrt{11}}$. The blue trapezes indicate the ortho-projections of 
tetrahedrons; each tetrahedron, in addition to $4$ blue edges, includes two non-adjacent/skewed edges,
one green and one red. The faint green/pink circles indicate insertions that repel the vertices of the 
tetrahedrons; the faint green circles mark the insertion sites located in $ T_{i,k+1}\cap\bbZ^3$ (there 
are $3$ of them), while the faint pink circles (another $3$) mark the insertion sites located in 
$T_{i,k+2}\cap\bbZ^3$. Both the faint green and faint pink insertions are of type (I). 

The above collection $\xi_0$ is minimal irreducible due to the following argument. The 
ortho-projection of any 
irreducible collection $\xi$ to $T_{i,k}$ has the boundary that is a continuous broken line,
possibly with self-intersections. In a standard way we can define the orientation of this line. 
It is not hard to check that the only way to make a turn along this line is by angle $\pi/6$, from a long
blue segment to a green or red one, or vice versa, from a green or red segment to a long blue one.
In other words, to make a turn, one needs to combine a triangle and a tetrahedron. The common site
$y$ of the pair $(\hbox{triangle, tetrahedron})$ has $E(y)=2/3$. The minimal amount of tetrahedrons
needed for a full turn is $6$, which yields $12$ sites $y$ with $E(y)=2/3$, i.e., the energy of $\xi$ is at 
least $8$. \ep %This completes the proof of Lemma \ref{lem:4.3}. 

\bl\label{lem:4.5}
Among all {\rm{PC}}s, $\vphi^{(5)}_{i,\{j_k\}}\in \sS^{(5)}_{\rm{per}}$ the maximal density  $1/9$ of exitations 
$\gamma_2^{\ast}$ of type {\rm{(IIa)}} is achieved %maximal among the periodic\\ {\rPC}s
on \ $\vphi^{(5)}_{i,\{j_k\}}\in\sH^{(5)}$.%, and it equals $1/9$.\el
\el

\bp
By construction, an excitation of type (IIa) is present in a layer $\tau$ iff its neighboring layers 
$\tau'$ and $\tau''$ are the same, i.e., are labeled by the same digit from $\{0, 1, 2\}$, which implies 
the assertion of the lemma. A direct calculation verifies that the density of excitations of type {\rm{(IIa)}} 
in $\vphi^{(5)}_{i,\{j_k\}}\in\sH^{(5)}$ equals $1/9$. 
\ep

To complete the proof of Theorem 4.1B, observe that Lemma \ref{lem:4.5}  implies that 
the PCs (equivalently PGSs) from $\sH^{(5)}$ are  dominant in the sense of \cite{BS}, pp. 111--112 
(which includes the
Peierls bound (2.9) from  \cite{BS}). Also, for $u$ large enough, (2.10b) from \cite{BS} 
holds true with $c=u^{-2}/18$. This completes the verification of conditions required in \cite{BS}. 

A direct calculation shows that $\sharp (\sH^{(5)})=4\cdot 3\cdot 3\cdot 2=72$. \hfill{\scriptsize 
$\blacksquare$}

%#########################################
%                                                                                             D^2=6
%#########################################

\section{Case $D^2=6$}

For $D^2=6$, an LRFF $\unf=\unf^{(6)}$ has  
\begin{equation}\label{fD=06} 
f(0)=1,\;\; f(1)=\dfrac{2}{3},\;\;f(2)=\dfrac{1}{3},\;\;f(3)=\dfrac{1}{8},\;\;f(4)=\dfrac{1}{6},\;\;f(5)=\dfrac{1}{24}.
\end{equation}  
Here the balls $\rB (x)=\rB_{\sqrt{6}}(x)$ contain $57$ sites. Each $\psi\in\sA_{\sqrt{6}}\big(\rB (x)\big)$ contains at most $7$ particles. Consider the potential function $U(\psi)=U^{(6)}(\psi)$:
\beq\label{eq:U6}U (\psi)=\dfrac{1}{12}\sum_{y\in B (x)}\psi (y)f\big(\rho (x,y)^2\big).
% \quad\hbox{where $\psi\in\sA_{\sqrt{6}}\big(\rB (x)\big)$.}
\eeq%{\mathbf 1}(\psi (y)=1)  
The normalizing constant $C$ from \eqref{eq:C} equals $12$. For any finite $\phi\in\sA_{\sqrt{6}}\big(\bbZ^3\big)$, we have $\sharp  (\phi )=\sum\limits_{x\in\bbZ^3}U (\phi\upharp_{\rB (x)})$, 
in agreement with \eqref{F}. 

The set $\sS^{(6)}$ comprises two families of layered PCs emerging from the constructions 
proposed in Section \ref{Sec2}; cf. \eqref{layered1}, \eqref{layered2}. One family consists of layered PCs 
$\onwl{\bigcup}\limits_{k\in\bbZ}\tau^{(6)}_{i,4k,j_k}$,  as in
\eqref{layered1}, with $q=6$, $h=4$ and $r=6$; here every 
$\tau^{(6)}_{i, 4k, j_k}$ is a triangular ${\sqrt 6}$-sub-mesh in $\bbZ^3\cap T_{i,4k}$ where $T_{i,4k}$ is 
a basic ${\sqrt{2/3}}$-mesh defined as in \eqref{eq:Tik1}.  For $i=0,1,2,3$ and $s_2=s_2(i), s_3=s_3(i)$ 
given by \eqref{eq:1s2s3} we set
\beq\label{eq:tau000}\tau^{(6)}_{i,\,0,\,0}:=\big\{m(1,-2s_2,s_3)+n(-1,-s_2,2s_3),\;m,n\in\bbZ\big\}.\eeq
%where $i=0,1,2,3$ and $s_1,s_2\in\{-1,1\}$ are as in \eqref{eq:1s2s3}. % \eqref{eq:is1s2},  \eqref{eq:s1s2i}. 
%The sub-lattice $\tau^{(6)}_{i,\,0,\,0}$ has 
Consider the following $6$ sites at distance ${\sqrt 6}$ from the origin:
$$\beac  w_{i,1}=(1,-2s_2,s_3),\;w_{i,2}=(-1,-s_2,2s_3),\;w_{i,3}=(-2,s_2,s_3),\\
w_{i,4}=-w_{i,1},\; w_{i,5}=-w_{i,2},\;w_{i,6}=-w_{i,3}.\ena$$
Additionally, set $w_{i,0}:=(0, 0, 0)$.
Define $7$ meshes $\gam_{i, j}\subset T_{i,0}\,$:
$$\gam_{i,\,j}:=\tau^{(6)}_{i,\,0,\,0}+\dfrac{1}{3}w_{i,j},\;\;j=0,\ldots ,6.$$

Next, for $k\in\bbZ$, we say that $j_k=j$ if 

\beq\label{eq:tau46}\beac\tau^{(6)}_{i,\,4k,\,j_k}:=
\dfrac{4k}{3}(1,s_2,s_3)+\gamma_{i,j}.\ena\eeq

The sequence $\{j_k\}$ satisfies the following conditions. First, $j_0=0$ and $j_k\neq j_{k+1}$. Further, 
$$\tau^{(6)}_{i,\,4(k+1),\,j_{k+1}} - \tau^{(6)}_{i,\,4k,\,j_{k}}= \diy\frac{1}{3}w_{i,j'},\qquad 
j'\in\{2,4,6\}. $$
%where either $j'\in\{1,3,5\}$ or $j'\in\{2,4,6\}$. The specific choice depends on whether the shift of a site from $\tau^{(6)}_{i,\,4k,\,j_{k}}$ by $\dfrac{4}{3}(1,s_2,s_3)+\dfrac{1}{3}w_{i,j'}$ leads to a site in $\bbZ^3$ or not. 

%put: 
%\beq\label{eq:tau46}\beac\tau^{(6)}_{i,\,4k,\,j_k}:=
%\dfrac{4k}{3}(1,s_2,s_3)+\bcs \tau^{(6)}_{i,\,0,\,0},&\hbox{for $j_k=0$, if $k=0\bmod 3$,}\\
%\gam_{i,j_k},&\hbox{for $j_k=2,4,6$, if $k=1\bmod 3$,}\\
%\gam_{i,j_k},&\hbox{for $j_k=1,3,5$, if $k=2\bmod 3$.}\ecs\ena\eeq
%Here, the allowed sequences $\{j_k\}$ have $j_k=0$ for $k=0\mod 3$, $j_k=2,4,6$ for $k=1\mod 3$
%and $j_k=1,3,5$ for $k=2\mod 3$.

The second family consists of layered PCs $\onwl{\bigcup}\limits_{k\in\bbZ}\alp^{(8,16)}_{i,3k,j_k}$,  
as in \eqref{layered2}, with $q_1=8$, $q_2=16$, $h=3$ and $t=2$, where every 
$\alp^{(8,16)}_{i, 3k, j_k}$ is a rhombic $({\sqrt 8},{\sqrt{16}})$-sub-mesh in $\bbZ^3\cap Q_{i,3k}$, with 
$Q_{i,3k}$ being a basic rectangular $1\times {\sqrt 2}/2$ mesh defined as in \eqref{eq:Qik1}.

Namely, for $i=0,1,2,3,4,5$ labeling the non-main diagonals and $s_1=s_1(i), s_2=s_2(i), s_3=s_3(i)$ 
given by \eqref{eq:s1s2i2} we set
\beq\label{eq:alp000}\alp^{(8,16)}_{i,0,0}:=\{m(2a + 2b)+n(2a - 2b):\;m,n\in\bbZ\},\eeq
where vectors $a$ and $b$ are defined in terms of $s_1, s_2, s_3$ in \eqref{a}, \eqref{b}.
%Here $i=0,1,2,3,4,5$ labels non-main diagonals and $s_1,s_2\in\{-1,1\}$ are connected to $i$ via \eqref{eq:s1s2i2}.  
The rhombic lattice $\alp^{(8,16)}_{i,0,0}$ lies in $Q_{i,0}$.% where $Q_{6,0}$ is the projection of $\bbZ^3$ to the plane $\rx_2-\rx_3=0$. 

%The sub-lattice $\alp^{(8,16)}_{i,0,0}$ has 
Further, consider the following $2$ sites at distance $\sqrt{3/2}$ from the origin:
 $$\beac%v_{i,j}\in Q_{i,0}$, $j=0,1,2,3$: 
v_{i,1}=a+b,\;v_{i,2}=a-b,%\;v_{i,3}=-a+b,\;v_{i,4}=-a-b.
\ena$$
Additionally, set $v_{i,0}:=(0, 0, 0)$. Define 3 meshes $\beta_{i,j}\subset Q_{i,0}$:
$$\beta_{i,j}=\alp^{(8,16)}_{i,0,0}+v_{i,j}, \qquad j=0,1,2.$$
Next, given $k\in\bbZ$, we say that $j_k=j$ if 
\beq\label{eq:alp8,16}\alp^{(8,16)}_{i, 3k,j_k}=\dfrac{3k}{2}(s_1,s_2,s_3)+\beta_{i,j}.
%\bcs &\hbox{ for $j=0,1,2$, if $k\geq 1$,}\\ &\hbox{ for $j=0,3,4$, if $k\leq -1$.}\ecs
\eeq

The sequence $\{j_k\}$ satisfies the following conditions. First, $j_0=0$ and $j_k\neq j_{k+1}$. Further, 
$$\alpha^{(8, 16)}_{i,\,3(k+1),\,j_{k+1}} - \alpha^{(8, 16)}_{i,\,3k,\,j_{k}}= v_{i,j'}, \qquad j'\in\{1,2\}.$$

\bigskip
\FigureFiveOne

\FigureFiveTwo

%The allowed sequences $\{j_k\}$ for this family have $j_k=0$ for $k=0$, $j_k=0,1,2$ for $k\geq 1$,
%$j_k=0,3,4$ for $k\leq -1$ and $j_k\neq j_{k+1}$ for all $k\in\bbZ$.

The set $\sS^{(6)}$ consists of all layered configurations $\onwl{\bigcup}\limits_{k\in\bbZ}\tau^{(6)}_{i,4k,j_k}$ 
and $\onwl{\bigcup}\limits_{k\in\bbZ}\alp^{(8,16)}_{i,3k,j_k}$, 
with the allowed sequences $\{j_k\}$ and $\bbZ^3$-shifts of these configurations. We also introduce the subset $\sS^{(6)}_{\rm{per}}$ consisting of PCs 
with periodic allowed sequences $\{j_k\}$, and $\bbZ^3$-shifts of these configurations.
%We also introduce the subset $\sS^{(6)}_{\rm{per}}$ consisting of PCs with
%periodic allowed sequences $\{j_k\}$ and their $\bbZ^3$-shifts.

\bthm
Set \ $\sS^{(6)}$ exhausts all \ {\rPC}s for $D^2=6$. The cardinality of \ $\sS^{(6)}$ is continuum. 
Set \ $\sS^{(6)}_{\rm{per}}$ is countable and exhausts all periodic\, {\rPC}s\ for $D^2=6$.
The particle density of any \ \rPC\ from \ $\sS^{(6)}$ equals $1/12$.
\ethm

\bp
The set of ACs $\psi$  on $\rB(x)$ which give $U(\psi) =1/12$
is partitioned into 8 subsets;
each subset is characterized by a fixed collection of values $f(\cdot)$ participating in the sum $U(\psi)$
\beq\beal\label{fD=6}
\{f(0)\},\; \{f(1),f(3),f(3),f(5),f(5)\},\;  \{f(1),f(3),f(5),f(5),f(5),f(5),f(5)\},\;\\\{f(1),f(4),f(5),f(5),f(5),f(5)\},\; \{f(2),f(2),f(2)\},\; 
\{f(2),f(2),f(4),f(4)\},\; \\
\{f(2),f(4),f(4),f(4),f(4)\},\; \{f(4),f(4),f(4),f(4),f(4),f(4)\}.\; 
\ena\eeq
Suppose that a PC $\varphi$ contains two particles at distance $\sqrt{9}$ from each other. The only corresponding 
subset is characterized by the collection $\{f(1),f(4),f(5),f(5),f(5),f(5)\}$. Any $\psi$ from this subset is congruent to the AC with occupied sites at 
$$(0,0,-1),\; (0,0,2),\; (2,1,0),\ (1,-2,0),\; (-2,-1,0),\;(-1,2,0).$$
Then the vacant site $(0,1,0)$ is at distances $\sqrt{2}$, $\sqrt{2}$, $\sqrt{4}$ and $\sqrt{5}$ from some of the 
particles above but there is no subset in \eqref{fD=6} which contains $\{f(2),f(2),f(4),f(5)\}$. Therefore, a PC $\varphi$ 
does not contain two particles at distance $\sqrt{9}$ from each other.  

Among the remaining ACs $\psi$ with $U(\psi) =1/12$, only the subset $\{f(2),f(2),f(2)\}$ does not contain a pair of particles at distance 
$\sqrt{8}$ from each other, but one cannot construct the entire PC solely with subset $\{f(2),f(2),f(2)\}$, as the regular 
$\sqrt{6}$-tetrahedron does not belong to $\bbZ^3$. Thus, some other subset from \eqref{fD=6} must be utilized. Consequently, 
a PC $\varphi$ contains a pair of particles, say $(2,1,1)$ and $(2,-1,-1)$, at distance $\sqrt{8}$ from each other. Then 
the vacant site $(2,0,0)$ is at distance $\sqrt{2}$ from both of them (it is located in the middle of the joining segment) 
and the only containing subset from \eqref{fD=6} is $\{f(2),f(2),f(4),f(4)\}$ with the only implementation containing the particles 
at $(0,0,0)$ and $(4,0,0)$. Next, the vacant site $(1,0,1)$ is at distance $\sqrt{2}$ from both $(0,0,0)$ and $(2,1,1)$. 
The corresponding subsets from \eqref{fD=6} are $\{f(2),f(2),f(2)\}$ and $\{f(2),f(2),f(4),f(4)\}$.

First, consider the case $\{f(2),f(2),f(4),f(4)\}$. The only possible implementation contains particles at $(1,0,3)$ and
$(1,-2,1)$, both at distance $\sqrt{4}$ from $(1,0,1)$. These two particles are at distance $\sqrt{8}$ from each other 
which implies the particles at $(-1,-1,2)$ and $(3,-1,2)$ by the earlier argument. The result is the parallelepiped with 
vertices 
$$(0,0,0),\; (2,1,1),\; (4,0,0),\; (2,-1,-1),\; (-1,-1,2),\; (1,0,3),\;(3,-1,2),\; (1,-2,1),$$
where two faces are rhombuses with diagonals $\sqrt{8}, \sqrt{16}$, two faces are rhombuses with  
diagonals $\sqrt{6}, \sqrt{18}$, and two faces are rhombuses with diagonals $\sqrt{10}, \sqrt{14}$.

Second, consider the case $\{f(2),f(2),f(2)\}$. The only possible implementation contains a particle at $(1,-1,2)$. Then 
consider a vacant site $(2,0,1)$ which is at distances $\sqrt{1}$, $\sqrt{3}$, $\sqrt{5}$, $\sqrt{5}$, $\sqrt{5}$ from 
other particles. The only corresponding subset from \eqref{fD=6} is $\{f(1),f(3),f(5),f(5),f(5),f(5),f(5)\}$, and its only 
implementation 
contains particles at $(3,0,3)$ and $(3,-2,1)$. Now, the vacant site $(3,-1,1)$ is at distances $\sqrt{1}$, $\sqrt{3}$, 
$\sqrt{5}$, 
$\sqrt{5}$, $\sqrt{5}$, $\sqrt{5}$ from other particles. This leads to the subset $\{f(1),f(3),f(5),f(5),f(5),f(5),f(5)\}$, and 
its only implementation contains a particle at $(5,-1,2)$. The result is  the parallelepiped with vertices 
$$(0,0,0),\; (2,1,1),\; (4,0,0),\; (2,-1,-1),\; (1,-1,2),\; (3,0,3),\; (5,-1,2),\; (3,-2,1)$$
where, as in the previous case, two faces are rhombuses with diagonals $\sqrt{8}, \sqrt{16}$, two faces are
rhombuses with diagonals $\sqrt{6}, \sqrt{18}$, and two faces are rhombuses with diagonals 
$\sqrt{10}, \sqrt{14}$.

Thus, a PC $\varphi$ is a concatenation of parallelepipeds congruent to the above one. Consider all pairs of 
parallelepipeds in $\varphi$ having a common $\sqrt{8}, \sqrt{16}$-rhombic face. They necessarily form disjoint 
double-infinite 
sequences which we call beams. If all parallelepipeds in a beam are not the shifts of each other (i.e. some pairs of them 
are symmetric reflections of each other) then each parallelepiped adjacent to the beam can be glued to the beam in a
unique way (via the common face), which keeps all $\sqrt{8}, \sqrt{16}$-rhombuses  parallel to each other. Continuing 
this process, we reconstruct the entire PC $\varphi$ which is a stack of parallel 
$\sqrt{8}, \sqrt{16}$-rhombic meshes  by construction.

If every next parallelepiped in a beam is a shift of the previous one then the beam contains two 
flat faces constructed 
solely from $\sqrt{6}$-equilateral triangles. A parallelepiped glued to such face does not necessarily have a common face 
with a parallelepiped from the beam. It may have one common triangle with two consecutive parallelepipeds from the 
beam. Nevertheless, the $\sqrt{10}, \sqrt{14}$-rhombic faces can be glued in a unique way,
which extends a single 
beam into a pair of parallel $\sqrt{6}$-triangular meshes. Finally, such pairs of meshes can be combined into the 
entire PC $\varphi$ which is a stack of parallel $\sqrt{6}$-triangular meshes by construction.
\ep

The rigorous analysis of EPGMs for $D^2=6$ large $u$ remains an open problem. %Here we offer the following
%conjecture. Similarly to the case $D^2=5$, we can 
%introduce the collection $\sH^{(6)}$ of HCP-type PCs: these are layered configurations
%$\onwl{\bigcup}\limits_{k\in\bbZ}\alp^{(8,16)}_{i,4k,j_k}$ where $\{j_k\}=\overline{01}$ or $\{j_k\}=\overline{02}$, 
%sub-script $j_k$ takes two values:
%\beq\label{eq:D^2=6HCP}j_0=0,\;\hbox{ and }\;\bcs j_k=&\hbox{$0,1$ for $k\geq 1$ and $0,3$ for $k\leq -1$, or}\\
%j_k=&\hbox{$0,2$ for $k\geq 1$ and $0,4$ for $k\leq -1$,}\ecs\;\hbox{with $j_k\neq j_{k+1}$,}\quad\eeq
%and their $\bbZ^3$-shifts. The cardinality of $\sH^{(6)}$ is $144=6\cdot 4\cdot 2\cdot 3$.\vskip .5cm

%{\bf Conjecture.} Let $u$ be large enough: $u\geq u^0(6)$. Then there are  
%$144$ {\rEPGM}s, i.e., $\sharp (\sE ({\sqrt 6},u))=144$, and each \rEPGM\ is generated by
%a \rPGS\ from $\sH^{(6)}$.

%#########################################
%                                                                                             D^2=8,9,12
%#########################################

\section{Cases $D^2=8,9,12$}
%#########################################
%                                                                                             D^2=8
%#########################################

\subsection{$D^2=8$.}\label{Subsec:D=8}  

%The case of PCs for $D^2=8$ is similar to $D^2=2$; this case can be 
%considered as a direct generalization of case  $D^2=2$.
For $D^2=8$, an LRFF family $\unf^{(8)}$ has 
\beq\label{fD=8} f(0)=1,\;\; f(1)=\dfrac{1}{2},\;\;f(2)=f(3)=\dfrac{1}{4},\;\;f(4)=\dfrac{1}{6},\;\;f(5)=f(6)=\dfrac{1}{8}.\eeq
Here the balls $\rB (x)=\rB_{\sqrt 7}(x)$ contain $81$ sites.
%$\rB (x)=\{y\in\bbZ^3:\;\rho (x,y)\leq{\sqrt 2}\}$. 
Each $\psi\in\sA_{\sqrt{8}}\big(\rB (x)\big)$ contains at most $6$ particles. Consider 
the potential function $U(\psi )=U^{(8)}(\psi )$:
\beq\label{eq:U8}U (\psi )=\dfrac{1}{16}\sum_{y\in B (x)}\psi (y)f\big(\rho (x,y)^2\big).
% \quad\hbox{where $\psi\in\sA_{\sqrt{8}}\big(\rB (x)\big)$.}
\eeq%{\mathbf 1}(\psi (y)=1)  
The normalizing constant $C$ from \eqref{eq:C} equals $16$. 
For any finite $\phi\in\sA_{\sqrt{8}}\big(\bbZ^3\big)$,
we have $\sharp  (\phi )=\sum\limits_{x\in\bbZ^3}U (\phi\upharp_{\rB (x)})$, in agreement with  \eqref{F}.

A direct calculation shows that the lattice
\beq\beal\label{eq:FCC8}
\vphi^{(8)}:=\big\{m\,(2,2,0) + n\,(2,0,2) + k\,(0,2,2) :\, m, n, k\in\bbZ\big\}=2\bbA_3\ena\eeq
is a PC. %: it is an FCC sub-lattice in $\bbZ^3$. 
The fundamental parallelepiped for $\vphi^{(8)}$
has volume $16$. Consider the collection $\sS^{(8)}$ of $\bbZ^3$-shifts of $\vphi^{(8)}$. 

\bthmi\label{thm:6.1A.} 
Set \ $\sS^{(8)}$ exhausts all \ {\rPC}s for $D^2=8$. The cardinality of \ $\sS^{(8)}$ is $16$. 
Every\ \rPC \ is periodic. The {\rPC}s\ are $\bbZ^3$-symmetric to each other. The particle density of any \ \rPC \ equals $1/16$.
\ethmi

\bp
The set of ACs $\psi$ on $\rB(x)$ which give $U(\psi)=1/16$ %contains $27$ items and 
is partitioned into $6$ subsets; each subset is characterized by a fixed collection  
of values $f(\,\cdot\,)$ participating in the sum $U(\psi)$
\beq\beac\label{f8list}\{f(0)\}, \{f(1), f(5), f(5), f(5), f(5)\}, \{f(1), f(6), f(6), f(6), f(6)\},\\
\{f(2),f(2), f(6), f(6),f(6),f(6)\}, \{f(3),f(3), f(3),f(3)\},\\ \{f(4),f(4),f(4),f(4),f(4),f(4)\}.\ena\eeq

Assume that a PC $\vphi$ contains two particles at distance $\sqrt{10}$. For definiteness, consider 
sites $(0,0,0)$ and $(3, 1, 0)$. Then the vacant site $(1, 1, 0)$ is at distance $\sqrt{2}$ and $\sqrt{4}$, 
respectively, from our chosen sites. The above list of subsets does not contain a subset which includes 
both $f(2)$ and  $f(4)$. This contradicts the fact that $\vphi$ is a PC. Therefore, there is no PC 
containing a pair of sites at distance $\sqrt{10}$. A direct enumeration shows that each of the ACs $\psi$ with $U(\psi)=1/16$ which does not have a pair of sites at distance $\sqrt{10}$ contains a pair of sites at distance $\sqrt{8}$.

Assume that a PC $\vphi$ contains two sites at distance $\sqrt{8}$. For definiteness, consider sites $(0,0,0)$ and $(2, 2, 0)$. Then the vacant site $(1, 1, 0)$ is at distance $\sqrt{2}$ from both chosen sites. There is only one subset in the list \eqref{f8list} containing $f(2)$ twice. Consequently, the only possibility to implement this subset is to place particles at sites $(2, 0, \pm 2), (0, 2, \pm 2)$. 

The above argument can be repeated for sites $(1, 0, \pm 1)$, $(0, 1, \pm 1)$, etc., and consequently  $\vphi$ is uniquely recovered.
%Take a PC $\vphi$ on $\bbZ^3$ and assume that $x\in\bbZ^3$ is occupied: $\vphi (x)=1$.
%Consider the site $y=x+(1,1,1)$. There is a unique way to get $U(\vphi\upharpoonright_{B (y)},y)=1$,
%namely, to put a particle in $x+(2,2,2)$ which correspond to the sub-set $\{f(3,f(3\}$.
%A similar argument asserts that we must put a particle in all $8$ sites $x+(\pm 2,\pm 2,\pm 2)$.
%We can repeat this step for every of these 8 sites, then for their ${\sqrt{12}}$-neighbors and so on. 
%This leads to a BCC-lattice $\vphi_{1,1}$ or its shift. 
\ep

%Equations:
%$$f(1)+4f(5)=1,\;2f(2)+4f(6)=1,\;f(3)=1/4,\;f(4)=1/6. $$

\bthmj\label{thm:6.1B.} 
Let $u$ be large enough: $u\geq u^0(8)$. Then there are 
$16$ {\rEPGM}s, i.e., $\sharp (\sE(\sqrt{8}, u))=16$, and each \rEPGM\ is generated by
a \rPGS\ from $\sS^{(8)}$.
\ethmj

\bp
Follows from Theorem 6.1A, the Peierls bound \eqref{hamexcitation} and the PS theory. 
\ep

The PC $\vphi^{(8)}$ is another example of a $D$-FCC sub-lattice in $\bbZ^3$. The cases $D^2=2, 8$ are also covered by general Theorems 8.1A %\ref{thm:9.2} 
and 8.1B. %\ref{thm:101}.  
The LRFF based proofs presented above are more elementary.

%#######################################
%                                                                                     D^2=9
%#######################################

\subsection{$D^2=9$.} 

For $D^2=9$, an LRFF family $\unf^{(9)}$ has  
\beq\label{fD=9} f(0)=1,\; f(1)=\frac{2}{3},\;f(2)=\frac{1}{2},\;
f(3)=\frac{1}{4},\;f(4)=f(5)=\frac{1}{6},\;f(6)=\frac{1}{12},\;f(8)=0.\eeq
%{\it A proper family.}
Here the balls $\rB (x)=\rB_{\sqrt 7}(x)$ contain $81$ sites.
%$\rB (x)=\{y\in\bbZ^3:\;\rho (x,y)\leq{\sqrt 2}\}$. 
Each $\psi\in\sA_{3}\big(\rB (x)\big)$ contains at most $6$ particles. Consider 
the potential function $U(\psi )=U^{(9)}(\psi )$:
\beq\label{eq:U9}U (\psi )=\dfrac{1}{20}\sum_{y\in\rB (x)}\psi (y)f\big(\rho (x,y)^2\big).
 %\;\hbox{where $\psi\in\sA_3\big(\rB (x)\big)$.}
\eeq%{\mathbf 1}(\psi (y)=1)  
The normalizing constant $C$ from \eqref{eq:C} equals $20$. 
For any finite $\phi\in \sA_3\big(\bbZ^3\big)$, we have $\sharp  (\phi )=\sum\limits_{x\in\bbZ^3}U (\phi\upharp_{\rB (x)})$, in agreement with \eqref{F}.

Consider the set $\sS^{(9)}$ formed by 6 congruent lattices $\vphi^{(9)}_{i,\,l}$,  
$i=1,2,3$, $l=0,1$, defined below, and their $\bbZ^3$-shifts. Set 
\beq\beal \vphi^{(9)}_{1,0}:=\Big\{m(0,3,1) + n(0,-1,3) + k(2,1,2) :\,m,n,k\in\bbZ \Big\},\\
\vphi^{(9)}_{1,1}:=\Big\{m(0,3,-1)+n(0,-1,-3)+k(2,1,-2):\;m,n,k\in\bbZ\Big\},\ena\eeq
with $\vphi^{(9)}_{1,1}$ obtained from $\vphi^{(9)}_{1,0}$ via the reflection about the plane $\rx_3=0$. %rotation by the angle $2\arctan (1/3)$ around the $\rx_1$-axis. 
Lattices 
$\vphi^{(9)}_{2,l}$ and   $\vphi^{(9)}_{3,l}$ are obtained from $\vphi^{(9)}_{1,l}$ via the rotation in 
$\bbR^3$ by $\pi/2$ about the $\rx_3$- and $\rx_2$-axis, respectively. The fundamental parallelepiped for
each lattice has volume $20$.

Observe that $\vphi^{(9)}_{1,0}=\onwl{\bigcup}\limits_{k\in\bbZ}\theta^{(10)}_{1,2k,j_k}$ where  
$$\theta^{(10)}_{1,0,0}=\Big\{m(0,3,1)+n(0,-1,3):\;m,n\in\bbZ\Big\}$$
is a square $\sqrt{10}$-lattice in the plane\ $\rx_1=0$ and                            %subsequent shifts 
$\theta^{(10)}_{1,2k,0}:=\theta^{(10)}_{1,0,0}+k(2,0,0)$ is the square $\sqrt{10}$-mesh in the affine plane 
$\rx_1=2k$, $k\in\bbZ$. Correspondingly, $\theta^{(10)}_{1,2k,1}:=\theta^{(10)}_{1,2k,0}+k(0,1,2)$, that is, the sites of $\theta^{(10)}_{1,2k,1}$ are located at the centers of the $({\sqrt{10}}\times{\sqrt{10}})$-squares of $\theta^{(10)}_{1,2k,0}$ and {\it vice versa}. Then $\vphi^{(9)}_{1,0}=\onwl{\bigcup}\limits_{k\in\bbZ}\theta^{(10)}_{1,2k,j_k}$, where $\{j_k\}=\overline{01}$. Cf. \eqref{layered3}, with $q=10$, $h=2$ and $u=1$.% the family of square ${\sqrt q}$-meshes reduced to a single
%item, $\theta^{(10)}_{i,k}\subset Z_{i,k}$, with $u=1$ and sub-scripts $j$ and  $\{j_k\}$ omitted. 
%Note that vector $(2,1,2)$, of length $3$, has the property that its projection $(0,1,2)$ to
%plane  $\rx_1=0$ gives the center of the ${\sqrt{10}}\times{\sqrt{10}}$-square of $\theta^{(10)}_{1,0}$.
 %The cardinality of set $\sS^{(9)}$ equals $3\cdot 2\cdot 10\cdot 2=120$, and each PC from 
%$\sS^{(9)}$ is periodic and has particle density $\dfrac{1}{20}$.

A representation of PC $\vphi^{(9)}_{1,1}$ as a union of square $(\sqrt{10}\times \sqrt{10})$-meshes can be 
constructed in a similar way. 

\bthmk\label{thm:6.2A.} 
Set \ $\sS^{(9)}$ exhausts all \ {\rPC}s for $D^2=9$. The cardinality of \ $\sS^{(9)}$ is $120$. 
Every \ \rPC \ is periodic. The {\rPC}s\ are $\bbZ^3$-symmetric to each other. The particle density of any\ \rPC\ equals $1/20$.
\ethmk

\bp The set of ACs $\psi$ on $\rB(x)$ which give 
$U(\psi)=1/20$ %contains $318$ items and 
is 
partitioned into $8$ subsets; each subset is characterized by a fixed collection  
of values $f(\,\cdot\,)$ participating in the sum $U(\psi)$
\beq\label{eq:distlist9}\beac \{f(0)\}, \{f(1),f(5),f(6),f(6)\},\{f(1),f(6),f(6),f(6),f(6)\},\\
\{f(2),f(3),f(5),f(6)\},\{f(2),f(4),f(5),f(5)\},\\
\{f(2),f(5),f(5),f(5)\},\;\{f(2),f(5),f(5),f(6),f(6)\},\\
\{f(4),f(4), f(5),f(5),f(5),f(5)\}. \ena\eeq

Assume that a PC $\vphi$ contains two particles at distance $\sqrt{12}$. For definiteness, consider sites $x_1=(0,0,0)$ and $x_2=(2, 2, 2)$. Then the vacant site $(1, 1, 1)$ is at distance $\sqrt{3}$ from both $x_1$ and $x_2$. However, the above list \eqref{eq:distlist9} does 
not contain collections 
with at least 2 distances between occupied sites equal to $f(3)$. 
This contradicts the fact that $\vphi$ is a PC. Therefore, there is no PC containing a pair of sites at distance $\sqrt{12}$. A direct enumeration shows that each of the ACs $\psi$ with $U(\psi)=1/20$ which does not have a pair of sites at distance $\sqrt{12}$ contains a pair of sites at distance $\sqrt{10}$.

Assume that a PC $\vphi$ contains two particles at distance $\sqrt{10}$. For definiteness, consider sites $x_1=(0,0,0)$ and 
$x_2=(1,3,0)$. Then the vacant site $(0,1,0)$ is at distance $1$ from $x_1$ and at distance ${\sqrt 5}$ from $x_2$. The only corresponding subset in \eqref{eq:distlist9} is $\{f(1),f(5),f(6),f(6)\}$; therefore  two sites $(-1,2,\pm 2)$ are 
occupied in $\vphi$. Now consider a vacant site $(0, 3, 0)$ which is at distance $1$ from $(1, 3, 0)$ and at distance $\sqrt{6}$ from sites  $(-1,2,\pm 2)$. The only possibility for site $(0, 3, 0)$ to implement one of the subsets from \eqref{eq:distlist9} is when there is an occupied site $(-2,4,0)$.

Repeating this argument 
for the vacant site $(-2,3,0)$ leads to the occupied site $(-3,1,0)$. Thus, we obtain a 
square by-pyramid whose $4$ base side-lengths are ${\sqrt{10}}$, and the remaining $8$ 
side-lengths are ${\sqrt 9}$. Iterating this construction yields three square ${\sqrt{10}}$-meshes 
in horizontal planes where $\rx_3= -2, 0, 2$. Iterating the construction, we get similar 
structures at levels $-4$ and $4$. And so on: as a result, PC $\vphi$ is 
identified as a member of collection $\sS^{(9)}$.   
\ep

\bthml\label{thm:6.2B.} 
Let $u$ be large enough: $u\geq u^0(9)$. Then there are 
$120$ {\rEPGM}s, i.e., $\sharp (\sE(\sqrt{9}, u))=120$, and each \rEPGM\ is generated by
a \rPGS\ from $\sS^{(9)}$.
\ethml

\bp
Follows from Theorem 6.2A, the Peierls bound \eqref{hamexcitation} and the PS theory. 
\ep

%#########################################
%                                                                                             D^2=12
%#########################################

\subsection{$D^2=12$.} 
For $D^2=12$, an LRFF family $\unf^{(12)}$ has  
\beq\label{fD=12}\beac f(0)=1, f(1)=1,f(2)=\dfrac{3}{4},
f(3)=f(4)=\dfrac{1}{2},f(5)=\dfrac{1}{4},f(6)=\dfrac{1}{8},\\
f(8)=f(9)=f(10)=f(11)=0.\ena\eeq
%{\it A proper family.}
Here the balls $\rB (x)=\rB_{\sqrt 7}(x)$ contain $81$ sites.
%$\rB (x)=\{y\in\bbZ^3:\;\rho (x,y)\leq{\sqrt 2}\}$. 
Each $\psi\in\sA_{\sqrt{12}}\big(\rB (x)\big)$ contains at most $4$ particles. Consider 
the potential function $U(\psi )=U^{(12)}(\psi )$:
\beq\label{eq:U12}U (\psi )=\dfrac{1}{32}\sum_{y\in B (x)}\psi (y)f\big(\rho (x,y)^2\big).
% \quad\hbox{where $\psi\in\sA_{\sqrt{12}}\big(\rB (x)\big)$.}
\eeq%{\mathbf 1}(\psi (y)=1)  
The normalizing constant $C$ from \eqref{eq:C} equals $32$. 
For any finite $\phi\in\sA_{\sqrt{12}}\big(\bbZ^3\big)$, we have $\sharp  (\phi )=\sum\limits_{x\in\bbZ^3}U (\phi\upharp_{\rB (x)})$, in agreement with \eqref{F}.

A direct calculation shows that the lattice
\beq\beal
\vphi^{(12)}:=\{m\,(4,0,0) + n\,(0,4,0) + k\,(2,2,2) :\, m, n, k\in\bbZ \} 
\ena\eeq
is a PC: it is a $\sqrt{12}$-BCC sub-lattice in $\bbZ^3$. The fundamental parallelepiped for $\vphi^{(12)}$ has volume $32$. The set 
$\sS^{(12)}$ is formed by the  $\bbZ^3$-shifts of $\vphi^{(12)}$. 

\bthmm\label{thm:6.3A.} 
Set \ $\sS^{(12)}$ exhausts all \ {\rPC}s for $D^2=12$. The cardinality of \ $\sS^{(12)}$ is $32$. 
Every \ \rPC \ is periodic. The {\rPC}s\ are $\bbZ^3$-symmetric to each other. The particle density of any \ \rPC\  equals $1/32$.
\ethmm

\bp The set of ACs $\psi$ on $\rB(x)$ which give 
$U(\psi)=\diy\frac{1}{32}$ %contains 31 items and 
is 
partitioned into $9$ subsets; each subset is characterized by a fixed collection  
of values $f(\,\cdot\,)$ participating in the sum $U(\psi)$
\begin{equation}\beac \label{distlist12}\{f(0)\}, \{f(1)\}, \{f(2),f(5)\},\{f(2),f(6),f(6)\},\{f(3),f(3)\},\{f(4),f(4)\},\\
\{f(4),f(5),f(5)\},\;\{f(4),f(5),f(6),f(6)\},\;\{f(5),f(5),f(5),f(5)\}.\ena
\end{equation}

Consider an occupied site in a PC and, for definiteness, assume that it is $(0, 0, 0)$. Take the vacant site 
$(1, 1, 1)$. The list \eqref{distlist12} contains a single subset containing $f(3)$; therefore the site $(2, 2, 2)$ is 
also occupied. 

A similar argument asserts that we must put a particle in all $8$ sites $(\pm 2,\pm 2,\pm 2)$.
We can repeat this step for every of these 8 sites, then for their ${\sqrt{12}}$-neighbors and so on.  
\ep

\bthmn\label{thm:6.3B.} 
Let $u$ be large enough: $u\geq u^0(12)$. Then there are 
$32$ {\rEPGM}s, i.e., $\sharp (\sE(\sqrt{12}, u))=32$, and each \rEPGM\ is generated by
a \rPGS\ from $\sS^{(12)}$.
\ethmn

\bp
Follows from Theorem 6.3A, the Peierls bound \eqref{hamexcitation} and the PS theory. 
\ep

The PCs $\vphi^{(3)}$ and $\vphi^{(12)}$ are examples of $D$-BCC sub-lattices in $\bbZ^3$. We think that $D^2=3, 12$ are the only cases where a $D$-BCC sub-lattice is a PC in $\bbZ^3$. 

%############################################
%                                                                                             D^2=10
%###########################################

\section{Case $D^2=10$}\label{SecD=10} 
For $D^2=10$, an LRFF $\unf^{(10)}$ has %uniquely defined and has  
\beq\label{fD=10}f(0)=1,f(1)=\frac{5}{6},f(2)=f(3)=\frac{1}{2},f(4)=\frac{1}{3},f(5)=f(6)=\frac{1}{6},f(8)=f(9)=0.
\eeq
%{\it A proper family.}
Here the balls $\rB (x)=\rB_{\sqrt 7}(x)$ contain $81$ sites.
%$\rB (x)=\{y\in\bbZ^3:\;\rho (x,y)\leq{\sqrt 2}\}$. 
Each $\psi\in\sA_{\sqrt{10}}\big(\rB (x)\big)$ contains at most $6$ particles. Consider 
the potential function $U(\psi )=U^{(10)}(\psi )$:
\beq\label{eq:U10}U (\psi )=\dfrac{1}{26}\sum_{y\in B (x)}\psi (y)f\big(\rho (x,y)^2\big).
% \quad\hbox{where $\psi\in\sA_{\sqrt{10}}\big(\rB (x)\big)$.}
\eeq%{\mathbf 1}(\psi (y)=1)  
The normalizing constant $C$ from \eqref{eq:C} equals $26$. 
For any finite $\phi\in\sA_{\sqrt{10}}\big(\bbZ^3\big)$, we have $\sharp  (\phi )=\sum\limits_{x\in\bbZ^3}U (\phi\upharp_{\rB (x)})$, in agreement with  \eqref{F}.

%{\bf $D^2=10$.} 
To identify the PCs, consider the following 8 congruent lattices $\vphi^{(10)}_{i,\,l}$,  
$i=0,1,2,3$, $l=0,1$
\beq\label{eq:vphirt10}\beacl
\vphi_{i,0}^{(10)}&:=\Big\{m\,(-1,-3s_2(i),4s_3(i))\\
\;&\qquad\qquad+n\,(3,-4s_2(i),s_3(i))+k\,(0 , 3s_2(i), -s_3(i)):\; m,n,k\in\bbZ\Big\},\\
\vphi_{i,1}^{(10)}&:=\Big\{m\,(-1, 4s_2(i), -3s_3(i))\\
\;&\qquad\qquad+n\,(3, s_2(i),-4s_3(i))+k\,(0, -s_2(i), 3s_3(i)):\; m,n,k\in\bbZ\Big\},\ena\eeq
where $s_2(i)$, $s_3(i)$ are given by (2.11), and $\vphi^{(10)}_{i,1}$ is obtained from $\vphi^{(10)}_{i,0}$ 
via the reflection about the plane $\rx_2=(-1)^i \rx_3$.
The fundamental parallelepiped of each sub-lattice $\varphi^{(10)}_{i,l}$ has volume 26.
As in previous cases, we denote by $\sS^{(10)}$ the set formed by the $\bbZ^3$-shifts of $\varphi^{(10)}_{i,l}$.
%+x$, $x\in\bbZ^3$.
%collection of these ACs; the cardinality $\sharp\;\sS^{(10)}=4\cdot 26 =104$. 

Observe that $\vphi^{(10)}_{i,0}=\onwl{\bigcup}\limits_{k\in\bbZ}\tau^{(26)}_{i,2k,j_k}$ where  
$$\tau^{(26)}_{i, 0, 0}=\Big\{m\,(-1,-3s_2(i),4s_3(i))+n\,(3,-4s_2(i),s_3(i)):\;m,n\in\bbZ\Big\}$$
is a triangular $\sqrt{26}$-lattice in the plane\ $\rx_1  +\rx_2 s_2+ \rx_3 s_3=0$, and                            
%subsequent shifts 
$$\tau^{(26)}_{1,2k,0}:=\tau^{(26)}_{1,0,0}+\dfrac{2k}{3}(1,s_2,s_3)$$ is the triangular $\sqrt{26}$-mesh in the 
plane $\rx_1  +\rx_2 s_2+ \rx_3 s_3=2k$, $k\in \bbZ$. Correspondingly, 
$$\tau^{(26)}_{i,2k,1}:=\tau^{(26)}_{i,2k,0}+\dfrac{1}{3}(-2, 7s_2, -5s_3),$$ 
that is, the sites of $\tau^{(26)}_{i,2k,1}$ are located at the centers of the ${\sqrt{26}}$-triangles of 
$\tau^{(26)}_{i,2k,0}$ and {\it vice versa}. Similarly, the sites of 
$$\tau^{(26)}_{i,2k,2}:=\tau^{(26)}_{i,2k,0}+\dfrac{1}{3}(-5, -2s_2, 7s_3)$$ 
are located at the centers of the ${\sqrt{26}}$-triangles of $\tau^{(26)}_{i,2k,1}$ and {\it vice versa}. Then 
$\vphi^{(10)}_{i,0}=\onwl{\bigcup}\limits_{k\in\bbZ}\tau^{(26)}_{i,2k,j_k}$, where $\{j_k\}=\overline{012}$. Cf. 
\eqref{layered1}, with $q=26$, $h=2$ and $r=2$.

A representation of PC $\vphi^{(10)}_{i,1}$ as a union of $\sqrt{26}$-triangular meshes can be constructed in a 
similar way. 

\bthmo\label{thm:7A.} 
Set \ $\sS^{(10)}$ exhausts all \ {\rPC}s for $D^2=10$. The cardinality of \ $\sS^{(10)}$ is $208$. 
Every\ \rPC \ is periodic. The\ {\rPC}s\ are $\bbZ^3$-symmetric to each other. The particle density of any\ \rPC\  equals $1/26$. \ethmo

\bp
It is convenient to take $\varphi^{(10)}_{0,0}$ as a representative for $\sS^{(10)}$.
% Given a $k\in\bbZ$, the 
% corresponding occupied sites listed in \eqref{eq:vphirt10} form a triangular $\sqrt{26}$-mesh 
% $\tau_{\,k}=\tau^{(26)}_{\,k}$ lying in the plane 
%$$P_{0, 2k}:\;\rx_1 +\rx_2 +\rx_3 = 2k$$ %\; {\dfrac{\sqrt 4}{3}}$$
%orthogonal to the $(1,1,1)$-diagonal, with $i(s_1,s_2)=0$. That is, $\tau^{(26)}_{\,k}=\vphi^{(10)}_{1,1}\cap P_{0, 2k}$, 
%and $\vphi^{(10)}_{1,1}=\onwl{\bigcup}\limits_{k\in\bbZ}\tau^{(26)}_{\,k}$. Cf. \eqref{layered1} where
%$q=26$, $h=2$ 
The distance between the planes containing meshes $T_{0, 2k}$ and $T_{0, 2(k+1)}$ is equal to $\sqrt{4/3}$. The 
occupied 
sites from meshes $\tau_{\,k-1}:=\tau^{(26)}_{0,2k-2,j_{k-1}}$ and  
$\tau_{\,k+1}:=\tau^{(26)}_{0,2k+2,j_{k+1}}$ 
are projected into centers of equilateral triangles from the mesh $\tau_{\,k}:=\tau^{(26)}_{0,2k,j_k}$. 
Moreover, for each occupied site from $\tau_{\,k}$ there are:

{(i)} 3 occupied sites at distance $\sqrt{10}$ from mesh $\tau_{\,k-1}$, 

{(ii)} 3 occupied sites at distance $\sqrt{10}$ from mesh $\tau_{\,k+1}$,  

{(iii)} 3 occupied sites at distance $\sqrt{14}$ from mesh $\tau_{\,k-2}$, 

{(iv)} 3 occupied sites at distance $\sqrt{14}$ from mesh $\tau_{\,k+2}$,   

{(v)} 1 occupied sites at distance $\sqrt{12}$ from mesh $\tau_{\,k-3}$,

{(vi)} 1 occupied sites at distance $\sqrt{12}$ from mesh $\tau_{\,k+3}$.   

\noindent
Here $\tau_{\,k\pm n}$ are defined similarly to  $\tau_{\,k\pm 1}$.

E.g., for the occupied site $(0, 0, 0)\in\tau_0$, the above collection (i) is  $I_\rbi=\{(1,0,-3)$,
$(0,-3,1),(-3,1,0)\}$, collection (ii) is $I_\rbii=\{(-1,0,3),(0,3,-1),(3,-1,0)\}$, collection
(iii) is $I_\rb3i=\{(1,-3,-2),(-3,-2,1),(-2,1,-3)\}$ collection (iv) is $I_\rbiv=\{(-1,3,2)$,
$(3,2,-1),(2,-1,3)\}$, the single-site collections  described in (v) and (vi) are\\
$I_\rbv=\{(-2,-2,-2)\}$ and $I_\rbvi=\{(2,2,2)\}$, respectively.

The vertices listed in $I_\rbi$, $I_\rb3i$ and $I_\rbv$, together with $(0, 0, 0)$, span a parallelepiped $\sP_{-}$
where all edges have length $\sqrt{10}$, all faces are rhombuses with diagonals of length $\sqrt{14}$ 
and $\sqrt{26}$, and the shortest diagonal of $\sP_{-}$ has length $\sqrt{12}$. A congruent parallelepiped,  $\sP_{+}$, is
spanned by $(0, 0, 0)$ and the vertices listed in $I_\rbii$, $I_\rbiv$ and $I_\rbvi$. Both  $\sP_{\pm}$ are 
fundamental parallelepipeds for $\vphi^{(10)}_{0,0}$. Each of  $\sP_{\pm}$ 
can be uniquely partitioned into 6 congruent tetrahedrons. For each of the tetrahedrons the corresponding 3 
pairs of opposite sides have the following lengths: $\sqrt{10}$ and $\sqrt{12}$,  $\sqrt{10}$ and $\sqrt{14}$,  
$\sqrt{10}$ and $\sqrt{14}$. Thus, the entire sub-lattice $\varphi^{(10)}_{0,0}$ is the union of congruent tetrahedrons. 
(Such a tetrahedron has the following dihedral angles: $\dfrac{\pi}{2}$, $\dfrac{\pi}{2}$, 
$\pi - 2\cdot \arccos\Bigg(\diy\sqrt{\frac{2}{7}} \Bigg)$, %1.00685369$
$\dfrac{\pi}{3}$, $\arccos\Bigg(\diy\sqrt{\frac{2}{7}} \Bigg)$, $\arccos\Bigg(\diy\sqrt{\frac{2}{7}} \Bigg)$; 
it is a member of the first infinite family of rational tetrahedrons 
from Theorem 1.8 in \cite{KKPR}.)

The set of ACs $\psi$ on $\rB(x)$ which give 
$U(\psi)=\diy\frac{1}{26}$ %contains 31 items and 
is 
partitioned into $15$ subsets; each subset is characterized by a fixed collection  
of values $f(\,\cdot\,)$ participating in the sum $U(\psi)$\newpage
%origin. (Equivalently it is specified by the values $f(\cdot)$ participating in the sum $U(\psi, o)$. Cf. 
%\eqref{eq:distlist10}.) These 15 sets are:
%The list of collections replacing \eqref{eq:distlist9} contains 14 items:
\beq\label{eq:distlist10}\beac \{f(0)\},\;\{f(1),f(5)\},\;\{f(1),f(6)\},\;\{f(2),f(4),f(6)\},\\
\{f(2),f(5),f(6),f(6)\},\;\{f(2),f(6),f(6),f(6)\},\\
\{f(3),f(3)\},\;\{f(3),f(4),f(5)\},\;\{f(3),f(4),f(6)\},\\ 
\{f(3),f(5),f(5),f(5)\},\;\{f(3),f(5),f(5),f(6)\},\\
\{f(3),f(5),f(6),f(6)\},\;\{f(3),f(6),f(6),f(6)\},\\
\{f(4),f(6),f(6),f(6),f(6)\},\{f(6),f(6),f(6),f(6),f(6),f(6)\}.\ena\eeq

Assume that a PC $\vphi$ contains two particles at distance $\sqrt{11}$. For definiteness consider 
sites $x_1 = (0,0,0)$ and $x_2 = (1,1,3)$. Then the vacant site $x_3 = (1,0,1)$ is at distance $\sqrt{2}$ 
from $x_1$ and $\sqrt{5}$ from $x_2$. Similarly, the vacant site $x_4 = (1,0,2)$ is at distance $\sqrt{5}$ 
from $x_1$ and at distance $\sqrt{2}$ from $x_2$. According to the list \eqref{eq:distlist10}, the only possible
subset containing those distances is $\{f(2),f(5),f(6),f(6)\}$. 
It is not hard to see that it is impossible to add two more sites at distance $\sqrt{6}$ from $x_3$ without 
breaking admissibility and the same is true for $x_4$. This contradicts the fact that $\vphi$ is a PC. 
Therefore, there is no PC containing a pair of sites at distance $\sqrt{11}$. 

A similar argument establishes that a PC $\varphi$ cannot contain occupied sites at distance 
$\sqrt{13}$ 
from each other. For this case $x_1 = (0,0,0)$, $x_2 = (2,3,0)$ and $x_3 = (1,1,0)$, $x_4 = (1,2,0)$.

Next, we verify that a PC $\varphi$ must contain a pair of occupied sites at distance $\sqrt{12}$.
Suppose that in $\vphi$ there is no pair of particles at distance $\sqrt{12}$ from each other. Then, a direct enumeration shows that, taking 
into account lattice symmetries, there are only $5$ distinct ACs $\psi\in\sA_{\sqrt{10}}(\rB (x))$ 
which do not contain a pair of particles at distances $\sqrt{11}, \sqrt{12}$ or $\sqrt{13}$: 
\beq\beac\label{eq:list2rt10} \psi_1 = \{(-2, -1,0),(1,0,0)\},\; \psi_2 = \{(-2, -1, -1),(0,0, 2),(1,0, -1)\},\\
\psi_3 = \{(-2, -1,0),\;(0, 2, -1),\;(1, -1, -1),\;(1,0, 2)\},\; \\
\psi_4 = \{(-2, -1, -1),\;(-1, -1, 2),\;(0, 2,0),\;(1, -1, -2),\;(2, -1, 1)\},\; \\
\psi_5 = \{(-2, -1, -1),(-1, -1, 2),(-1, 2, 1),(1, -2, -1),(1,1,-2),(2, 1, 1)\}.\ena\eeq
The corresponding collections of values $f(\,\cdot\,)$ are, respectively:
$$\beac\{f(1),f(5)\},\; \{f(2),f(4),f(6)\},\; \{f(3),f(5),f(5),f(5)\},\; \\ \{f(4),f(6),f(6),f(6),f(6)\},\; \{f(6),f(6),f(6),f(6),f(6),f(6)\}.
\ena$$

It turns out that for each of the last 4 ACs $\psi_i$, $i=2,3,4,5$, in \eqref{eq:list2rt10} there is a 
vacant site $x_i$ such 
that it is impossible to have $\phi\supset \psi_i$ such that  $U(\phi\upharp_{\rB(x_i)}) =1/26$. Namely, 
$x_5 = (0,-1,0)$, $x_4 = (0,-1,0)$, 
$x_3 = (0,0,1)$, $x_2 = (-1,0,0)$. Note that there is a possibility for $U(\phi\upharp_{\rB(x_2)}) = \diy\frac{1}{26}$ 
if the site $\bar x = (-2,2,0)\in \phi$. However,  in this case the distance between occupied sites $(-2,2,0)$ and $(0,0,2)$ 
is $\sqrt{12}$ which 
contradicts the assumption. %For the case of $\psi_1$ in \eqref{eq:list2rt10} 
The last possibility is an AC $\phi$ which contains $\psi_1$ and does not contain any of $\psi_i$, $i=2, 3, 4, 5$. Such 
a configuration must contain $8$ occupied sites forming a parallelepiped congruent to  $\sP_{\pm}$. But  $\sP_{\pm}$
contains a pair of particles at distance $\sqrt{12}$ which again contradicts the assumption. 

Finally, consider two occupied sites $x_1$ and $x_2$ in a PC $\varphi$ at distance $\sqrt{12}$. For 
definiteness consider sites $x_1 = (0,0,0)$ and $x_2 =(-2,2,2)$. Take the vacant site $x_3=(0,2,1)$. For 
$\vphi\ni x_1, x_2$, the only possibilities
to have $U(\vphi\upharp_{\rB(x_3)}) = \diy\frac{1}{26}$ are when either the site $(1,3,0)$ or $(1,3,2)$ is 
occupied since the subset $\{f(5),f(5)\}$ implies either the subset $\{f(3),f(5),f(5),f(5)\}$ or the subset 
$\{f(3),f(5),f(5),f(6)\}$, both of which contain $f(3)$. Due to symmetry, it suffices to consider the site $(1, 3, 0)$ only. 
The unique way to obtain the subset 
$\{f(3),f(5),f(5),f(6)\}$ is to place a particle at site $(1, 1, 3)$. However, a particle at $(1, 1, 3)$ is at distance 
$\sqrt{11}$ from $x_1$ which is impossible in a PC. The unique way to obtain the subset $\{f(3),f(5),f(5),f(5)\}$ 
is to place a particle at site $(1, 2, 3)$. 

Repeating the argument with $x_3=(0,1,2)$ in place of $(0,2,1)$, we obtain another occupied site $(0,-1,3)$. 
Continuing this process with $x_3= (-1,0,2)$, $(-2,0,1)$, $(-2,1,0)$  and $(-1,2,0)$, we add other 
four occupied sites $(-3,-1,2)$, $(-3,0,-1)$, $(-2,3,-1)$ and $(1, 3, 0)$. These 8 occupied sites form 
the vertices of a parallelepiped congruent to  $\sP_{\pm}$, where all sides have length $\sqrt{10}$.

Now take the vacant site $\ovx=(1,1,1)$. The only possibility to have $U(\vphi\upharp_{\rB(\ovx)}) = 1/26$ 
is when the site $(3,0,1)$ is occupied, as the set $\{f(3),f(5),f(5)\}$ 
implies either the set $\{f(3),f(5),f(5),f(5)\}$ or the set $\{f(3),f(5),f(5),f(6)\}$; the latter combination is 
again impossible without breaking admissibility. 
This produces another pair of occupied sites $(1,2,3)$ and $(3,0,1)$ at distance $\sqrt{12}$ from each other. 
Therefore, the previous construction can be repeated, to recover another parallelepiped congruent to  $\sP_{\pm}$,
adjacent to the first one. This process can be iterated further, until all of $\varphi$ is recovered.
That is, we obtain that any PC $\vphi$ is a member of $\sS^{(10)}$. 
\ep

\bthmp\label{thm:7B.} Let $u$ be large enough: $u\geq u^0(10)$. Then there are 
$208$ {\rEPGM}s, i.e., $\sharp (\sE(\sqrt{10}, u))=208$, and each \rEPGM\ is generated by
a \rPGS\ from $\sS^{(10)}$.
\ethmp

\bp
Follows from Theorem 7A, the Peierls bound \eqref{hamexcitation} and the PS theory. 
\ep

%%%%%%%%%%%%%%%%%%%%%%%%%%%%%%%%%
%                                                         D^2=11

%%%%%%%%%%%%%%%%%%%%%%%%%%%%%%%%%%%%%%%
\section{Case $D^2=11$}\label{SecD=11}

The case $D^2=11$ is interesting because it is the first case where the Voronoi cell of the minimal 
volume does not tessellate the space (it is true for $D$ large enough). To identify the PCs 
for $D=\sqrt{11}$  we develop a new technique that was not used in previous sections. It is an 
analytic argument combined with a computer enumeration. A part of the difficulty was to make such  
enumeration efficient enough to be completed in a reasonable time. 

A part of the technique is the concept of a {\it discrete Voronoi cell} (DVC); it is meaningful for any 
exclusion distance $D$. Given $y\in\bbZ^3$ and an AC $\phi \in\sA_{D}\big(\bbZ^3\big)$, 
$\phi\neq\varnothing$, we calculate 
$$m(y) (=m(y,\phi)):= \min \Big[\rho(x' ,y),\;x'\in\phi\Big].$$ 
Define the DVC of site $x\in\phi$ as the set
$$\DVC(x) (=\DVC(x,\phi )):=\Big\{y \in \bbZ^3:\; \rho(y,x) = m(y)\Big\}.$$ 
Further, let $\sharp(y)$ denote the cardinality of the set $\Big\{x'\in\phi:\; \rho(x' ,y) = m(y)\Big\}$.
The {\it volume} of DVC $\DVC(x)$ is defined as
$$v(\DVC(x)) := \sum_{y \in\DVC(x)} {1 \over \sharp(y)}.$$
Observe that the intersection $\DVC(x') \cap \DVC(x'')$ can be non-empty for some $x'\not = x''$ but for 
$\phi\in\sA_{D}\big(\bbZ^3\big)$ we have a formal equality
$$\sum_{x \in \phi} v(\DVC(x)) \equiv \sharp(\bbZ^3),$$
where $\sharp(\bbZ^3)$ stands for the number of lattice sites in $\bbZ^3$.

Due to the discreteness, the {\it minimal} DVC volume 
$$v_\circ :=\min \,\Big[\;v(\DVC(x, \phi)):\;x\in\phi, \, \phi\in \sA_{D}\big(\bbZ^3\big)\Big]$$
is well-defined. If there exists a configuration $\varphi \in\sA_{D}\big(\bbZ^3\big)$ such that for all 
$x\in\varphi$
$$v(\DVC(x)) = v_\circ$$
then $\varphi$ is a PC. 
%It is not hard to see that $\varphi$ is a (canonical) PC in terms of properly defined m-potential.

Indeed, %{\color{blue} let $B_{r}(y) = \{x \in \bbZ^3:\; |x-y| \le r\}$ before this notation stands for the open ball.} 
for any saturated %admissible saturated configuration $\phi\upharpoonright_{B_{4D}(y)}= \{x_i\}$ {\color{blue} Is it 
$\phi\in\sA_{D}\big(\bbZ^3\big)$ define the potential
$$
U(\phi\upharp_{\rB_{4D}(y)}) := -\dfrac{1}{\sharp (y)}\;\sum_{x\in\phi:\; \DVC (x) \ni y} 
\dfrac{1}{v(\DVC(x))}.$$
Then for any $\phi\in\sA_{D}\big(\bbZ^3\big)$ we have a formal equality
$$-\sum_{y \in \bbZ^3} \; U(\phi\upharp_{\rB_{4D}(y)}) 
= \sum_{y \in \bbZ^3} \; {1 \over \sharp(y)}\;\sum_{x:\; \DVC(x) \ni y} {1 \over v(\DVC(x)) } 
= \sum_{x \in \phi}{1 \over v(\DVC(x))} \;\sum_{y \in \DVC(x)} {1 \over \sharp(y)}
= \sharp(\phi),$$
i.e., $U$ is a potential counting the number of particles in $\phi$. Clearly, 
$$U(\phi\upharp_{B_{4D}(y)}) \ge -\dfrac{1}{v_\circ}\,,$$
and therefore the particle density in $\phi\in\sA_{D}\big(\bbZ^3\big)$ does not exceed \ $\dfrac{1}{v_\circ}$.
%and therefore
%$$\sharp(\phi) := \sum_{x_i \in \phi} 1 =  -\sum_{y \in \bbZ^3} \; U(\phi\upharpoonright_{B_{4D}(y)}) 
%\le {\color{blue}{|\bbZ^3| \over v}}.$$

For $D=\sqrt{12}$ the minimal DVC volume is equal to $32$, and there exists a unique PC 
$4\bbZ^3 \cup \big((2,2,2) + 4\bbZ^3\big)$, up to $\bbZ^3$-shifts and $\bbZ^3$-symmetries. Correspondingly, 
the maximal particle density is \  
$\dfrac{1}{32}$. Our aim is to show that for $D=\sqrt{11}$ the maximal particle density is also \ $\dfrac{1}{32}$. 
Unfortunately, for $D=\sqrt{11}$ there exist DVCs $\DVC(x)$ having $v(\DVC(x)) < 32$, which prevents a direct 
application of the above argument based on minimal DVCs. However, we will show that a neighborhood of an 
exceptionally small DVC $\DVC(x)$ always contains one or more DVCs $\DVC(x')$ with $v(\DVC(x'))$ large enough
to compensate for the volume deficiency $32-v(\DVC(x))$. 

To establish these facts we need an appropriate lower bound for the DVC volume. More specifically, it is enough 
to consider $57$ sites $y\in \ov\rB_{\sqrt{5}}(x)$, where $\ov\rB_{\sqrt{5}}(x)$ is the closed lattice ball of radius 
$\sqrt{5}$ centered at site  $x\in\bbZ^3$. The intersection 
$$\BDVC(x)(=\BDVC(x,\phi )) := \DVC(x) \cap \ov\rB_{\sqrt{5}}(x)$$ 
is called a {\it bounded} \ DVC (BDVC).  The maximal possible BDVC volume is equal to $1+6+12+8+6+24 = 57$, 
where we counted the number of sites $y$ with $\rho(x, y)^2 = 0,1,2,3,4,5$, respectively.  Note that 
$$v(\BDVC(x)) \leq v(\DVC(x)).$$
Therefore, the bound 
%$$\sum_{x_i \in \phi} |c(x_i)| \le |\bbZ^3|.$$ Therefore,  
$$v(\BDVC(x)) \ge 32, \qquad x\in\phi ,$$
implies that for $D=\sqrt{11}$ the particle density in $\phi\in\sA_{\sqrt{11}}(\bbZ^3)$ \ is \ $\leq\dfrac{1}{32}$. 

In the arising context, two particles 
$x', x'' \in \phi$ are called {\it neighbors} if $\BDVC(x') \cap \BDVC(x'') \not = \es$. Later on, we use a more 
restrictive notion of a `true neighbor'.

\begin{lemma}\label{L:BDVC}%\label{{\bf Lemma~1.}}
For $D=\sqrt{11}$ there exist, up to $\bbZ^3$-symmetries and $\bbZ^3$-shifts, $38$ {\rm{BDVC}}s of
 volume 
\ $\dfrac{380}{12}$, $106$ {\rm{BDVC}}s of volume $\dfrac{382}{12}$, and $14$ {\rm{BDVC}}s of volume 
$\dfrac{383}{12}$. All these {\rm{BDVC}}s are listed in the output of program \ {\tt{BDVC.java}}. 
%BDVC\_11\_5.txt}. {\color{blue}shrift, .java ???}
\end{lemma}

\bp
 Without loss of generality we investigate only BDVCs centered at the origin $o=(0,0,0)$. It is not hard to 
see that all $24$ 
$\bbZ^3$-symmetric images of $x = (2,2,3)$ do not affect $v(\BDVC(o))$ as for any contributing $y$ the distance 
$\rho(o, y) < \rho(x, y)$. The same is true for all $24$ $\bbZ^3$-symmetric images of $x = (1,3,3)$, as well as of 
any $x$ with $\rho(o, x)^2 > 20$. Thus, we only need to consider $x$ which are $\bbZ^3$-symmetric images of
$$ (1,1,3),\;(2,2,2),\;(0,2,3),\;(1,2,3),\;(0,3,3),\;(0,0,4),\;(0,1,4),\;(1,1,4),\;(0,2,4).$$ 
The rest of the proof uses an exhaustive computer enumeration taking into account the above remarks. 

Together with the origin $o$, an admissible configuration of $n-1$ neighbors forms an $n$-site {\it tuple} 
(a term used in programs {\tt{BDVC.java}} and {\tt{BDVCNeighbors.java}}; see below). A 
part of the proof is that we enumerate such tuples 
recursively, exhaustively and without repetitions. Any tuple with the DVC volume $<32$ must have 
at least one particle at distance $\sqrt{11}$ from the origin. Without loss of generality we assume that this is a 
particle at $(-3,-1,-1)$. Accordingly, we initialize the BDVC enumeration by fixing the $2$-tuple
$$(0,0,0),\;(-3,-1,-1).$$
A search for an admissible tuple of a larger size is confined to sites inside the lattice ball $\ov\rB_{\sqrt{20}}(o)$. 
To streamline the search, we index all sites in $\ov\rB_{\sqrt{20}}(o)$ so that the two above sites,
$(0,0,0)$ and $(-3,-1,-1)$, have indices $0$ and $1$, 
respectively. To eliminate repetitions (encountering a tuple more than once),  we
list the elements of a tuple (particles) in an increasing order of site indices. 

We begin enumerating the admissible tuples with adding a third particle to the initial $2$-tuple. 
The search for such a particle is done by sequential trying all sites in $\ov\rB_{\sqrt{20}}(o)$, beginning with the 
site with index $3$. Afterwards, we add a fourth particle in a similar manner and so on, until it becomes impossible 
to add one more particle without breaking admissibility. 

In the latter situation we remove the last 2 particles from the obtained tuple and attempt to add an additional particle 
whose site index is larger than the site index of the second to the last between the removed particles. To be more 
specific, if the indices of the removed particles are $i''$ and $i' > i''$ then we start our search from the site with index 
$i'' + 1$. If this search is unsuccessful, we remove another (the third from the last) particle having site index $i''' < i''$  
and again attempt a forward search starting with site index $i''' + 1$. If the consecutive searches fail and we end up 
removing all particles except for the initial two then all possible admissible tuples are enumerated. 

An opposite situation is when, after removing several particles from the tuple, we are able to add a new particle to it. 
Subsequently, we proceed as in the beginning by adding more and more particles until the particle addition process gets 
stuck. At that moment we again perform the removal of particles from the tuple and so forth.

The total amount of different tuples encountered during the exhaustive enumeration equals $7758631864$. 
Each tuple with the corresponding BDVC volume $< 32$ is printed out as soon as it is discovered. The 
corresponding counts 
%statistics calculated over the corresponding printout {\tt BDVC\_11\_5.txt} 
match those listed in the lemma.  
\ep

In the corollary below we extend the assertion of Lemma \ref{L:BDVC} to DVCs.

\begin{corollary}%{\bf Corollary~1.} 
For $D=\sqrt{11}$, up to $\bbZ^3$-symmetries and $\bbZ^3$-shifts, there exist $38$ {\rm{DVC}}s of volume 
$\dfrac{380}{12}$, $106$ {\rm{DVC}}s of volume $\dfrac{382}{12}$ and $14$ {\rm{DVC}}s of volume 
$\dfrac{383}{12}$. All these {\rm{DVC}}s are listed in the output of \ {\tt{BDVC.java}}.%{\tt BDVC11\_5.txt}.
\end{corollary}

\bp
It is a direct calculation to verify that each BDVC from Lemma \ref{L:BDVC} is actually a DVC.
\ep

\def\ovB{\overline B}

Next, we establish that, among neighbors of a particle $x\in\phi$ with $v(\BDVC(x))<32$, there always 
exist one or several particles such that the total volume of their BDVCs is large enough to compensate for the 
volume deficiency $32-v(\BDVC(x))$. To this end, consider the following definition. Given a configuration 
$\phi\in\sA_{\sqrt{11}}(\bbZ^3)$ and a particle $x'\in\phi$, we say 
that $x'$ is a {\it true neighbor} of particle $x\in\phi\setminus\{x'\}$ if 
either $\BDVC(x',\phi)$ contains a site $y\in\bbZ^3$ with $\rho(y, x') = \rho(y, x)$, or ball 
$\ov\rB_{\sqrt{5}}(x)$ contains a site $y\in\bbZ^3$ with $\rho (y, x') < \rho(y, x)$ and $\rho(y, x) < \rho(y, x'')$ 
for all $x''\in\phi\setminus\{x,x'\}$. The property of being a true neighbor means that removing $x'$ from $\phi$ 
changes BDVC $\BDVC (x)$. As was mentioned at the beginning of the proof of Lemma \ref{L:BDVC}, each true 
neighbor $x'$ of $x$ belongs to $\ovB_{\sqrt{20}}(x)$. Consequently, the true neighbor property is determined
by the restriction $\phi\upharp_{\ovB_{\sqrt{20}}(x)}$.

%{\cre Note that the property of being a true neighbor is non-symmetric: $x'$ may be a true neighbor of $x$ but not vice versa.}

A particle $x\in \phi$ with $v(\BDVC(x)) > 32$ is called a {\it donor}. A particle $x\in \phi$ with $v(\BDVC(x)) < 32$ 
is called an {\it acceptor}. For a particle $x\in\phi$ denote by $n(x)$ the number of true acceptor neighbors of $x$. 
Given a donor $x$, we wish to distribute the excess volume 
$v(\BDVC(x))-32$ evenly among its true acceptor neighbors. To distribute evenly means that a true acceptor neighbor
$x'$ receives an increment in the volume $v(\BDVC(x'))$ equal to $\dfrac{v(\BDVC(x))-32}{n(x)}$ and referred to as 
an excess donation (from $x$ to $x'$).
After a donation procedure is performed for every donor, we obtain a {\it re-distributed} BDVC volume 
$\widehat{v}(\BDVC(x))$ for every particle $x\in\phi$; if $x$ is an acceptor (from one or more donors)
then $\widehat{v}(\BDVC(x))$ will be $>v(\BDVC(x))$. By construction, $\widehat{v}(\BDVC(x))=32$ for each 
non-acceptor $x\in \phi$.

%Given donor $\bx''$ the corresponding amount $n$ of acceptors is always smaller than $16$ as a particle can have at most 16 neighbors.

\begin{lemma}\label{re-distrib}%{\bf Lemma~2.} 
Let $\phi\in\sA_{\sqrt{11}}\big(\bbZ^3\big)$ and $x\in\phi$ be an acceptor particle. Then
$$\widehat{v}(\BDVC(x)) \ge 32.$$
%for every $x_i\in\phi$.
\end{lemma} \def\unv{\underline v} \def\unphi{\underline\phi} \def\untau{\underline\tau} \def\ovB{\overline B}

\bp The presented proof is computer assisted. Its analytical argument yields an algorithm implemented in program {\tt{BDVCNeighbors.java}}. Observe that all possible acceptor BDVCs are listed in Lemma \ref{L:BDVC}. 
%Consequently, let $\phi$ be an AC containing such a BDVCs. 
We analyze tuples
formed by acceptor particles $x$, their true neighbors $x'_j$ and true neighbors of the latter. More
precisely, the first element of such a tuple 
is an acceptor particle $x$ with $v(\BDVC(x) )< 32$; the remaining particles are the true neighbors $x'_j$ of $x$ and the true neighbors $x''_{k(j)}$ of $x'_j$s. Here 
$x'_j\in\ovB_{\sqrt 20}(x)$, $x''_{k(j)}\in\ovB_{\sqrt 20}(x'_j)$ and particles $x'_j$ and $x''_{k(j)}$ bear the
indices inherited from an enumeration of sites in ball $\ovB_{\sqrt{20}}(o)$ by means of a lattice shift. 

%From now on the configuration $\phi$ and an acceptor particle $x\in\phi$ are supposed to be fixed.
%Consequently, the true neighbors $x'_j$ of $x$ are fixed, as well as the true neighbors $x''_{k(j)}$
%of particles $x'_j$.

For each true neighbor $x'_j$ we 
calculate the {\it conditionally minimal} BDVC volume and {\it conditionally minimal} BDVC excess volume this true neighbor is capable to donate to $x$.
%$\unv (x'_j)=\unv (x'_j,\untau)$ 
%and identify the 
%corresponding conditionally minimal BDVC (or BDVCs if there are several of them.) 
The minimization is done under 
the condition that the following particles are present in the tuple: $x$, $x'_j$ and all other true 
neighbors $x'_{j'} \not=x'_j$. 
%In other words, we perform 
%the minimization of the volume $v(\BDVC(x'_j))$ over the admissible tuples $\tau'$ containing $x$ and 
%all true neighbors of $x$. ??  
In particular, this allows us to identify 
the true neighbors $x'_j$ for which the conditionally minimal BDVC volume exceeds $32$. 
Such a neighbor will never be an acceptor. A remaining true neighbor $x'_{j}$ has a possibility to become an acceptor 
at least in some tuples. We treat such $x'_j$s as acceptors in our lower estimates below. 

The sum over $j$ of all obtained conditionally minimal BDVC excess volume donations gives a lower bound for 
the total excess volume donated to the acceptor's BDVC $\BDVC(x)$. Computationally, the above individual 
minimization (for each $x'_j$ separately) is considerably less massive than a simultaneous minimization (for all $x'_j$ collectively).

For each identified donor's conditional BDVC $\BDVC(x'_j)$ the exact number $n(x'_j)$ of its neighbor's BDVCs 
$\BDVC(x''_{k(j)})$ with $v(\BDVC(x''_{k(j)}))<32$ cannot be known without another massive enumeration. 
Instead, we estimate $n(x'_j)$ from above. 
Given a true neighbor $x'_{j}$ of particle $x$, we enumerate tuples consisting of $x$, $x'_j$, all other true 
neighbors $x'_{j'} \not=x'_j$ of $x$, and the neighbors  $x''_{k(j)}$ of $x_j$. Among the $x''_{k(j)}$ we identify the true neighbors of $x_j$, and they all are counted as acceptors 
(which gives an upper estimate for the number of acceptor true neighbors of $x'_j$). Among all other true 
neighbors $x'_{j'} \not=x'_j$ we count as acceptors only those particles which have been classified earlier as possible acceptors (see above). The particle $x$ is always counted as an acceptor true neighbor of $x'_j$. 
The total estimated number of acceptors is denoted by $N(x'_j)$. Clearly, $N(x'_j) \ge n(x'_j)$, and 
%by the number $N(\bx)$ of all neighbors. Recall, that $c(\bx)\cap c(\bx'') \not = \es$ for a neighbor $\bx''$. Correspondingly, 
instead of the actual excess donation $\dfrac{v(\BDVC(x'_j))-32}{n(x'_j)}$ we use its lower bound 
$\dfrac{v(\BDVC(x'_j))-32}{N(x'_j)}$.  

The implemented algorithm (see program {\tt{BDVCNeighbors.java}}) relies on a classification of $v(\BDVC(x'_j))$ 
(as potential acceptors or non-acceptors) which has been performed in advance. However, for the validity of the 
argument the algorithm double-checks that this classification is correct.

In the resulting output of \ {\tt{BDVCNeighbors.java}} one can see that for every considered tuple the accumulated 
lower-bounded excess volume is larger than the deficiency of $v(\BDVC(x))$. Thus, unlike the DVC volume $v(\DVC(x))$, the re-distributed BDVC volume 
$\widehat{v}(\BDVC(x))$ is not smaller than $32$ for any particle $x$.
\ep

\begin{corollary}%{\bf Corollary~2.} 
For $D=\sqrt{11}$ the particle density in any admissible configuration $\phi$ is not larger than \ $\dfrac{1}{32}$.
\end{corollary}

\bp The proof is straightforward. \ep
%The statement follows from the estimate
%$$32 \sum_{x_i \in \phi} 1 \le \sum_{x_i \in \phi} ||c(x_i)|| = \sum_{x_i \in \phi} |c(x_i)| \le |\bbZ^3|, \qquad \phi\in\sA_D\big(\bbZ^3\big)$$
%which was discussed earlier.
%\ep

\medskip
Lemma \ref{re-distrib} allows us to use the minimal re-distributed BDVC volume ($=32$) instead of the minimal BDVC volume ($<32$) to recover the existence of a PC, e.g. $\varphi = 4\bbZ^3 \cup \big((2,2,2) + 4\bbZ^3\big)$. Contrary to the case of $D=\sqrt{12}$, for $D=\sqrt{11}$ there exist other PCs not taken into each other by $\bbZ^3$-symmetries and $\bbZ^3$-shifts. In fact, for $D=\sqrt{11}$ the cardinality of the set of PCs is continuum.

%{\color{blue}Recall that a {\it sliding} is an ability to construct a finite perturbation of a perfect configuration such that this perturbation has an arbitrarily large diameter but, at the same time, the corresponding amount of removed particles is bounded from above by an absolute constant. The typical scenario for sliding is when some straight line of occupied sites in a perfect configuration can be shifted (along itself) without breaking the admissibility of the configuration (and consequently resulting in another perfect configuration). }

\begin{theorem}
For $D=\sqrt{11}$ the corresponding hard-core model exhibits sliding.
\end{theorem}

\bp Consider the BCC configuration $\varphi = 4\bbZ^3 \cup \big((2,2,2) + 4\bbZ^3\big)$ as the initial periodic PC. In particular, $\varphi$ contains particles $x_k = (2k, 2k, 2k)$, $k\in \bbZ$ belonging to the lattice main diagonal.
If we shift these particles along this main diagonal by the vector $(1,1,1)$ then instead we obtain the particles at sites $(2k+1, 2k+1, 2k+1)$, $k\in \bbZ$.
It is not hard to see that the resulting configuration is still an admissible one. Indeed, a particle at site $(2k+1, 2k+1, 2k+1)$ has 6 neighbors 
$$\begin{array}{ll}
(2k, 2k, 2k+4),\; (2k+2, 2k-2, 2k+2),\; (2k+4, 2k, 2k),\\
(2k+2, 2k+2, 2k-2),\; (2k, 2k+4, 2k),\; (2k-2, 2k+2, 2k+2) \end{array} $$
at squared distance $11$ from it and 2 neighbors 
$$(2k-1, 2k-1, 2k-1)\; (2k+3, 2k+3, 2k+3)$$
at squared distance $12$ from it.
All other particles in $\varphi$ are further away from $(2k+1, 2k+1, 2k+1)$.
Similarly, one can slide in $\varphi$ any line of particles parallel to any of main diagonals. This constitutes sliding.
\ep

%%%%%%%%%%%%%%%%%%%%%%%%%%%%%%%%%%%%%
%                                                         D^2=2l^2

%%%%%%%%%%%%%%%%%%%%%%%%%%%%%%%%%%%%%%%

\section{Case $D^2=2\ell^2$}\label{SecD=2llEPGM}

The case $D^2=2\ell^2$, $\ell\in\bbN$ relies on results about dense-packing configurations in $\bbR^3$ established in \cite{Ha1, Ha2, LHF}. These results allow us to identify the PCs and establish the corresponding Peierls bound. Our argument utilizes the {\it scoring} function ${\ov\sigma}({\ov\phi}\upharp_{\overline{\rB}(x)})$, $x\in {\ov\phi}$, which has been constructed in \cite{Ha1}, see (1.4) and Definition 5.12. Here and below, ${\ov\phi}$ is a $1$-AC in $\bbR^3$ and ${\ov\rB}(x)$ is the closed ball in $\bbR^3$ centered at $x\in\bbR^3$: 
\beq\beac
{\ov\phi}\subset\bbR^3,\;\rho (y,y')\geq 1\;\;\forall\;\;y\in
{\ov\phi}\;\hbox{ and }\;y'\in{\ov\phi}\setminus\{y\},\\ 
 {\ov\rB}(x)={\ov\rB}_{2.51}(x):=\{y\in \bbR^3: \rho(x,y)\leq 2.51 \}.
\ena\eeq
The over-line symbol in this notation stresses that the objects under consideration are in $\bbR^3$.

We work with a modified scoring function ${\ov\ups} =-{\ov\sigma} +\dfrac{16\pi}{3}$ which has the following properties. 
\begin{description}
\item (I) ${\ov\ups}({\ov\phi}\upharp_{\overline{\rB}(x)})$ is shift-invariant: 
${\ov\ups}({\ov\phi}\upharp_{\overline{\rB}(x)})
={\ov\ups}\left(({\ov\phi}+u)\upharp_{\overline{\rB}(x+u)}\right)$, \, $\forall$ $u\in\bbR^3$ 
and $x\in\ov\phi$.   
\item (II) For a $1$-FCC configuration $\ov\varphi$ and any $x\in\ov\varphi$ 
$${\ov\ups}({\ov\vphi}\upharp_{{\ov\rB}(x)})={\ov\ups}^*\in (0,\infty ).$$ 
\item (III) For any $1$-AC $\ov\phi$ in $\bbR^3$ and $x\in{\ov\phi}$,
$${\ov\ups}({\ov\phi}\upharp_{\overline{\rB}(x)})\leq{\ov\ups}^*.%\in (0,\infty ),
%\;\hbox{ where }\;{\ov\sigma}^*:={\ov\sigma}({\ov\vphi}\upharpoonright_{{\ov\rB}(x_0)}),\;x_0\in\ov\vphi.
$$ 
%Here $\ov\vphi$  is an $1$-FFC- or an $1-$HCP-configuration in  $\bbR^3$. (For such $\ov\vphi$,
%the value ${\ov\sigma}({\ov\vphi}\upharpoonright_{{\ov\rB}(x_0)})$ does not depend on the choice of $x_0\in\ov\vphi$.) 
Cf. Theorems 1.7, 6.1 and Corollary 6.3 in \cite{Ha1}.
\item (IV) Suppose a 1-AC $\ov\phi$ in $\bbR^3$ coincides with a 1-FCC configuration $\ov\varphi$ 
outside a closed cube $\overline{C}_r(o)\subset \bbR^3$ of side-length $r$ centered 
at the origin $o$. Assume $r$ is large enough: $r \gg 2.51$. Then
$$\sum_{x \in{\ov\phi}\,\cap\, \overline{C}_{2r}(o)}{\ov\ups}({\ov\phi}\upharp_{{\ov\rB}(x)}) 
= \sum_{x \in{\ov\varphi}\,\cap\,{\ov C}_{2r}(o)}{\ov\ups}({\ov\varphi}\upharp_{\overline{\rB}(x)}).$$
Cf. Lemma 5.10 and Theorem 5.11 in \cite{Ha1}.
\end{description}

For a discrete $\phi\in\sA_D\big(\bbZ^3\big)$, we define 
$$\ups(\phi\upharp_{\rB_{r}(x)}):=\overline{\ups}(\overline{\phi}\upharp_{\overline{\rB}(x)}), 
\qquad \overline{\phi}:=\dfrac{1}{D}\phi,$$
where $r=2.51D$, and $\phi$ and its $\dfrac{1}{D}$-scaled version $\overline{\phi}$ are understood as 
sets of points in space. The condition $D^2=2\ell^2$ is necessary and sufficient to guarantee the existence of $D$-FCC 
sub-lattices in $\bbZ^3$. Correspondingly, for this case the function $\ups$ inherits properties (I)-(IV) 
from $\overline{\ups}$. 

It is convenient for us to use the relative form of \eqref{eq:ener} 
\beq\label{eq:ener1} H (\phi )=\big(-\ln u\big)\cdot ( \sharp (\phi)-\sharp( \varphi)),\;\;\phi\in\sA_D
\big(\bbZ^3\big), \; \varphi \;\hbox{is a PGS},\eeq 
which generates the same Gibbs measures on $\bbZ^3$ as \eqref{eq:ener}. Let 
$$\zeta(\phi\upharp_{B_r(x)}):=\dfrac{\ups (\phi\upharp_{B_r(x)})}{\ups^*} -1.$$
Then for a $D$-FCC sub-lattice $\varphi$ and $\phi\in\sA_D\big(\bbZ^3\big)$ such that $\varphi$ and $\phi$ 
coincide outside of $B_s(o)$, where $s\gg r=2.51D$ and $o$ is the origin, we have
$$\sharp (\varphi) -\sharp (\phi) = \sum_{x\in \phi} \zeta(\phi\upharp_{B_r(x)}),$$ 
as follows from the properties (II) - (IV) above. Consequently, 
\beq\label{m-potU}H(\phi )=(\ln u)\sum_{x\in \phi} \zeta(\phi\upharp_{B_r(x)}).\eeq The above representation is similar (but not equivalent) to \eqref{hamiltonian}. Consequently, in analogy with condition \eqref{m-pot}, a configuration $\varphi\in\sA_D(\bbZ^3)$ is called a PC if $\zeta(\varphi\upharp_{B_r(x)})=0$ for all $x\in\varphi$.

According to \cite{Ha1, Ha2, LHF}, the only dense-packing 
configurations in $\bbR^3$ are FCC, HCP and their layered mixtures, up to Euclidean motions. 

\bl\label{denseR3}
If a $D$-scaled version of a dense-packing configuration in \ $\bbR^3$ exists in\ $\bbZ^3$ then all such $D$-scaled configurations existing in\ $\bbZ^3$ exhaust the set \ $\sS^{(D^2)}$ of \ $D$-{\rPC}s in $\bbZ^3$. 
\el

\bp
The assertion follows from Claim 1.16 and Theorem 6.1 in \cite{Ha1} or Theorem 8.1, p. 138 in \cite{LHF}. 
\ep
%If, for a given attainable $D$, such structures exist, in a scaled version, in 
%lattice $\bbZ^3$ then, obviously, they give all $D$-PCs in $\bbZ^3$. 

Clearly, the $D$-scaled versions of these dense-packing configurations are $D$-FCC, $D$-HCP configurations, their layered mixtures $\vphi^{(D^2)}_{i, \{j_k\}}$ and their $\bbZ^3$-shifts. For the necessary and sufficient conditions for their existence and for the detailed description of their structures see Appendix A. %Adding results from \cite{Ha1, Ha2, LHF} leads us to the following theorems. In particular, for $D^2=2\ell^2$, constructions from \cite{Ha1, Ha2, LHF} lead to an m-potential representation of Hamiltonian $H(\phi)$ in (1.1) via the scoring function. See the definitions and properties of $\ov \sigma$ and its $\bbZ^3$-counterpart $\sigma$ in the next section. 
%The results from Appendix, together with the m-potential representation \eqref{m-potU}, lead to the following theorems. 

\def\uhr{\upharpoonright}

Now assume that $D^2=2\ell^2$ where $\ell\neq 0\mod 3$. In this case the set $\sS^{(D^2)}$ 
consists of $D$-FCC sub-lattices and their $\bbZ^3$-shifts.
To apply the PS theory, it remains to establish a suitable Peierls bound for contours. The definition 
of a contour can be given as a direct generalization of that in Sect. 3.1 of \cite{MSS2}.
More specifically, a rhombic template $F_{k,j}$ from \cite{MSS2} is replaced by a parallelepiped with
congruent rhombic faces, and the number of occupied sites in a $D$-PC $\vphi$ within such a 
parallelepiped is re-calculated in a straightforward manner. A contour 
$\Gam =\left({\rm{Supp}}\,(\Gam ), \phi\upharpoonright_{\rSp\,(\Gam )}\right)$ and its support 
${\rm{Supp}}\,(\Gam)$ are defined as  in Sect 3.1 of \cite{MSS2}. The statistical weight $w(\Gam )$
is defined as 
\beq\label{eq:Stweight}w(\Gam )=u^{\sharp (\phi\uhr_{{\rm{Supp}}\,(\Gam )})-
\sharp (\vphi\uhr_{{\rm{Supp}}\,(\Gam )})};\eeq
cf. (3.9) from \cite{MSS2}.

For a saturated AC $\phi$ the Peierls bound for  $w(\Gam )$ is 
established similarly to Lemma 5.4 from \cite{MSS2}. The analog of this lemma is 

\bl\label{PeierlsBound} (The Peierls bound) 
There exists a constant $p=p(D) > 0$ 
such that for any contour \ $\Gam =({\rm{Supp}}\,(\Gam ),\phi\upharpoonright_{{\rm{Supp}}\,(\Gam)})$ 
we have 
\be\label{(6)}
w(\Gam ) = \prod_{x \in\phi\upharpoonright_{{\rm{Supp}}}\,(\Gam )} u^{- \zeta(\phi\upharpoonright_{B_r(x)})}\leq  u^{-p(D)\|\rSp\,(\Gam )\|}\ee
where $\|\rSp\,(\Gam )\|$ stands for the number of templates (see \cite{MSS2}) in  $\rSp\,(\Gam )$.
\el

\bp The equality in \eqref{(6)} is the result of substituting \eqref{m-potU} in \eqref{eq:Stweight}.
Note that the contribution into the product in \eqref{(6)} comes only from 
sites $x$ where $\ups (\phi\uhr_{\rB (x)}) >\ups^*$; otherwise
(i.e., when $\ups =\ups^*$) site 
$x$ does not contribute into \eqref{(6)}. Observe that
$$\hbox{if }\;\ups (\phi\uhr_{\rB (x)}) - \ups^* \ge\ups^*\;\hbox{ then }\;
\ups (\phi\uhr_{\rB (x)}) -\ups^* \ge\frac{1}{2}\ups (\phi\uhr_{\rB (x)}).$$
On the other hand, due to discreteness
$$\hbox{if }\;\ups (\phi\uhr_{\rB (x)})-\ups^* < \ups^*\;\hbox{ then }\;
\ups (\phi\uhr_{\rB (x)}) -\ups^* \ge \del (D) \ge\frac{\del (D)}{2\ups^*} 
\ups (\phi\uhr_{\rB (x)}) $$
where $\del (D)>0$.
According to the definition of a $\vphi$-correct template, for a saturated AC $\phi$
we have an inequality
$$\sum\limits_{x \in\phi\upharpoonright_{{\rm{Supp}}\,\Gam} :\,\ups (\phi\uhr_{\rB (x)})>\ups^*} 
\ups (\phi\upharp_{\rB (x)}) \ge\diy\frac{1}{27 D^3} |\rSp\,(\Gam )|\,.$$ 
Also,
$||\rSp\,(\Gam )|| =\kappa (D)|\rSp\,(\Gam )|$ where $\kappa (D)>0$ and $|\rSp\,(\Gam )|$
denotes the number of sites in $\rSp\,(\Gam )$. Thus, we can take
\be\label{(6A)}\diy p (D)=\diy \frac{\kappa (D)}{27 D^3} \min \left(\frac{1}{2}, \frac{\del (D)}{\ups^*}
\right).\ee\ep

%The the representation \eqref{m-potU}. 
An extension of the Peierls bound to non-saturated ACs is straightforward because each template 
in $\phi\upharp_{{\rm{Supp}}\,\Gam}$ where one can add a particle contributes a factor
$\leq u^{-1}$ into the statistical weight $w(\Gam )$. 

The results from Appendix A, together with Lemmas \ref{denseR3} and \ref{PeierlsBound},
lead to the following theorems. 

\bthmr\label{thm:9.2}%{\bf Theorem 9.2.} {\sl 
Suppose that $D^2=2^{2n+1}$ where $n\in\bbN \cup \{0\}$. Let \ $\sS^{(D^2)}$ be the collection consisting of $D$-\rFCC \ sub-lattice $2^n\bbA_3$ and its $\bbZ^3$-shifts. Set \ $\sS^{(D^2)}$ exhausts all \ {\rPC}s. 
The cardinality of \ $\sS^{(D^2)}$ is $2^{3n+1}$. The particle density of any \rPC \ 
equals $1/2^{3n+1}$. All \ {\rPC}s are $D$-\rFCC \ sub-lattices, and they form a single equivalence class. 
\ethmr

\bp
It is well-known that equation \eqref{DiofEqn} has only trivial solutions (where two of the numbers $m, n, k$ 
vanish), iff $\ell=2^n$. See Theorem 5 p. 79, in \cite{Gross}. With this at hand, the assertion of the theorem 
follows from Lemma \ref{denseR3} and Theorem \ref{thm:9.1}.
\ep

\bthmx\label{thm:101} 
%Suppose that $D^2=2^{2n+1}$ where $n\in\bbN \cup \{0\}$. There exists $u^0=u^0 (D)\in (0,\infty )$
%with the following property. For $u\in (u^0,\infty )$, 
%each \ $\vphi\in\sS^{(D^2)}$ generates an \ \rEPGM, and each \ \rEPGM \ is generated by some 
%$\vphi\in\sS^{(D^2)}$. Consequently, $\sharp\sE (\sqrt{D}, u) =2^{3n+1}$.
Suppose that $D^2=2^{2n+1}$ where $n\in\bbN \cup \{0\}$. Let $u$ be large enough: $u\geq u^0(D^2)$. Then there are 
$2^{3n+1}$ {\rEPGM}s, i.e., $\sharp (\sE(D, u))=2^{3n+1}$, and each \rEPGM\ is generated by
a \rPGS\ from $\sS^{(D^2)}$.
\ethmx

\bp
The structure of PGSs for values of $D$ under consideration is given in Theorem 9.1A %\ref{thm:9.2} 
and consists of a single equivalence class. The  Peierls bound \eqref{(6)} allows us 
to complete the proof via the PS theory.  
\ep

\bthmt\label{thm:9.3}%{\bf Theorem 9.3.} {\sl 
Suppose that $D^2=2\ell^2$ where $\ell\in\bbN$, $\ell\neq 2^n $ and $\ell\neq 0 \hskip-2pt\mod 3$. Let \ $\sS^{(D^2)}$ be the collection consisting of all $D$-\rFCC \ sub-lattices and their $\bbZ^3$-shifts. Set \ $\sS^{(D^2)}$ exhausts all \ {\rPC}s. In total, there are finitely many \ {\rPC}s. The particle density of any \rPC \ 
equals $1/2\ell^3$. All \ {\rPC}s are $D$-\rFCC \ sub-lattices, and form more than one equivalence class. 
%Then the \ {\rPC}s  
%are the $\bbZ^3$-shifts of $D$-\rFCC \ sub-lattices in $\bbZ^3$; each sub-lattice gives rise to
%$2\ell^3$ shifted \ {\rPC}s (itself included). All \ {\rPC}s \ are periodic. The particle density in any \ \rPC \ equals 
%$1/2\ell^3$. 

The number of equivalence classes of $D$-\rFCC \ sub-lattices and the 
$\bbZ^3$-symmetries for each class (and hence the cardinality of the class) depend on the rational 
prime decomposition of $\ell$, as detailed in the Appendix A. 
\ethmt

\bp
The assertion of the theorem follows from Lemma \ref{denseR3} and Theorem \ref{thm:9.1}.
\ep

\bthmu\label{thm:102} Suppose that $D^2=2\ell^2$ where $\ell\in\bbN$, $\ell\neq 2^n$  and $\ell\neq 0 \hskip-2pt\mod 3$. Let $u$ be large enough: $u\geq u^0(D^2)$. Then there exists at least one dominant \rPGS-equivalence class. Each \ \rPGS \ $\vphi$ from a dominant class 
generates an \rEPGM\ $\mu_\vphi$. Conversely, every \ \rEPGM \ $\mu$  is generated by a \ \rPGS \ from 
some dominant class.
\ethmu

\bp
The structure of PGSs for values of $D$ under consideration is given in Theorem 9.2A  and consists of two or more equivalence classes.  The  Peierls bound \eqref{(6)} allows us 
to complete the proof via the PS theory.
\ep

\bthmv\label{thm:9.4}%{\bf Theorem 9.4.} {\sl 
Suppose $D^2= 2\ell^2$ where $\ell\in\bbN$ and $\ell=0 \hskip-2pt\mod 3$. Let $\sS^{(D^2)}$ be the set consisting of\
layered \ {\rAC}s: $\vphi^{(D^2)}_{i, \{j_k\}}$ for $i=0,1,2,3$ and allowed sequences $\{j_k\}$, all rotations of \ {\rAC}s $\vphi^{(D^2)}_{i, \{j_k\}}$ inscribed in $\bbZ^3$, 
and the $\bbZ^3$-shifts of such \ {\rAC}s. Set \ $\sS^{(D^2)}$ exhausts all \ {\rPC}s. The cardinality of \ $\sS^{(D^2)}$ is continuum. The particle density 
in a\ \rPC \ equals \ $1/2\ell^3$. The subset \ $\sS^{(D^2)}_{\rm{per}}$ consisting of periodic 
layered \ {\rAC}s \ from \ $\sS^{(D^2)}$ is countable and exhausts 
all periodic \ {\rPC}s. % or equivalently, \;{\rPGS}s.
Depending on the rational prime decomposition of $\ell$, the sub-lattices in $\sS^{(D^2)}_{\rm{per}}$ are partitioned into more than one but finitely many equivalence classes. 
\ethmv

\bp
The assertion of the theorem follows from Lemma \ref{denseR3} and Theorem \ref{thm:9.1}.
\ep

It is natural to expect that some analog of Theorem 9.2B %\ref{thm:102} 
holds true also for $D^2=2\ell^2$ when $\ell=0\hskip-2pt\mod 3$, and it can 
be proved by using methods from \cite{BS}. The first step here is the identification of the 
PGSs, and it is completed in Theorem 9.3A. %\ref{thm:9.4}. 
The next step is the identification of the lowest order of the perturbation theory in which the 
infinite degeneracy of PGSs is removed. It turns out that this order equals $2$ (equivalently 
the statistical weight equals $u^{-2}$), and the smallest excitation removing the degeneracy 
is described in the following way. 

%The next %(and final) excitation of energy $2$, $\gamma_2^{\ast}$, is constructed as follows. 
Consider three subsequent meshes $\tau':=\tau^{(D^2)}_{i,2(k-1),j_{k-1}}, \tau:=\tau^{(D^2)}_{i,2k,j_k}, 
\tau'':=\tau^{(D^2)}_{i,2(k+1),j_{k+1}}$ and assume that 
in the middle mesh $\tau$ there is a triangle $\triangle_0$ and in meshes $\tau', \tau''$ there are 
triangles $\triangle_{\pm 1}$, and the centers of $\triangle_{\pm 1}$ are projected to the center of
$\triangle_0$. Then, if we place a particle at the center of $\triangle_0$ and remove the particles 
from the vertices of $\triangle_0$, we obtain the desired excitation.

The maximal density of the above $u^{-2}$-excitations is achieved at the $D$-HCP configurations 
$\vphi^{(D^2)}_{i, \overline{01}}$, $\vphi^{(D^2)}_{i, \overline{02}}$ with  $i=0,1,2,3$, their rotations and, subsequently, $\bbZ^3$-shifts. 
Consequently, only the $D$-HCP configurations are expected to be dominant. The first difficulty in
completing the proof of this claim lies in the verification that there is no other non-trivial excitation of order $2$. 
Second, we need to identify, among all classes of $D$-HCP configurations, the dominant one. 
In Theorem 4.1B we verified it for a similar case where $D=\sqrt{5}$, but the proof of the claim for a general 
$D^2=2\ell^2$ with $\ell=0 \hskip-2pt\mod 3$ is not
known. Nevertheless, we put forward  the following conjecture.

\begin{nnconjecture} 
Suppose $D^2=2\ell^2$ where $\ell\in\bbN$ and $\ell=0 \hskip-2pt\mod 3$. All \ {\rEPGM}s
are generated by {\rPC}s from a single equivalence class. This class has cardinality $16\ell^3$ and consists of \
$8$ $D$-\rHCP\ configurations and their \ $\bbZ^3$-shifts. Consequently,  $\sharp\sE({\sqrt 2}\ell, u) =16\ell^3$.

%The equivalence class, 
%of cardinality $16\ell^3$, containing $8$ $D$-\rHCP\ configurations $\vphi^{(D^2)}_{i, \overline{01}}$, 
%$\vphi^{(D^2)}_{i, \overline{02}}$, with $i=0,1,2,3$, and their $\bbZ^3$-shifts is the only one dominant 
%i.e., generating \ {\rEPGM}s). Consequently,  $\sharp\sE({\sqrt 2}\ell, u) =16\ell^3$.
\end{nnconjecture}

%%%%%%%%%%%%%%%%%%%%%%%%%%%%%%%%%%%%
%                                                                                                                  D^2=2l^2
%%%%%%%%%%%%%%%%%%%%%%%%%%%%%%%%%%%%%

\section{Appendix A: $D$-FCC sub-lattices in $\bbZ^3$ and corresponding layered structures}\label{AppendixA}

It is known that a $D$-FCC sub-lattice in $\bbZ^3$ exists iff $D^2=2\ell^2$ with $\ell\in \bbN$, 
see Proposition 12 in \cite{Iona2007}, and in this section we focus on such values of $D^2$.
A canonical example of a $D$-FCC sub-lattice is the scaled lattice $\bbA_3$:
\beq\label{eq:A3ii}\beacl \ell\bbA_3:=\big\{m\,(\ell,\ell,0) + n\,(\ell,0,\ell) + k\,(0,\ell,\ell) :\, m, n, k\in\bbZ\big\}.  %\\
\ena\eeq
Cf. \eqref{eq:A3i}. Depending upon 
the rational prime decomposition of $\ell$, there may exist other $D$-FCC sub-lattices in $\bbZ^3$; see below.
We also verify that the whole collection of layered structures emerging in $\bbR^3$ exists in $\bbZ^3$  iff $\ell=0 \hskip-2pt\mod 3$,
in which case  they fit the construction described in Section \ref{Sec2} and give PCs
of the form \eqref{layered1}.

%Given $k \in \bbZ$ and $i = 0,1,2,3$, denote by $T^{(2)}_{i, k}$ the intersection of $\bbZ^3$ and the affine plane 
%$$\rx_1+s_2\rx_2+s_3\rx_3=k$$
%where $s_l=s_l(i)\in\{-1,1\}$, $l=2,3$; cf. \eqref{eq:is1s2}, \eqref{eq:s1s2i}. Then
%each $T^{(2)}_{i, k}$ is a triangular $\sqrt{2}$-mesh:  
%\beq\label{eq:Tik2}\beal T^{(2)}_{i, k}=\big\{m\left(1,-s_2,0\right)
%+n\left(1,0,-s_3\right),\;m,n\in\bbZ\big\}+(k,0,0).\ena\eeq
%Cf. \eqref{eq:Tik1}. By construction:
%$$\bbZ^3 = \bigcup_{k \in \bbZ} T^{(2)}_{i, k}, \qquad \bbA_3 = \bigcup_{k \in \bbZ} T^{(2)}_{i, 2k}\quad
%\hbox{ and }\quad \ell\bbA_3 = \bigcup_{k \in \bbZ} \ell T^{(2)}_{i, 2k}\,.$$

%It is instructive to note a property of periodicity of triangular ${\sqrt 2}\ell$-meshes $\ell T^{(2)}_{i, 2k}$: 
%given $l=0,1,2$, the meshes
%$\ell T^{(2)}_{i,2k}$ have the same projection to plane $\rx_1+s_2\rx_2+s_3\rx_3=0$ whenever 
%$k=l\bmod 3$:
%$$\ell T^{(2)}_{i,2k} -\dfrac{2k\ell}{3}(1,s_1,s_3)=\bcs \ell T^{(2)}_{i,0},&k=0\bmod 3,\\
%\ell T^{(2)}_{i,0}+\dfrac{2\ell}{3}(2,-s_1,-s_2),&k=1\bmod 3,\\
%\ell T^{(2)}_{i,0}-\dfrac{2\ell}{3}(2,-s_1,-s_2),&k=2\bmod 3.\ecs$$
%Geometrically, any two of the three meshes $\ell T^{(2)}_{i,2k} -2k\ell(1,s_1,s_3)/3$ have points 
%at the centers of the third one.
%This allows us to construct, for $\ell =0\bmod 3$, a continuum of layered $D$-ACs $\vphi^{(D^2)}_{i,\{j_k\}}$
%that will be PCs for $D^2=2\ell^2$.

%Suppose $\ell = 3t$ andL

Let $\tau^{(D^2)}_{i, 2\ell k, j}$ be a triangular $\sqrt{2}\ell$-mesh as in \eqref{layered1}, with $q=2\ell^2$ and $h=2\ell$, $r=2$. The dependence on $j=0,1,2$ is given by:  
%subset $T^{(2)}_{i,2k\ell}$, $j=0,1,2$, be triangular 
%$\sqrt{2}\ell$-meshes: 
\beq\label{layered4}\beac\tau^{(D^2)}_{i, 2\ell k, 0}:=\{m\ell (1,-s_2,0)+n\ell (1,0,-s_3):\;m,n\in\bbZ\}
+\dfrac{2\ell k}{3} (1,s_2,s_3)\\%(=\ell T^{(2)}_{i,2k}),\\
\tau^{(D^2)}_{i, 2\ell k,1}:=\tau^{(D^2)}_{i, 2\ell k, 0}+\dfrac{\ell}{3}(-2,s_2,s_3),\quad 
\tau^{(D^2)}_{i, 2\ell k,2}:=\tau^{(D^2)}_{i, 2\ell k, 0}+\dfrac{\ell}{3}(2,-s_2,-s_3),\ena\eeq
where $s_2=s_2(i)$ and $s_3=s_3(i)$ are determined in (2.11).
%Consider a double-infinite sequence $\{j_k,\;k \in \bbZ\}$, such that $j_k\in\{0,1,2\}$, 
% $j_0=0$ and 
%$j_k \not = j_{k+1}$. 
Next, set $\vphi^{(D^2)}_{i, \{j_k\}}:=\onwl{\bigcup}\limits_{k \in \bbZ}\tau^{(D^2)}_{i, 2\ell k, j_k}$, with a double-infinite sequence $\{j_k\}$ such that $j_0=0$ and 
$j_k \not = j_{k+1}$. 
%with an allowed sequence $\{j_k\}$
%is the $\bbZ^3$-implementation of the corresponding layered dense-packing configuration in $\bbR^3$.
%These configurations fit the pattern in \eqref{layered1} with $q=2\ell^2$ and $h=2$. 

Observe that for any $\ell\in \bbN$, the layered ACs $\vphi^{(D^2)}_{i, \overline{012}}$ and $\vphi^{(D^2)}_{i, \overline{021}}$ are FCC $\ell$-sub-lattices in $\bbZ^3$. Moreover, all remaining layered ACs $\vphi^{(D^2)}_{i, \{j_k\}}$ also belong to $\bbZ^3$ iff $\ell=0 \hskip-2pt\mod 3$.  
This includes HCP $\ell$-configurations $\vphi^{(D^2)}_{i, \overline{01}}$ and $\vphi^{(D^2)}_{i, \overline{02}}$. 

%As in $\bbR^3$, a layered ACs $\vphi^{(D^2)}_{i, \{j_k\}}$ is an FCC- and HCP-mixture. It is periodic when 
%sequence $\{j_k\}$ is periodic, i.e., is constructed from its period $j_0j_1...j_p$; 
%such a sequence is referred to as $\overline{j_0j_1...j_p}$. In particular, the FCC $\ell$-sub-lattice 
%$\ell\bbA_3$ emerges as $\overline{012}$ or $\overline{021}$. 
 
A finite number of additional FCC $\ell$-sub-lattices $\rR \ell\bbA_3$, obtained from $\ell\bbA_3$ by 
non-trivial rotations $\rR$, 
may exist for a given $\ell$; this depends on the rational prime decomposition of $\ell$. If 
$\ell=0 \hskip-4pt\mod 3$ then the corresponding layered ACs $\rR \vphi^{(D^2)}_{i, \{j_k\}}$ 
are also inscribed in $\bbZ^3$.

For a given $\ell$, the identification of all FCC $\ell$-sub-lattices in $\bbZ^3$ is equivalent to the identification 
of cubic $\ell$-sub-lattices as they are in a 1-1 correspondence.

%In one direction, the above 1--1 correspondence 
It is straightforward that 
a cubic sub-lattice in $\bbZ^3$ with 
basis $\{x_1, x_2, x_3 \}$ contains an FCC sub-lattice with the basis 
\beq\label{eq:bases}
y_1=x_1+x_2, \quad y_2=x_2+x_3, \quad y_3=x_1+x_3.\eeq
Furthermore, all FCC 
$\ell$-sub-lattices of $\bbZ^3$ can be obtained in this way. See Corollaries 2.2 and 2.3 in \cite{Iona2016}.
%It is less obvious that any FCC sub-lattice of $\bbZ^3$ can be obtain in this way. 

The latter fact originates from the connections between both types of sub-lattices and the ring %$\bbH (\bbZ)$ 
of integer quaternions. Every non-zero integer quaternion 
$z = a + b\cdot\ri + c\cdot\rj + d\cdot\rk\neq 0$, with $a,b,c,d\in\bbZ$, defines a 
non-trivial rotation of $\bbR^3$ given by the ortho-normal  Euler-Rodrigues matrix:
\beq\label{EuRod}\rR(z)=\frac{1}{\ell} 
\bpma
a^{2}+b^{2}-c^{2}-d^{2} & 2bc-2ad & 2bd+2ac \\
2bc+2ad & a^{2}-b^{2}+c^{2}-d^{2} & 2cd-2ab \\
2bd-2ac & 2cd+2ab & a^{2}-b^{2}-c^{2}+d^{2} \\\
\epma,\eeq %\quad R^{-1}=R^T$$
where 
\beq\label{eq:labcd} \ell=\ell (z)=||z||^2=a^{2}+b^{2}+c^{2}+d^{2}.\eeq  
The rotation angle $\alpha$ is given by
$$\alp =2\arccos\Bigg(\frac{a}{a^{2}+b^{2}+c^{2}+d^{2}}\Bigg), $$ 
and the rotation axis is along the vector $(b,c,d)$. 
The rows of the matrix $\ell R(z)$, $\ell=\|z\|^2$, form a basis of a cubic $\ell$-sub-lattice of $\bbZ^3$, 
which we denote by $\bbZ^3(z)$, and all such 
sub-lattices can be obtained in this way \cite{Ca, Cr, FV, Sp}. Each basis vector $m, n, k$ of $\bbZ^3(z)$ 
represents a Pythagorean quadruple with
\begin{equation}\label{DiofEqn} m^2+n^2+k^2=\ell^2.
\end{equation}
The columns of the matrix $\ell R(z)$ form a basis of the conjugated sub-lattice $\bbZ^3(\ov z)$ 
corresponding to the conjugated quaternion $\ov z$, that is, to the rotation along the same axis but
in the opposite direction. The lattice $\bbZ^3(z)$ coincides with the original lattice $\bbZ^3$ iff $\|z\|=1$.

If the row-vectors of $\ell\rR(z)$ are $x_1, x_2, x_3$ then $y_1, y_2, y_3$ defined as in \eqref{eq:bases}
form a basis of the corresponding FCC $\ell$-sub-lattice of $\bbZ^3$, denoted by $\bbA_3(z)$. Note that 
the length of vectors $y_i$ is $\sqrt{2}\ell$; together with the origin, vectors $y_i$ give the 
vertices of a regular tetrahedron.  The lattice $\bbA_3(z)$ coincides with the original lattice $\bbA_3$ 
iff $\|z\|=1$. More generally, given an integer quaternion 
$z$, we can consider %meshes $T^i_k(z)$, 
%sub-lattices $\bbA_3(z)$ and 
layered ACs $\vphi^{(D^2)}_{i, \{j_k\}}(z)$, as images of
%$T^i_k$, $\ellT^i_k$, $\bbA_3$, $\ell\bbA_3$ and 
$\vphi^{(D^2)}_{i, \{j_k\}}$ % $\ell\bbL^{(i)}_{\{j_k\}}$, 
under the rotation $\rR (z)$ generated by $z$.

%\section{Cubic sub-lattices of $\bbZ^3$}

The rest of this section is devoted to the description of all cubic $\ell$-sub-lattices of $\bbZ^3$ including 
their number and symmetries. Cf. \cite{IO}. The symmetries of cubic $\ell$-sub-lattices and hence FCC 
$\ell$-sub-lattices are important for our considerations because if a sub-lattice generates an EPGM then every $\bbZ^3$-symmetric image of this sub-lattice also generates an EPGM. An example of a cubic 
$\ell$-sub-lattice is $\ell\bbZ^3$, with the basis
\begin{equation}\label{1basis} \{(\ell,0,0), (0,\ell,0), (0,0,\ell)\},\end{equation}
where $\ell\in \bbN$. In general, we say that a sub-lattice is a cubic 
$\ell$-sub-lattice if it has a basis formed by three mutually orthogonal integer vectors of length $\ell$. 

The facts collected in this section are consequences of classical algebraic number theory, but we were not able 
to find a single source containing them in the desired form. We present these facts as a series of propositions
accompanied with proofs when no direct reference is available. 

\begin{proposition}\label{prop:lZ3} %{Prop3.1}
A cubic $\ell$-sub-lattice of \ $\bbZ^3$ exists iff\; $\ell$ is a positive integer.
\end{proposition}

\bp If $\ell \in \bbZ$ then $\ell\bbZ^3$ is the desired sub-lattice. If $\ell \not\in\bbZ$ but there exists a 
cubic $\ell$-sub-lattice of $\bbZ^3$ then $\ell = \sqrt{d}$ where $d\in \bbN$ as all basis vectors of the 
sub-lattice have integer coordinates. Moreover, the vector product of two basis vectors is a vector of 
length $\ell^2$ with integer coordinates which is collinear to the third basis vector, also with integer coordinates. 
The ratio of lengths of these two collinear vectors is equal to $\ell^2/\ell=\ell$, and it should also be equal to the 
ratio of the corresponding first coordinates. As both coordinates are integers, their ratio is rational and 
cannot be equal to an irrational $\sqrt{d}$.
\ep   

Since the HC model under consideration is $\bbZ^3$-symmetric, we are interested only 
in the equivalence classes of sub-lattices with respect to $\bbZ^3$-symmetries. We use the standard notation 
$O_\rh$ for the group of $\bbZ^3$-symmetries which is of order $48$. The group $O_h$ consist of:
\beq\label{description}\beal
\hbox{(i) the identity,}\\
\hbox{(ii) six rotations by $\pm\dfrac{\pi}{2}$ with respect to one of the coordinate axis,}\\
\hbox{(iii) three rotations by $\pi$ with respect to one of the coordinate axis,}\\
\hbox{(iv) six rotations by $\pi$ with respect to one of the coordinate plane diagonals,}\\
\hbox{(v) eight rotations by $\pm\dfrac{2\pi}{3}$ with respect to one of the main  diagonals,}\\
\hbox{(vi) twenty four composition of the inversion (central symmetry)}\\ 
\hbox{with each of the previous elements.}\ena\eeq

\noindent
The $24$ symmetries listed in (i-v) form the subgroup $S_4$ of $O_\rh$. Item (vi) recognizes the fact 
that $O_\rh\simeq Z_2\times S_4$ where the group $Z_2$ consists of the identity and the inversion.

Each cubic $\ell$-sub-lattice of $\bbZ^3$ is invariant under the inversion, hence under the action of $O_\rh$ every stabilizer of a cubic $\ell$-sub-lattice 
contains $Z_2$, and every equivalence class (identified with an orbit of this sub-lattice under the action of $O_\rh$) contains at most
$24$ sub-lattices. In the generic case, the stabilizer group is exactly $Z_2$, and the sub-lattices
in the class are obtained from each other by symmetries listed in (ii-v). Cf. Proposition \ref{prop:24class} below. %Consequently, 

All non-generic cases are where a cubic $\ell$-sub-lattice is invariant under a larger sub-group.
It turns out (cf. Propositions \ref{prop:1class}--\ref{prop:12class} below) that the possible stabilizer groups are $Z_2\times Z_2$, $Z_6$, ${\rm{Dih}}_4$, ${\rm{Dih}}_6$ 
and $O_\rh$, of orders $4$, $6$, $8$, $12$ and $48$ respectively. As all listed orders are different, the size of the 
equivalence class is determined by the type of the corresponding stabilizer. Each of the propositions below characterizes the class corresponding to each stabilizer and exhausts the list of all possible stabilizers.

\begin{proposition}\label{prop:1class} %{Prop3.2}
A class with a single cubic $\ell$-sub-lattice is formed only by $\ell\bbZ^3$, with the basis \eqref{1basis},
where $\ell\in \bbN$. This sub-lattice is
invariant under all symmetries from $O_\rh$. %Such a class is unique for a given $\ell$.
\end{proposition}

\bp Suppose we have a cubic $\ell$-sub-lattice invariant under all elements of $O_h$ and different from $\ell\bbZ^3$. Assume it has a basis vector $(m,n,k)$ of length $\ell$ with non-zero components $m,n,k$. Take 
vector $(n,-m,k)$ obtained via the rotation by $\pi/2$ around the vertical axis;
it  has length $\ell$ and belongs to the sub-lattice. The scalar product of these two vectors equals $mn-mn+k^2=k^2>0$, whereas it should be either $0$ or $-\ell^2$.  
This is a contradiction.  

Assume  the sub-lattice contains vector $(m,n,0)$ of length $\ell$ with non-zero $m,n$.
Then vector  $(0,m,n)$, of length $\ell$, belongs to the sub-lattice as it is obtained via rotation 
by $2\pi/3$ around the main diagonal collinear with the vector $(1,1,1)$ (referred to 
as the $(1,1,1)$-diagonal). The scalar product of these two vectors equals $mn$, but as before it should be $0$ or $-\ell^2$. We again get a contradiction.

The remaining possibility is that all $6$ vectors of length $\ell$ in our $\ell$-sub-lattice have two
components $0$. Then it coincides with $\ell\bbZ^3$.   \ep

\begin{proposition}\label{prop:4class} %{Prop3.3}
{\rm{(i)}} A class with $4$ cubic $\ell$-sub-lattices exists iff $\ell=3t$, $t\in \bbN$, and is unique 
for a given $t$. Such a class is formed by the 
sub-lattices obtained via the rotation of $\ell\bbZ^3$ by the angle $\pi$ about each of the $4$ 
main diagonals. These sub-lattices are spanned by the following $4$ bases  
\beq\label{4basis}\beal
\{(-t, 2t, 2t), \,(2t, -t, 2t), \,(2t, 2t, -t)\};\;\,\,\,\,\,\, \{(t, 2t, 2t), \,(-2t, -t, 2t), \,(-2t, 2t, -t)\};\\
\{(-t, -2t, 2t), \,(2t, t, 2t), \,(2t, -2t, -t)\};\;\,\, \{(-t, 2t, -2t), \,(2t, -t, -2t), \,(2t, 2t, t)\}.\ena\eeq

{\rm{(ii)}} Each cubic $\ell$-sub-lattice from the class is invariant under the rotations by $2\pi/3$ about the 
corresponding main 
diagonal, the reflection about any diagonal plane containing this main diagonal and the inversion, which generate the 
dihedral stabilizer subgroup ${\rm{Dih}}_3\times Z_2 \simeq {\rm{Dih}}_6 < O_\rh$ of order~$12$.
%invariant under rotations by $\pm 2\pi/3$ about the corresponding main diagonal and the reflections about 
%the emerging $3$ diagonal planes (the planes passing through the main diagonal and the coordinate axes).  
%That is, the stabilizer of an $\ell$-sub-lattice is a dihedral sub-group ${\rm{Dih}}_6 < O_\rh$ of order $12$. 
\end{proposition}

\bp It is a direct calculation to verify that the class containing 4 cubic $\ell$-sub-lattices \eqref{4basis} satisfies the properties in (ii), and each sub-lattice from \eqref{4basis} does not have additional symmetries. Therefore, it 
remains to verify the inverse: if a cubic $\ell$-sub-lattice is invariant under symmetries listed in statement (ii) 
then it is one of the sub-lattices \eqref{4basis}.

Suppose a cubic $\ell$-sub-lattice is invariant under the rotation by $\pm 2\pi/3$ about a main 
diagonal. For definiteness, choose the  $(1,1,1)$-diagonal and consider the rotation angle 
$2\pi/3$. Let a vector $(m,n,k)\not= (\ell,\ell,\ell)$ of length $\sqrt{3} \ell$ be collinear to a main diagonal of 
our sub-lattice. Then its image after the rotation, $(k,m,n)$, must give another main diagonal of the sub-lattice. 
Consequently, the $\cos$ of the angle between these two diagonals, $(km + mn + nk)/3\ell^2$, must be 
equal to $1/3$. This implies that
$$(m+n+k)^2 = (m^2+n^2+k^2) +2(km + mn + nk) = 3\ell^2 + 2\ell^2=5\ell^2,$$
which is impossible with integer $m, n,k ,\ell$, due to irrationality of $\sqrt{5}$. Thus, the only remaining 
possibility is $(m,n,k)= (\ell,\ell,\ell)$, i.e., that the cubic sub-lattice is obtained by a rotation of $\ell \bbZ^3$ around 
the $(1,1,1)$-diagonal. Let the angle of rotation be $\alpha$. 

Suppose that the sub-lattice is also invariant under the 
reflection about 
a diagonal plane. Then it is not hard to see that this diagonal plane must contain vector $(1,1,1)$, and 
the angle $\alpha$ must be $\pi$. Moreover, the standard orthogonal basis of $\ell \bbZ^3$ is mapped, under 
the rotation by $\pi$ around the $(1,1,1)$-diagonal, into $\{(-t, 2t, 2t), \,(2t, -t, 2t), \,(2t, 2t, -t)\}$, where $t=\ell/3$. The corresponding three vectors belong to $\bbZ^3$ only if $\ell=0 \hskip-2pt\mod 3$. 
    
The remaining triples in \eqref{4basis} correspond to the other three main diagonals. \ep

\begin{proposition}\label{prop:6class} %{Prop3.4}
{\rm{(i)}} A class with $6$ cubic $\ell$-sub-lattices exists iff $\ell$ has a prime factor $p=1\hskip-2pt\mod 4$.  
The sub-lattices forming such a class are obtained by rotating $\ell\bbZ^3$ by the  
angles $\pm2 \arctan\Big(\diy\frac{b}{a}\Big)$ %non-multiple of $\pi/4$ 
around each of the $3$ coordinate axes, and they are spanned by
the bases 
\beq\label{6basis}\beac\{(\ell,0,0),\; (0, n, k),\; (0,-k, n)\}, \quad \{(\ell,0,0),\; (0, k, n),\; (0,n, -k)\},\\
\{(0,\ell,0),\; (k,0, n),\; (n,0, -k)\}, \quad \{(0 ,\ell,0),\; (n, 0, k),\; (-k,0,n)\},\\
\{(0,0,\ell),\; (n, k,0),\; (-k, n,0)\}, \quad \{(0,0,\ell),\; (k, n, 0),\; (n, -k, 0)\}.\ena\eeq
Here $a,b,n,k,t\in\bbN$ are such that 
\beq\label{Pytha}a>b,\;{\rm{gcd}}\,(a,b)=1,\; \ell =(a^2+b^2)t,\; n =(a^2-b^2)t,\; k=2abt,\; n^2+k^2=\ell^2.\eeq
The above pair $a,b\in\bbN$ is identified with the pair of conjugated Gaussian integers 
$a\pm b\cdot\ri$ in the ring $\bbZ\left[{\sqrt{-1}}\right]$. Each conjugated pair from $\bbZ\left[{\sqrt{-1}}\right]$ generates $2$ sub-lattices in each row of \eqref{6basis}.
Different triples $a,b, t\in\bbN$ with $a>b$, ${\rm{gcd}}\,(a,b)=1$, $t(a^2+b^2)=\ell$ determine different classes. 

{\rm{(ii)}} When the rotation axis is fixed, each of the emerging $2$  sub-lattices (listed in the 
respective row in \eqref{6basis}) is invariant under 
the rotation by $\pi/2$ about the chosen coordinate axis and
the inversion, generating the dihedral stabilizer subgroup ${\rm{Dih}}_4 < O_\rh$ of order $8$.
 
{\rm{(iii)}} If $\ell$ contains distinct prime factors $p_i=1\hskip-2pt\mod 4$ with multiplicities $\rho_i \ge 0$ then
the corresponding number of classes of cardinality $6$ equals 
\beq\label{STwo}\frac{1}{2}s_2(\ell) := \frac{1}{2}\left[\prod_{i} (2\rho_i+1) -1\right].\eeq
\end{proposition}

\bp (i, ii) It is a direct calculation to verify that the class containing 6 sub-lattices \eqref{6basis} satisfies 
properties (ii), and each sub-lattice from \eqref{6basis} does not have additional symmetries. 
The next step is to check the inverse: if a cubic $\ell$-sub-lattice is invariant under symmetries listed in statement 
(ii) then it is one of sub-lattices \eqref{6basis}.

Suppose 
a cubic $\ell$-sub-lattice is invariant under the rotation by $\pm\dfrac{\pi}{2}$ around a
coordinate axis. Then the sub-lattice is also invariant under the rotation by $\pi$ around the 
same axis. %As a result, all nine $\bbZ^3$-symmetries listed in (ii) and (iii) from
%\eqref{description} are analyzed in a similar way. So, 
For definiteness, assume that the sub-lattice is invariant 
under the rotation by $\pi$ around the vector $(1,0,0)$. 

Take the collection of $6$ vectors 
of length $\ell$: three forming an orthogonal basis of the sub-lattice plus their opposites obtained via 
the inversion.  Let $(m,n,k)$ 
be a vector from the collection forming the smallest angle $\alpha$ with $(1,0,0)$. By 
construction, the vector $(m,-n, -k)$ obtained via the rotation must also belong to the sub-lattice. This 
may happen only if $\alpha = 0$, $\alpha=\dfrac{\pi}{4}$ or $\alpha =\dfrac{\pi}{2}$, as two vectors of length $\ell$ from the  
sub-lattice must be either collinear or orthogonal. The case $\alpha = \dfrac{\pi}{2}$ is impossible as it implies that all 6 vectors belong to the half-space $m \le 0$. The case  $\alpha = \dfrac{\pi}{4}$ implies that the angle between $(m,n,k)$ and $(m,-n,-k)$ is $\dfrac{\pi}{2}$ and therefore $m^2=n^2+k^2$  and, consequently, $l = \sqrt{2}m$, violating the requirement $l \in N$. Thus, 
$\alpha = 0$, i.e $(\ell,0,0)$  is the basis vector of our sub-lattice, which occurs only if the sub-lattice 
is $\ell \bbZ^3$ or is obtained by a rotation of $\ell \bbZ^3$ around $(\ell,0,0)$. 

This implies that
the basis of the sub-lattice is $\{(\ell,0,0),(0, n, k),(0,-k, n)\}$
which is listed first in the top line in \eqref{6basis}. Here $n, k$ are integers such that $n^2+k^2=\ell^2$. 
The second triple in the top line in \eqref{6basis} corresponds to the conjugated sub-lattice obtained via the 
rotation by the same angle in the opposite direction. The remaining triples in \eqref{6basis} correspond to 
the other coordinate axes. 

The next step is to verify representation \eqref{Pytha}. 
The already established condition $n^2+k^2=\ell^2$ in \eqref{Pytha} means that $n,k,\ell$ form a Pythagorean triple. 
%and the problem is reduced to the identification of $\ell$-quadratic sub-lattices of $\bbZ^2$. 
The structure of such triples is associated with the rotation matrices  
$$\rR=\frac{1}{\sqrt{a^2 + b^2}}\begin{pmatrix}a & b \\ -b & a \end{pmatrix},\quad
\rR^{-1}=\frac{1}{\sqrt{a^2 + b^2}}\begin{pmatrix}a & -b \\ b & a \end{pmatrix}.$$ 
The rows of the orthogonal matrix 
$$t(a^2 + b^2)\rR^2=t\begin{pmatrix}a^{2}- b^{2} & 2ab \\ 
-2ab  & a^{2}- b^{2} \end{pmatrix}$$
parametrize Pythagorean triples in  \eqref{Pytha} through orthogonal pairs $(n,k)$ and $(-k, n)$. It is well-known that this 
parametrization can be derived from the identification of the matrices $R$ and $R^{-1}$ with the pair of 
conjugated Gaussian integers $a\pm b\cdot\ri$.

Given $\ell \in\bbN$, the non-trivial representation of $\ell^2$ as the sum of two positive squares, i.e. as the 
norm of a Gaussian integer, exists iff the rational prime 
decomposition of $\ell$ has a factor $p=1$ mod $4$. Here non-trivial means different from the representation 
$\ell^2+0^2=\ell^2$. A prime $p=1$ mod $4$ is the product of two conjugated 
Gaussian primes $a\pm b\cdot\ri$, where $a>b>0$ with $\gcd(a,b)=1$ are uniquely defined by $p$. 
A rational prime $p=3$ mod $4$ is always a Gaussian prime (for example see \cite{Gross, BSh}). Denote by $p_i$, 
$i=1,2, \ldots$ the set of distinct primes of the form $4s+1$ entering the rational prime decomposition of $\ell$ 
with multiplicities $\rho_i \ge 1$. For each $i$ we have $p_i = (a_i+ b_i\cdot\ri)(a_i-b_i\cdot\ri)$, where 
integers $a_i > b_i>0$ are uniquely defined by $p_i$. Thus, one can form a multitude of products
\beq\label{Prod1}\prod_{i=1} (a_i \pm b_i\cdot\ri)^{\alpha_i} , \quad 0 \le \alpha_i \le \rho_i, 
\quad \sum_{i=1} \alpha_i \ge 1\eeq
by varying the selection of $\alpha_i$ and the sign in front of $b_i$. Expanding the product in 
\eqref{Prod1}, we end up with a Gaussian integer $a + b\cdot\ri$ satisfying (after a multiplication by a 
unit in $\bbZ\left[{\sqrt{-1}}\right]$)
\beq\label{Cond2}a,|b|\in\bbN,\;a>|b|,\; \gcd(a,b)=1,\;\ell = (a^2+b^2)t,\;t\in\bbN.\eeq

Note that in \eqref{Cond2}, the value $b$ can be negative, and the triple $a$, $b$, $t$ defines a unique 
quadratic $\ell$-sub-lattice of $\bbZ^2$. Changing the sign of $b$ implies a conjugated sub-lattice (see the 
second triple in each row of \eqref{6basis}). This completes the verification of \eqref{Pytha} and therefore 
the proof of assertions (i) and (ii).

(iii) To establish \eqref{STwo}, observe that
there are $\rho_i$ possibilities to choose $\alpha_i >0$, and for each of them there are 2 possibilities to 
choose the sign between $a_i$ and  $ b_i$. An additional possibility is $\alpha_i =0$ which amounts to 
$2\rho_i+1$ choices in total. Since the choices for different $i$ are done independently, the quantity
\beq\label{eqns1}s_2(\ell) = \prod_{i} (2\rho_i+1) - 1\eeq
counts the total amount of possibilities,  excluding the case where all  $\alpha_i=0$. This is exactly the number of distinct square $\ell$-sub-lattices in $\bbZ^2$ different from $\ell\bbZ^2$, because a
pair of conjugated square $\ell$-sub-lattices of $\bbZ^2$ 
generates cubic  $\ell$-sub-lattices of $\bbZ^3$ from the same class. Hence, the number of classes corresponding to $\ell$ 
is equal to $\dfrac{s_2(\ell)}{2}$, which establishes (iii).
\ep

\bremnn {\rm The possibilities $a=b$ or $b=0$ excluded by \eqref{Pytha} give a class with
a single sub-lattice considered in Proposition \ref{prop:1class}.}
\eremnn

\begin{proposition}\label{prop:8class}%\label{Prop3.5}
{\rm{(i)}}
A class formed by $8$ cubic $\ell$-sub-lattices exists iff $\ell$ has a prime factor  $p=1\,{\rm{mod}}\,3$. 
The sub-lattices forming such a class are obtained by rotating $\ell\bbZ^3$ by the  
angles $\pm 2\arctan \Bigg(\dfrac{b{\sqrt 3}}{2a-b}\Bigg)$
around each of the $4$ main diagonals, and they are spanned by
the bases 
\beq\label{8basis}\beal
\{ (m, n ,k), \; (k ,m, n), \; (n, k , m) \},\; \{(m, k , n), \; (n , m, k), \; (k, n, m)\},\\
\{ (n, -m ,k), \; (m, -k, n), \; (k, -n , m)\},\; \{ (k, -m ,n), \; (m, -n, k), \; (n, -k , m)\},\\
\{ (-m, -n ,k), \; (-k ,-m, n), \; (-n, -k , m) \},\\ \hskip170pt \{ (-m, -k ,n), \; (-n ,-m, k), \; (-k, -n , m) \},\\
\{(-n, m ,k), \; (-m, k, n), \; (-k, n , m)\},\; \{(-k, m ,n), \; (-m, n, k), \; (-n, k , m)\}.\ena\eeq
Here $a,b,m,k,n,t\in\bbN$ are such that  
\beq\label{Eise1}\beal
a>2b,\;{\rm{gcd}}\,(a,b)=1,\; (a^2+b^2-ab)t=\ell,\; m = (a^2 - ab)t,\; n=abt,\\ k = (b^2 - ab)t,
\quad (m-k)^2 + (n-k)^2 - (m-k)(n-k) =\ell^2.\ena\eeq
%and
%\beq\label{Eise2} l^2= m^2+n^2+k^2 = (m+n+k)^2,\quad mk+nm+kn=0, \eeq
%and different pairs $a,b\in\bbN$ with $a>2b$, $a^2+b^2-ab=\ell$ determine different classes. 
The above pair $a,b\in\bbN$ is identified with the pair of conjugated Eisenstein integers 
$a + b\cdot\omega$ and $a + b\cdot\overline\omega$ in the ring $\bbZ\left[\omega\right]$, where 
$\omega =(-1+{\sqrt 3}\cdot \ri)/2$. Each conjugated pair from $\bbZ\left[\omega\right]$ generates $2$ 
sub-lattices in each row of \eqref{8basis}.
Different triples $a,b, t\in\bbN$ with $a>2b$, ${\rm{gcd}}\,(a,b)=1$, $t(a^2+b^2-ab)=\ell$ determine different classes. 

{\rm{(ii)}} When the rotation axis is fixed, each of the emerging $2$  sub-lattices (listed in the 
respective row in \eqref{8basis}) is invariant under 
the rotation by $2\pi/3$ about the chosen main diagonal and
the inversion, generating the stabilizer subgroup $Z_3\times Z_2 \simeq Z_6 < O_\rh$ of order~$6$.

{\rm{(iii)}} If $\ell$ contains distinct prime factors $p_i=1\,{\rm{mod}}\,3$ with multiplicities $\rho_i \ge 0$ then
the corresponding number of classes of cardinality $8$ equals 
\beq\label{ShTwo}\frac{1}{2}{\wh s}_2(\ell) := \frac{1}{2}\left[\prod_{i} (2\rho_i+1) -1\right].\eeq
\end{proposition}

\bp (i, ii) It is a direct calculation to verify that the class containing 8 sub-lattices \eqref{8basis} satisfies 
properties (ii), and each sub-lattice from \eqref{8basis} does not have additional symmetries. The next 
step is to check the inverse: if a cubic $\ell$-sub-lattice is invariant under symmetries listed in statement 
(ii) then it is one of sub-lattices \eqref{8basis}.

Suppose a cubic $\ell$-sub-lattice is invariant under the rotation by $\pm\dfrac{2\pi}{3}$ about a main 
diagonal. For definiteness, choose the $(1,1,1)$-diagonal and consider the rotation 
angle $\dfrac{2\pi}{3}$. Let a vector $(m,n,k)\not= (\ell,\ell,\ell)$ of length $\sqrt{3}\ell$ define a main diagonal of our cubic $\ell$-sub-lattice. Then its image after rotation, $(k,m,n)$, must give another main diagonal of the sub-lattice. 
Consequently, the $\cos$ of the angle between these two diagonals, $(km + mn + nk)/3\ell^2$, must be 
equal to $1/3$. This implies that
$$(m+n+k)^2 = (m^2+n^2+k^2) +2(km + mn + nk) = 3\ell^2 + 2\ell^2=5\ell^2,$$
which is impossible with integer $m, n,k ,\ell$ due to irrationality of $\sqrt{5}$. Thus, the only remaining 
possibility is $(m,n,k)= (\ell,\ell,\ell)$, i.e., that the cubic $\ell$-sub-lattice is obtained by a rotation of $\ell \bbZ^3$ around 
the $(1,1,1)$-diagonal. 

This implies that
the basis of the sub-lattice is $\{(m, n ,k), \;(k ,m, n), \;(n, k , m)\}$,
which is listed first in the top line in \eqref{8basis}. Here $m, n, k$ are integers such that 
\beq\label{Eise3}(m-k)^2 + (n-k)^2 - (m-k)(n-k) = \ell^2\eeq
which is the last condition in \eqref{Eise1}.
To see the necessity of \eqref{Eise3}, denote by $\bbA_2$
the intersection of $\bbZ^3$ and the plane $\rx_1+\rx_2+\rx_3=0$. Then $\bbA_2$ is a triangular lattice with distance 
$\sqrt{2}$ between nearest sites. For definiteness, select vectors $e_1 = (1,-1,0) = (1,0,0) - (0,1,0)$ 
and $e_2 = (0,1,-1) = (0,1,0) - (0,0,1)$ as a basis in $\bbA_2$. 
Obviously, $\ell\bbZ^3\cap \bbA_2=\ell\bbA_2$ is a triangular $\ell$-sub-lattice of $\bbA_2$. The rotation of 
$\ell\bbZ^3$ around the $(1,1,1)$-diagonal corresponds to the rotation of $\ell\bbA_2$ by the same angle  
around the origin. 
If under this rotation the vector $(\ell,0,0)$ is mapped to the vector $(m,n,k)$ then vectors $(0,\ell,0)$ and $(0,0,\ell)$ 
are mapped into $(k ,m, n)$ and $(n, k , m)$, respectively as a cyclic permutation of coordinates corresponds to 
a rotation by $2\pi/3$ mapping one basis vector into another. Consequently,  the vector $(\ell, -\ell,0)$ is 
mapped into the vector $(m-k, n-m, k-n)= (m-k)e_1 + (n-k)e_2$. Thus, \eqref{Eise3} defines an Eisenstein triple which is an $\bbA_2$-analogue of a Pythagorean triple $m^2+n^2 = \ell^2$ in $\bbZ^2$ considered in \eqref{Pytha}. 
The second triple in the top line in \eqref{8basis} corresponds to the conjugated sub-lattice obtained via the 
rotation by the same angle in the opposite direction. The remaining triples in \eqref{8basis} correspond to 
the other main-diagonals. 

It remains to verify representation \eqref{Eise1}. The problem of identifying triangular 
$\ell$-sub-lattices of $\bbA_2$ is similar to that for square $\ell$-sub-lattices of $\bbZ^2$, which was
discussed in the proof of the previous proposition. The only difference is that, instead of Gaussian integers, one 
needs to work with Eisenstein integers forming the ring $\bbZ\left[\omega\right]$. 
%Condition \eqref{Eise3} means that $(m-k),(n-k),\ell$ form an  $\bbA_2$-analogue of a Pythagorean 
%triple (an Eisenstein triple). 
The structure of Eisenstein triples in $\bbA_2$ is associated with the rotation matrices   
$$\rR=\frac{1}{\sqrt{a^2 + b^2-ab}}\begin{pmatrix}a & b \\ -b & a-b \end{pmatrix},\quad
\rR^{-1}=\frac{1}{\sqrt{a^2 + b^2-ab}}\begin{pmatrix}a-b & -b \\ b & a \end{pmatrix}$$ 
given in $\bbA_2$-coordinates $e_1$ and $e_2$. The rows of the matrix 
$$t(a^2 + b^2-ab)\rR^2=t\begin{pmatrix}a^{2}- b^{2} & 2ab-b^2 \\ 
b^2-2ab  & a^{2}-2ab \end{pmatrix}$$
parametrize the solutions $(m-k), (n-k)$ to  \eqref{Eise3}. This parametrization can be derived from the identification 
of the matrices $R$ and $R^{-1}$ with the pair of conjugated Eisenstein integers $a+b\cdot\omega$ and
 $a+b\cdot\overline\omega$.

Given $\ell \in\bbN$, the non-trivial representation of $\ell^2$ in the form \eqref{Eise3}, i.e. as the norm of an 
Eisenstein integer, exists iff the rational prime 
decomposition of $\ell$ has a factor $p=1 \mod3$. Here non-trivial means different from the representations 
$\ell^2+0^2- \ell\cdot 0=\ell^2$ and $\ell^2 + \ell^2 - \ell\cdot\ell = \ell^2$. 
A prime $p=1 \mod 3$ is the product of two conjugated 
Eisenstein primes $a+b\cdot\omega$ and $a+b\cdot\overline\omega$, where $a>|2b|>0$ with $\gcd(a,b)=1$ 
are uniquely defined by $p$. (Note, that for $p=3$, $a=2$ and $b=1$.) A rational prime $p=2 \mod 3$ is always an Eisenstein prime. Denote by $p_i$, $i=1,2, \ldots$ the set of distinct primes of the form $3s+1$ entering the rational prime 
decomposition of $\ell$ with multiplicities $\rho_i \ge 1$. For each $i$ we have $p_i = (a_i+ b_i\cdot\omega)
(a_i+b_i\cdot\overline\omega)$, 
where integers $a_i > |2b_i|$ are uniquely determined by $p_i$. Thus, one can form a multitude of products
\beq\label{Prod2}\prod_{i=1} (a_i+b_i\cdot\wh\omega)^{\alpha_i} , \quad 0 \le \alpha_i \le \rho_i, 
\quad \sum_{i=1} \alpha_i \ge 1\eeq
by varying the selection of $\alpha_i$ and the selection of $\wh\omega$ as either $\omega$ or $\overline\omega$. 
Expanding the product in \eqref{Prod2}, 
we end up with an Eisenstein integer $a + b\cdot\omega$ satisfying (after a multiplication by a unit in 
$\bbZ\left[{\omega}\right]$)
\beq\label{Cond3}a,|b|\in\bbN,\;a>|2b|,\; \gcd(a,b)=1,\;\ell = (a^2+b^2-ab)t,\;\hbox{where}\;t\in\bbN.\eeq

Note, that in \eqref{Cond3} the value $b$ can be negative, and the triple $a$, $b$, $t$ defines a unique 
triangular $\ell$-sub-lattice of 
$\bbA_2$. Changing the sign of $b$ and replacing $a$ with $a-b$ implies a conjugated sub-lattice (see the 
second triple in given row of \eqref{8basis}).
This completes the verification of \eqref{Eise1}. To finish the proof of assertions (i) and (ii), we only need to 
observe that the site $a e_1+b e_2 \in \bbA_2$ can be written as $\sqrt{2}\left(a - \dfrac{b}{2}, 
b \dfrac{\sqrt{3}}{2}\right)$ 
in Cartesian coordinates $e'_1 =\dfrac{1}{\sqrt{2}}e_1$, $e'_2 = \dfrac{2}{\sqrt{6}} e_2
- \dfrac{1}{\sqrt{6}}e_1$ implying the desired value for the tangent of the rotation angle. 
%This finishes the proof of assertions (i) and (ii).

(iii) To establish \eqref{ShTwo}, observe that there are $\rho_i$ possibilities to choose $\alpha_i >0$ 
and for each of them there are 2 possibilities to choose between $\omega$ and  $\overline\omega$. 
An additional possibility is $\alpha_i =0$ which amounts to $2\rho_i+1$ choices in total. Since the 
choices for different $i$ are done independently, the quantity
\beq\label{eqns2}\wh s_2(\ell) = \prod_{i=1} (2\rho_i+1) - 1\eeq
counts the total amount of possibilities,  excluding the case where all  $\alpha_i=0$. This is exactly 
the number of distinct triangular $\ell$-sub-lattices in $\bbA_2$ different from $\ell\bbA_2$ and its rotation 
by $\dfrac{\pi}{6}$ (the case of $a=2b$ treated in the~Proposition \ref{prop:4class}). Because a pair of 
conjugated triangular $\ell$-sub-lattices of $\bbA_2$ generates cubic $\ell$-sub-lattices of $\bbZ^3$ from 
the same class, the number of corresponding classes is equal to $\dfrac{\wh s_2(\ell)}{2}$. This 
yields~(iii).~\ep

\bremnn {\rm The case $a=2b$ excluded by \eqref{Eise1} gives rise to a class with
$4$ sub-lattices considered in Proposition \ref{prop:4class}. The case $b=0$ gives rise to a
class with a single sub-lattice from Proposition \ref{prop:1class}.}
\eremnn

\begin{proposition}\label{prop:12class}%\label{Prop3.6}
{\rm{(i)}} A class formed by $12$ cubic $\ell$-sub-lattices exists iff $\ell$ has a prime factor of the form 
$8s+1$ or $8s+3$. The sub-lattices forming such a class are obtained by rotating $\ell\bbZ^3$ by the  
angles $\pm 2\,{\rm{arctan}}\,\diy\Bigg(\frac{\sqrt{2}b}{a}\Bigg)$ around each of the $6$ non-main 
diagonals, and they are spanned by bases
\beq\label{12basis}\beal
\{(m,n,k),(n, m, -k),(-k,k,m-n)\},\{(m, n,-k),(n, m, k),(k, -k, m-n)\},\\
\{(n,-m,k),(m,-n,-k),(k,k,m-n)\},\{(n, -m,-k),(m, -n, k),(-k, -k, m-n)\},\\
\{(k,m,n),(-k, n,m), (m-n,-k,k) \},\{(-k, m, n),(k, n, m),(m-n, k, -k)\},\\
\{(k, n, -m),(-k, m, -n),(m-n, k, k) \},\{(-k, n, -m),(k, m, -n), (m-n, -k, -k)\},\\
\{(n,k,m),(m, -k, n),(k, m-n, -k) \},\{(n, -k, m),(m, k, n), (-k, m-n, k)\},\\
\{(-m, k, n),(-n, -k, m),(k, m-n, k) \},\{(-m, -k, n),(-n, k, m),(-k, m-n, -k)\}.\\
\ena\end{equation}
Here $a,b,m,n,k, t\in\bbN$ are such that  
\beq\label{12:cond}\beal a\neq b, \;\;a\neq 2b,\;{\rm{gcd}}\,(a,b)=1,\; (a^2+2b^2)t=\ell,\;\;m=a^2 t,\; n =2b^2 t,\;
k=2abt,\\ (m-n)^2 + 2k^2=\ell^2.\ena\eeq
The above pair $a,b\in\bbN$ is identified with the pair of conjugated algebraic integers 
$a\pm b\cdot\sqrt{2}\ri$ in the ring $\bbZ\left[{\sqrt{-2}}\right]$. Each conjugated pair from $\bbZ\left[{\sqrt{-2}}\right]$ generates $2$ sub-lattices in each row of \eqref{12basis}.
Different triples $a,b, t\in\bbN$ with ${\rm{gcd}}\,(a,b)=1$, $t(a^2+2b^2)=\ell$ determine different classes.

{\rm{(ii)}} Each sub-lattice in a class is invariant under the rotation by $\pi$ about the corresponding 
 non-main diagonal and the inversion, generating the stabilizer subgroup $Z_2\times Z_2 \simeq 
V_4 < O_\rh$ of order $4$. 

{\rm{(iii)}} If $\ell$ has distinct prime factors $p_i$ of the form $8s+1$ or $8s+3$ with multiplicities 
$\rho_i$ then the number of classes of cardinality $12$ equals 
\beq\label{tSThree}\frac{1}{2}{\wt s}_2(\ell) := \frac{1}{2}\left[\prod_{i} (2\rho_i+1) -1\right].\eeq
\end{proposition}

\bp
(i, ii) It is a direct calculation to verify that the class containing 12 sub-lattices \eqref{12basis} satisfies 
properties (ii), and each sub-lattice from \eqref{12basis} does not have additional symmetries. The next 
step is to check the inverse: if a cubic $\ell$-sub-lattice is invariant under symmetries listed in statement 
(ii) then it is one of sub-lattices \eqref{12basis}.

Suppose a cubic $\ell$-sub-lattice is invariant under the rotation by $\pm\pi$ about a non-main 
diagonal. For definiteness, choose the $(1,1,0)$-diagonal and consider the rotation 
angle $\pi$. Take the collection of $6$ vectors 
of length $\ell$: three forming an orthogonal basis of the sub-lattice plus their opposites obtained via 
the inversion.  Let $(m,n,k)$ 
be a vector from the collection, forming the smallest angle %$\alpha$ 
with $(1,1,0)$. By 
construction, the vector $(n, m, -k)$ obtained via the rotation of $(m,n,k)$ must also belong to the sub-lattice. This 
may happen only if the angle $\alpha$ between $(m,n,k)$ and $(n, m, -k)$ is $0$ or $\dfrac{\pi}{2}$, as two vectors of length $\ell$ from the  
$\ell$-sub-lattice must be either collinear or orthogonal. The case $\alpha = 0$ is impossible because in that 
case $n = m$, $k = 0$ implying $2m^2=\ell^2$, which contradicts irrationality of $\sqrt{2}$. Thus, 
$\alpha =\dfrac{\pi}{2}$, and $(\ell,\ell,0)$ is a diagonal of the considered cubic $\ell$-sub-lattice, i.e. the sub-lattice is 
obtained by a rotation of $\ell \bbZ^3$ around $(1,1,0)$-diagonal.

This implies that the basis of the sub-lattice is $\{(m,n,k),(n, m, -k),(-k,k,m-n)\}$, which is listed 
first in the top line in \eqref{12basis}. Indeed, by construction the length of $(m,n,k)$ and $(n,m,-k)$ is $\ell$ and the angle between them is $\dfrac{\pi}{2}$, that is, $m^2+n^2+k^2=\ell^2$ and $2mn-k^2=0$. Hence, the vector $(-k,k,m-n)$ is orthogonal to both of them and has length $\ell$.

Thus, $m, n, k$ are integers such that 
\beq\label{12:mnkcond} (m-n)^2 + 2k^2= \ell^2, \quad (-2k)^2+ 2 (m-n)^2 = 2\ell^2\eeq
which is the last condition in \eqref{12:cond}. To see the necessity of \eqref{12:mnkcond}, denote by $\bbL$
the intersection of $\bbZ^3$ and the plane $\rx_1+\rx_2=0$. Then $\bbL$ is a rectangular lattice 
$\bbZ\times\sqrt{2}\bbZ$. For definiteness, select vectors $e_1 = (0,0,1)$, $e_2 = (-1,1,0) = (0,1,0) - (1,0,0)$ 
as a basis in $\bbL$. 
Obviously, $\ell\bbZ^3\cap \bbL=\ell\bbL$ is a rectangular sub-lattice of $\bbL$. The rotation of 
$\ell\bbZ^3$ around the $(1,1,0)$-diagonal corresponds to the rotation of $\ell\bbL$ by the same angle  
around the origin. 
If under this rotation the vector $(\ell,0,0)$ is mapped into the vector $(m,n,k)$ then vector $(0, \ell,0)$ is mapped 
into $(n, m,-k)$ obtained from $(m,n,k)$ via rotation by $\pi$ about the $(1,1,0)$-diagonal. Consequently, 
the vector $(0, 0, \ell)$ is mapped into the vector $(-k, k, m-n)=\dfrac{1}{m+n} \Big((m,n,k)\times(n,m,-k)\Big)= 
(m-n)e_1 + ke_2$. Accordingly, the vector $(-\ell, \ell, 0)$ is mapped into $(n-m, m-n, -2k) =  -2k e_1 + (m-n)e_2$. 
Thus, \eqref{12:mnkcond} is an $\bbL$-analogue of Pythagorean and Eisenstein  triples considered in \eqref{Pytha} and \eqref{Eise1}, respectively. 
The second triple in the top line in \eqref{12basis} corresponds to the conjugated sub-lattice obtained via the 
rotation by the same angle in the opposite direction. The remaining triples in \eqref{12basis} correspond to 
the other non-main diagonals. 

It remains to verify representation \eqref{12:cond}. The problem of identifying rectangular 
sub-lattices of $\bbL$ congruent to $\ell\mathbb{L}$ is similar to that for square $\ell$-sub-lattices of $\bbZ^2$, which was
discussed in the proof of Proposition \ref{prop:6class}. The only difference is that, instead of Gaussian 
integers, one needs to work with algebraic integers from the ring $\bbZ\left[\sqrt{-2}\right]$ belonging to 
the norm-Euclidean quadratic field $\bbQ\left[\sqrt{-2}\right]$. %Condition \eqref{12:mnkcond} means that 
%$k, m-n, \ell$ form an $\bbL$-analogue of a Pythagorean 
%triple. 
The structure of triples \eqref{12:mnkcond} is associated with the rotation matrices   
$$\rR=\frac{1}{\sqrt{a^2 + 2b^2}}\begin{pmatrix}a & b \\ -2 b & a \end{pmatrix},\quad
\rR^{-1}=\frac{1}{\sqrt{a^2 + 2b^2}}\begin{pmatrix}a & -b \\ 2b & a \end{pmatrix}$$ 
given in $\bbL$-coordinates $e_1$ and $e_2$. The rows of the matrix 
$$t(a^2 + 2b^2)\rR^2=t\begin{pmatrix}a^{2}- 2b^{2} & 2ab \\ -4ab  & a^{2}- 2b^{2} \end{pmatrix}$$
parametrize solutions $m-n, k$  and $-2k, m-n$ to the pair of equations \eqref{12:mnkcond}. This 
parametrization can be derived from the identification of the matrices $R$ and $R^{-1}$ with the pair of 
conjugated algebraic integers $a\pm b\cdot \sqrt{2}\ri \in \bbZ\left[\sqrt{-2}\right]$.

Given $\ell \in\bbN$, the non-trivial representation of $\ell^2$ in the form \eqref{12:mnkcond}, i.e. as the norm 
of an algebraic integer from $\bbZ\left[\sqrt{-2}\right]$, exists iff the rational prime decomposition of $\ell$ 
has a factor $p=1$ mod $8$ or $p=3$ mod $8$. Here non-trivial means different from the representations 
$\ell^2+2\cdot 0^2 =\ell^2$, $t^2 + 2(2t)^2=(3t)^2$ and $(2t)^2 + 2(4t)^2=(6t)^2$.
A rational prime $p$ is the product of two conjugated primes 
$a\pm b\cdot\sqrt{2}\ri$ in $\bbZ\left[\sqrt{-2}\right]$, where $a, b>0$ with $\gcd(a,b)=1$ are uniquely 
defined by $p$.  A rational prime $p=5$ mod $8$ or $p=7$ mod $8$ is always an algebraic prime in 
$\bbZ\left[\sqrt{-2}\right]$. Denote by $p_i$, $i=1,2, \ldots$ the set of distinct rational primes 
of the form $8s+1$ or $8s+3$ entering the rational prime decomposition of $\ell$ with multiplicities 
$\rho_i \ge 1$. For each $i$ we have $p_i = (a_i+ b_i\cdot\sqrt{2}\ri)(a_i-b_i\cdot\sqrt{2}\ri)$, where 
integers $a_i, b_i>0$ are uniquely defined by $p_i$. Thus, one can form a multitude of products
\beq\label{Prod3}\prod_{i=1} (a_i \pm b_i\cdot\sqrt{2}\ri)^{\alpha_i} , \quad 0 \le \alpha_i \le \rho_i, 
\quad \sum_{i=1} \alpha_i \ge 1\eeq
by varying the selection of $\alpha_i$ and the sign in front of $b_i$. Expanding the product in 
\eqref{Prod3}, we end up with an algebraic integer $a + b\cdot\sqrt{2}\ri$ satisfying (after a multiplication 
by a unit in $\bbZ\left[{\sqrt{-2}}\right]$)
\beq\label{Cond4}a,|b|\in\bbN,\;\gcd(a,b)=1,\;\ell = (a^2+2b^2)t,\;t\in\bbN.\eeq

Note, that in \eqref{Cond4} the value $b$ can be negative, and the triple $a$, $b$, $t$ defines a 
unique rectangular sub-lattice of 
$\bbL$ congruent to $\ell\bbL$. Changing the sign of $b$ implies a conjugated sub-lattice (see the 
second triple in given row of \eqref{12basis}).
This completes the verification of \eqref{12:cond}. To finish the proof of assertions (i) and (ii), we only 
need to observe that the site $a e_1+b e_2 \in \bbL$ can be written as $\left(a, \sqrt{2} b \right)$ 
in Cartesian coordinates $e'_1 =e_1$, $e'_2 = \sqrt{2} e_2
$ implying the desired value for the tangent of the rotation angle. 
%This finishes the proof of assertions (i) and (ii).

(iii) To establish \eqref{tSThree}, observe that
there are $\rho_i$ possibilities to choose $\alpha_i >0$, and for each of them there are 2 possibilities to 
choose the sign between $a_i$ and  $ b_i$. An additional possibility is $\alpha_i =0$ which amounts to 
$2\rho_i+1$ choices in total. Since the choices for different $i$ are done independently, the quantity
\beq\label{eqns3}{\wt s}_2(\ell) = \prod_{i=1} (2\rho_i+1) - 1\eeq
counts the total amount of possibilities,  excluding the case where all  $\alpha_i=0$. 
This is exactly the number of distinct rectangular sub-lattices in $\bbL$ congruent to $\ell\bbL$ but different from $\ell\bbL$ and the rotations of $\ell\bbL$ by $2\arctan \big(\sqrt{2}\big)$ or $2\arctan \Big(\sqrt{\frac{1}{2}}\Big)$ (the cases of $a=b$ or $a=2b$ treated in the~Proposition \ref{prop:4class}). Because a
pair of conjugated rectangular sub-lattices of $\bbL$ 
generates cubic  $\ell$-sub-lattices of $\bbZ^3$ from the same class, the number of corresponding classes 
is equal to $\dfrac{{\wt s}_2(\ell)}{2}$ which establishes~(iii). 
\ep

\bremnn {\rm The cases $a=b$ and $a=2b$ excluded in \eqref{12:cond} give rise to a class with
$4$ sub-lattices considered in Proposition \ref{prop:4class}. The case $b=0$ gives rise to a
class with a single sub-lattice from Proposition \ref{prop:1class}.}
\eremnn

Propositions \ref{prop:lZ3} - \ref{prop:12class} are dealing with solutions to the Diophantine equation \eqref{DiofEqn} which defines Pythagorean quadruples. This equation is a special case of the classical equation 
\beq\label{mnkl}m^2+n^2+k^2=t.\eeq  
The number of solutions $(m,n,k)$ to \eqref{mnkl} is usually denoted by $r_3(t)$ and depends on the rational prime decomposition of $t$. If an integer $\ell$ has rational prime representation $\ell=2^{\rho_2}\prod\limits_{p_j\geq 3}p_j^{\rho_j}$ then the 
number $r_3(\ell^2)$ is given by (see \cite{CH}, formula (3.1)): 
\beq\label{RlTwo}r_3(\ell^2)=6\prod_{p_j\geq 3}\left[\frac{p_j^{\rho_j+1}-1}{p_j-1}-(-1)^{(p_j-1)/2}
\frac{p_j^{\rho_j}-1}{p_j-1}\right].\eeq

\begin{proposition} \label{prop:24class} %{Prop3.7}
{\rm{(i)}} %A class formed by $24$ sub-lattices exists iff $\ell$ is of the form $\ell=a^2+b^2+c^2+d^2$, where 
%$a, b, c, d\in \bbZ_+$ are such that either $a=0$ and $1\leq b < c < d$ or $1\leq a, b, c, d$ and at most 
%two numbers among $a, b, c, d$ coincide. 
A class formed by $24$ cubic $\ell$-sub-lattices exists iff $r(\ell)>0$, where 
\begin{equation}\label{r}
r(\ell):=r_3(\ell^2) - 30  + 24(\ell^2\hskip -10pt \mod 3) - 12 s_2(\ell)
 - 24 {\wh s}_2(\ell) - 36{\wt s}_2(\ell);
\end{equation}
the number $r(\ell)$ is computed from the prime decomposition of $\ell$ according to \eqref{STwo},   \eqref{ShTwo}, 
\eqref{tSThree}, \eqref{RlTwo}. 

{\rm{(ii)}} Each sub-lattice in a class is invariant under the inversion, generating the stabilizer subgroup  $Z_2
 <O_\rh$.

{\rm{(iii)}} The quantity $\diy\frac{r(\ell)}{144}$ gives a lower bound for the number of classes with $24$ sub-lattices.
 It is only a lower bound since sub-lattices from two different classes may share a basis vector.
\end{proposition}

\bp
Each of Propositions \ref{prop:1class} - \ref{prop:12class} describes a special collection of solutions to 
\eqref{DiofEqn}. Each $\ell$-sub-lattice found in these propositions determines $6$ solutions to \eqref{DiofEqn} 
which are the basis of the sub-lattice and its inverse. Each of these $6$ solutions is obtained by a rotation 
of one of the basis vectors from $\ell\bbZ^3$. Each basis vector belongs to a coordinate axis, and rotations
of a coordinate axis trace some cones in $\bbR^3$.

The following cones are traced when we rotate a coordinate axis in accordance with Propositions 
\ref{prop:6class} - \ref{prop:12class}.
\begin{description}
\item{(i)} The coordinate axes (Proposition \ref{prop:6class}).
\item{(ii)} The coordinate planes (Proposition \ref{prop:6class}).
\item{(iii)} The diagonal planes (Proposition \ref{prop:12class}).
\item{(iv)} Four cones having one of the main diagonals as the rotation axis

$$(\pm x \pm y + z)^2 - (x^2 + y^2 + z^2) = 0\;\hbox{ (Proposition \ref{prop:8class}).}$$
 
\item{(v)} Six cones having one of the non-main diagonals as the rotation axis
$$\beal
(\pm x + y)^2 - (x^2 + y^2 + z^2) = 0\\
(\pm x + z)^2 - (x^2 + y^2 + z^2) = 0\\
(\pm y + z)^2 - (x^2 + y^2 + z^2) = 0\ena
\;\hbox{ (Proposition \ref{prop:12class}).}$$
\end{description}
 
\medskip\noindent
Possible intersections between cones from Proposition \ref{prop:6class} and from Propositions 
\ref{prop:8class} or \ref{prop:12class} are only the coordinate axes, which corresponds to the case from Proposition \ref{prop:1class}. Possible intersections between cones from 
Proposition \ref{prop:8class} and from Proposition \ref{prop:12class} are only the lines along the vectors
$$\bma
(-t, 2t, 2t),(2t, -t, 2t),(2t, 2t, -t),(t, 2t, 2t),(-2t, -t, 2t),(-2t, 2t, -t),\\
(-t, -2t, 2t),(2t, t, 2t),(2t, -2t, -t),(-t, 2t, -2t),(2t, -t, -2t),(2t, 2t, t),\ema$$
emerging in Proposition \ref{prop:4class}, which is possible only for $\ell$ divisible by 3. This observation 
allows us to count distinct solutions to \eqref{DiofEqn} originating from different propositions. 

Proposition \ref{prop:1class} describes $6$ solutions of \eqref{DiofEqn}. Proposition \ref{prop:4class} 
describes $6$ solutions for each of the $4$ sub-lattices, resulting in $24$ solutions in total. Each sub-lattice in 
Proposition \ref{prop:6class} contains $2$ solutions included in Proposition \ref{prop:1class} 
plus $4$ new solutions. This amounts to $4\cdot 6\cdot\dfrac{1}{2}s_2(\ell)$ solutions. Each sub-lattice in 
Proposition \ref{prop:8class} describes $6$ new solutions. This amounts to 
$6\cdot 8 \cdot\dfrac{1}{2}{\wh s}_2(\ell)$ solutions. Each sub-lattice in Proposition \ref{prop:12class} 
describes $6$ new solutions. This amounts to $6\cdot 12 \cdot\dfrac{1}{2}{\wt s}_2(\ell)$ solutions. If there 
exists at least one solution different from the ones listed above then it necessarily belongs to a class 
with $24$ sub-lattices. The RHS of \eqref{r} evaluates the amount of such solutions. This completes 
the proof of assertion (i). The proofs of (ii) and (iii) are straightforward as each class of $24$ sub-lattices 
has $24\cdot 6= 144$ solutions to \eqref{DiofEqn}. 
\ep

Due to the 1-1 correspondence \eqref{eq:bases} between the cubic $\ell$-sub-lattices and FCC 
$\ell$-sub-lattices in $\bbZ^3$ we arrive at the following theorem.

\bthm\label{thm:9.1}%{\bf Theorem 9.1.} {\sl 
The $D$-\rFCC\ sub-lattices in $\bbZ^3$ exist iff $D^2=2\ell^2$ where $\ell \in\bbN$. For $D^2=2\ell^2$,
these \rFCC\ 
$\ell$-sub-lattices are grouped into a finite number of disjoint classes, where each class contains 
$1,4,6,8,12$ or $24$ sub-lattices. The sub-lattices in a given class are obtained from each other via 
$\bbZ^3$-symmetries. The number of classes of a given cardinality and the coordinate representation 
of the basis of each sub-lattice in the class depend on  the rational prime decomposition of $\ell$ as 
detailed in Propositions \ref{prop:lZ3} - \ref{prop:24class}. Each \ \rFCC\ $\ell$-sub-lattice is $\bbA_3(z)$ for an integer
quaternion $z$ with $\|z\|^2=\ell$. 
\ethm

\bp
Follows directly from Propositions \ref{prop:lZ3} - \ref{prop:24class}.
\ep

%################################################################
%                                                                                                                        PS for hard-core potential
%####################################################################

\section{Appendix B: Hard-core potentials in the Pirogov-Sinai theory}\label{AppendixB}

The original Pirogov-Sinai theory (\cite{PiS}, \cite{Si}, \cite{Za}) and its various extensions (e.g., \cite{BS}) consider lattice 
models with a finite spin space and a finite potential of a finite range. Nevertheless, it is straightforward to extend this theory 
to hard-core models of a finite exclusion radius (which are models with an infinite potential).

Standard notions of a ground state and a periodic ground state are applicable to both finite-potential models and hard-core 
models of the above type. %The same is true for a stronger notion of a perfect configuration, which is applicable to $m$-potentials \cite{HS}, again of both types. The concept of an m-potential, without using this name, has been introduced in $\cite{HS}$ where the potential is assumed to be finite. Allowing the value $+\infty$ for 
The first fundamental assumption of the PS theory is that the model has a finite number of 
PGSs, which remains the requirement for the hard-core models under consideration.

Based on the above requirement, one can partition the entire lattice, say $\bbZ^n$, into cubes of side-length $l$ 
such that $l$ is larger than the interaction radius, and every PGS can fit into this cube considered as a torus. This 
construction is applicable to both finite-potential models and hard-core models.

Denote by $S$ the spin space of a model and by $C$ the lattice cube of side-length $l$. A convenient, though not required, 
step is to replace the spin space $S$ with the space $S^C$, which reduces the model to the one with a nearest-neighbor 
interaction. Here the nearest neighbors of a given lattice site are the sites at a distance not larger than $\sqrt{n}$ from this site. 
The constructed nearest-neighbor interaction takes finite values if the initial model has a finite-value potential. The 
original hard-core requirement translates into forbidding some pairs of spins at nearest-neighbor sites. Nevertheless, by 
construction, the PGSs (which are constant configurations in the reduced model) remain admissible. Without loss of generality 
we assume from now on that the spin space is a finite subset of $\bbN$, and the PGSs are constant configurations with the 
spin values $1,2,\ldots, k$ respectively (i.e., there are $k$ of them).

The next step in the PS theory is a definition of a {\it $q$-correct} point (\cite{Za}, p. 561) in a spin configuration $\phi$,
with $q\in\{1,\ldots ,k\}$. 
Note that after the change of the spin space the interaction radius equals $\sqrt{n}$. Consequently, the above definition 
implies that a $q$-correct point $x$ is a lattice site for which not only $\phi(x)= q$, but $\phi(x')= q$ for all nearest 
neighbors $x'$ of $x$ as well. This definition remains applicable to hard-core models. The same is true for the definition 
of a {\it contour} in a given configuration $\phi$, which is a connected component (in the sense of nearest neighbors) 
of non-correct points (cf. \cite{Za},  p. 561). Note that such a contour is a pair consisting of a connected component of 
points called a {\it support} of a contour and a spin configuration in this support. 

Any spin configuration in a finite volume with a PGS boundary condition is in a one-to-one correspondence with a finite 
collection of finite contours (called a {\it boundary} in \cite{Za}, p. 561). In particular, a configuration mapped into a 
single {\it $q$-contour} has a constant PGS-value $q$ in the {\it exterior} of the contour and some (possibly different) 
constant PGS-values $q_i$ in the $i$-th connected component of the {\it interior} of the contour. Here the exterior and 
the interior are defined as connected components of the complement of the contour support. (See (1.5) in \cite{Za}.)

Under this construction, a {\it relative energy} of a $q$-contour is defined as the difference between the energy of the 
PGS $q$ in the considered volume and the energy of the above configuration containing a single contour (see (1.6) in 
\cite{Za}). Without loss of generality one can assume that the pair potential is equal to $0$ for any pair of neighboring 
sites having the same PGS-values, and it is positive for any other pair of spins at neighboring sites (the value $+\infty$ 
is also allowed). Consequently, a  relative energy of a contour is simply the sum of pair interactions over the pairs of 
neighboring sites belonging to the support of the contour. An energy of a contour represents a {\it contour functional} 
(Section 1.6. in \cite{Za}) which, in turn, defines a {\it statistical weight} of a contour.
The rest of the argument in \cite{Za} can be applied verbatim. 

Essentially, the hard-core potential has an impact only upon the statistical weight of a contour. Namely, some contours 
have statistical weight $0$. Moreover, the arguments in \cite{Za} utilize only upper bounds on the absolute values of 
statistical weights of contours (see (1.9), (1.23), (1.50), (2.2), (2.28) in \cite{Za}). Finally, everything is reduced to 
the analysis of a polymer model (Section 2.1 in \cite{Za}) where a polymer is a 
connected lattice set, and the associated statistical weight is the sum over statistical weights of the contours having 
this set as their support. In the hard-core case not all configurations in a given support contribute to this sum, which 
only makes the sum smaller.

%###########################################################

%#################################################################
\bigskip
{\bf Acknowledgements}. We are grateful to Thomas Hales for answering a number of questions related 
to the proof of Kepler's conjecture. We thank Peter Sarnak for elucidating us on some important number 
theoretical results. IS and YS thank the Math Department, Penn State University, for hospitality
and support. A part of the work was carried out while IS was a Beaufort Visiting Fellow at St John's College, Cambridge. IS thanks St John's College, Cambridge and DPMMS, University of Cambridge, for the warm hospitality and support. YS thanks St John's College, Cambridge, for long-term support.

\end{document}